\documentclass[usenatbib]{mn2e}
\usepackage{graphicx}
\newcommand{\fracb}[2]{\left(\frac{#1}{#2}\right)}

\newcommand{\mean}[1]{\langle{#1}\rangle}
\newcommand{\text}[1]{\quad\mbox{#1}\quad}

\newcommand{\aap}{A\&A}

\newcommand{\gtrsim}{\lower.7ex\hbox{$\;\stackrel{\textstyle>}{\sim}\;$}}
\newcommand{\lesssim}{\lower.7ex\hbox{$\;\stackrel{\textstyle<}{\sim}\;$}}

\title[Impulsive Magnetic Relativistic Acceleration]
{Impulsive Acceleration of Strongly Magnetized Relativistic Flows}

\author[Granot et al.]
{Jonathan Granot$^{1}$ 
\thanks{j.granot@herts.ac.uk},
Serguei S.~Komissarov$^{2}$ 
\thanks{serguei@maths.leeds.ac.uk}
and Anatoly Spitkovsky$^{3}$
\thanks{anatoly@astro.princeton.edu}
\\
$^{1}$Centre for Astrophysics Research, University 
  of Hertfordshire, College Lane, Hatfield, Herts, AL10 9AB, UK
\\
$^{2}$Department of Applied Mathematics, The 
  University of Leeds, Leeds, LS2 9GT, UK
\\
$^{3}$Department of Astrophysical Sciences, Peyton Hall, 
  Princeton University, Princeton, NJ 08544, USA
}

\begin{document}
\date{Received/Accepted}
\maketitle

\begin{abstract}

The strong variability of magnetic central engines of AGN and GRBs may
result in highly intermittent strongly magnetized relativistic
outflows. We find a new magnetic acceleration mechanism for such
impulsive flows that can be much more effective than the acceleration
of steady-state flows. This impulsive acceleration results in
kinetic-energy-dominated flows that are conducive to efficient
dissipation at internal MHD shocks on astrophysically relevant
distances from the central source.  For a spherical flow, a discrete
shell ejected from the source over a time $t_0$ with Lorentz factor
$\Gamma\sim 1$ and initial magnetization $\sigma_0 =
B_0^2/4\pi\rho_0c^2\gg 1$ quickly reaches a typical Lorentz factor
$\Gamma \sim \sigma_0^{1/3}$ and magnetization
$\sigma\sim\sigma_0^{2/3}$ at the distance $R_0\approx ct_0$.  At this
point the magnetized shell of width $\Delta \sim R_0$ in the lab frame
loses causal contact with the source and continues to accelerate by
spreading significantly in its own rest frame. The expansion is driven
by the magnetic pressure gradient and leads to relativistic relative
velocities between the front and back of the shell. While the
expansion is roughly symmetric in the center of momentum frame, in the
lab frame most of the energy and momentum remain in a region (or
shell) of width $\Delta\sim R_0$ at the head of the flow. This
acceleration proceeds as $\Gamma \sim (\sigma_0R/R_0)^{1/3}$ and
$\sigma \sim \sigma_0^{2/3} (R/R_0)^{-1/3}$ until reaching a coasting
radius $R_c \sim R_0\sigma_0^2$ where the kinetic energy becomes
dominant: $\Gamma \sim \sigma_0$ and $\sigma
\sim 1$ at $R_c$. Then the shell starts coasting and spreading
(radially), its width growing as $\Delta\sim R_0(R/R_c)$, causing its
magnetization to drop as $\sigma\sim R_c/R$ at $R>R_c$.  Given the
typical variability time-scales of AGN and GRBs, the magnetic
acceleration in these sources is a combination of the
quasi-steady-state collimation acceleration close to the source and
the impulsive (conical or locally quasi-spherical) acceleration
further out.  The interaction with the external medium, which can
significantly affect the dynamics, is briefly addressed in the
discussion.

\end{abstract}


\section{Introduction}
\label{sec:introduction}

The first questions raised by the discovery of astrophysical jets are
how they are powered, collimated, and accelerated.  Most of them --
jets from young stars, Active Galactic Nuclei (AGN), Galactic X-ray
Binaries, and Gamma Ray Bursts (GRBs), are associated with disk
accretion\footnote{The only exceptions are the jets of Pulsar Wind
Nebulae as there are no indications of accretion disks around their
pulsars.  These jets are most likely not produced directly by the
pulsars but instead form downstream of the termination shock of pulsar
winds~\citep{L02,KL04}.}, and this suggests that accretion disks are
essential for jet production. The astrophysical jets seem to be highly
supersonic as many of their features are nicely explained by internal
shocks.  In the laboratory, highly collimated supersonic jets are
normally produced when a high pressure (and temperature) gas escapes
from a chamber via a finely designed nozzle.  However, it seems highly
unlikely that such refined ``devices'' are formed naturally in
astrophysical systems. They would require cold and dense gas to form
the walls of the chamber with a massive compact object in the center
\citep{BR74}, but such configurations are highly unstable 
\citep{NSSW81,SSNW83}. This has lead to the idea that the collimation 
of astrophysical jets may have a completely different mechanism
involving a strong magnetic field.  Although this magnetic field still
needs to be confined within a channel, the conditions on its geometry
are less restrictive. If this field is anchored to a rotating object,
such as an accretion disk, then it naturally develops an azimuthal
component. The hoop stress associated with this magnetic field
component creates additional collimation of the flow within the
channel. Moreover, this leads to a magnetic torque being applied to
the rotating object and thus a natural way of powering outflows
by tapping the rotational energy of the central object.

In order to produce a relativistic flow this way, the magnetic energy
per particle must exceed its rest energy. Thus, the jet plasma must be
highly rarefied. Such rarefied plasma is naturally produced only in
the magnetospheres of black holes and neutron stars.  Moreover, the
strong magnetic field shields these magnetospheres and prevents them from being
contaminated by the much denser surrounding plasma.  In contrast,
young stars can eject a lot of mass from their surface and this seems
to explain why their jets are not relativistic.  Magnetospheres of
accretion disks are likely to be heavily mass-loaded and are not able
to produce relativistic jets for the same reason.

It has to be stressed that magnetic flows must still be collimated
externally until they become super-fast-magnetosonic. The magnetic hoop
stress can result in self-collimation of the inner core but cannot
prevent sideways expansion of the outer sheath. However, when the flow
becomes super-fast-magnetosonic, the speed of this lateral expansion
becomes smaller than the flow speed along the jet direction, and the
jet remains collimated.  For non-relativistic jets the condition of
passing through the fast-magnetosonic surface also implies almost
completed acceleration of the flow (50\% conversion of magnetic energy
into kinetic energy). In contrast, the relativistic jets still remain
Poynting-flux dominated at this point and the acceleration process may
continue well into the super-fast-magnetosonic regime.

The issue of the efficiency of energy conversion (from magnetic to
kinetic form) is related to the issue of subsequent energy
dissipation, which is required in order to explain the observed
electromagnetic emission from both the jets and the structures they
create when they collide with the external medium. Traditionally, one
of the most favorite channels of dissipating the energy of supersonic
flows has been the formation of shock waves. However, in the case of
relativistic flows this mechanism can be much less efficient if the
flow is Poynting-flux dominated. First of all, it is the kinetic
energy of the flow that is dissipated\footnote{This applies to fast
magnetosonic shocks. At a slow magnetosonic shock, the magnetic energy
dissipates as well and the kinetic energy can actually
increase. However, slow shocks are much less robust and harder to
generate compared to the fast ones.}, and if only a small fraction of
the total energy is in the kinetic form then this already severely
limits the efficiency of dissipation. Secondly, the compression ratio
and hence the fraction of kinetic energy that dissipates also decrease
with increasing magnetization. Thus, in order to dissipate a
significant fraction of the available energy the flow should not only
become super-fast-magnetosonic, but it should also become dominated by
kinetic energy before it is shocked
\citep{Leis05,MGA09,MA10}.
             
The magnetic acceleration of relativistic flows has been the subject
of theoretical research for decades. The main focus of this research
has been on the models of steady-state axisymmetric dissipation-free
flows (the ``standard model''). The main reason behind this is
simplicity. Only in this case was there a hope of building a rigorous
theory. Yet, even this idealized model is rather complex, and
solutions could be found only if an additional symmetry,
e.g. self-similarity, or other simplifying condition was introduced
\citep[e.g.][]{BL92,VK03,BN06}.  More recently the problem was
approached using numerical methods
\citep{KBVK07,KVKB09}. 

There are a number of problems with the standard model, which are most
severe in the case of a spherical wind. In this case the theory
predicts an asymptotic Lorentz factor of $\Gamma \sim \sigma_0^{1/3}$,
where $\sigma_0 = B_0^2/4\pi\rho_0c^2 \gg 1$ is the initial
magnetization parameter, which determines the maximum possible Lorentz
factor corresponding to a total conversion of the Poynting flux into
the bulk motion kinetic energy in a steady-state flow
\citep[e.g.,][]{GJ70}. This is in conflict with the observations of
many astrophysical sources. In particular, the high observed values of
$\Gamma$ in many sources would require an extremely large initial
magnetization $\sigma_0$ that would in turn imply a very high
asymptotic magnetization, $\sigma \sim \sigma_0^{2/3} \gg 1$, making
it impossible to achieve efficient shock dissipation within the
outflow.

A potential way to overcome this problem is by resorting to collimated
outflows. This can increase the asymptotic value of $\Gamma$ and
reduce that of $\sigma$ by up to a factor of $\sim\theta_{\rm
jet}^{-2/3}$, where $\theta_{\rm jet}$ is the asymptotic half-opening
angle of the jet.  The collimation has to be strong enough to preserve
causal connectivity across the flow (in the lateral direction). The
faster the flow and the higher its fast-magnetosonic Mach number
becomes, the smaller its opening angle should be. By the time one
half of the Poynting flux is converted into kinetic energy
($\sigma\sim 1$), the jet half-opening angle $\theta_{\rm jet}$ should
not exceed $\theta_{\rm max} = 1/\Gamma$, where $\Gamma \sim \sigma_0$
is the jet Lorentz factor at that time.  Observations of AGN jets do
indeed show that $\theta_{\rm jet}< 1/\Gamma$ ~\citep{P09}.  However,
for GRB jets with $\Gamma\simeq 400$ (or $10^2\la\Gamma \la 10^{3.5}$)
this constraint gives $\theta_{\rm max}\simeq 0.14^\circ$ (or
$0.018^\circ\la\theta_{\rm max} \la 0.57^\circ$), which is much
smaller compared to generally accepted values of the half-opening
angle, $2^\circ \la\theta_{\rm jet} \la 30^\circ$ \citep{FWK00,PK01}.

In addition, the standard theory of GRB afterglow emission can explain
the jet-break in their light curves only if $\theta_{\rm jet}\Gamma
\gg 1$ \citep{Rhoads99,SPH99}. Although the Swift observations show 
that clear jet breaks are not as common as we used to think
\citep[e.g.,][]{Liang08}, this might be at least partly due to 
observational selection effects (Swift GRBs are dimmer on average as
Swift is more sensitive than previous missions), and there are still
some clear cases for jet breaks in the Swift era. Finally, late time
radio afterglow observations, when the flow becomes sub-relativistic,
provide fairly robust (no longer susceptible to strong relativistic
beaming) lower limits \citep[e.g.,][]{EW05} on the true energy that
remains in the afterglow blast wave at that time, of a few to several
times $10^{51}\;$ergs \citep{BKF04,Frail05}. Such a large true energy,
together with the inferred energy per solid angle in the prompt
gamma-ray emission and in the afterglow shock at early times imply
that the initial jet half-opening angle cannot be too small (typically
not much less than a few degrees).

It turns out that a transition from laterally confined to ballistic
flow is accompanied by a relatively short phase of acceleration of a
different kind \citep{KVK09,TNM09}.  Such a transition may occur in
the collapsar model at the stellar surface.  A sudden loss of lateral
pressure support causes a sideways expansion of the jet.  If the jet
is highly relativistic at the stellar surface the corresponding
increase in the jet opening angle is negligible.  However, a
rarefaction wave propagates into the jet and brings it out of lateral
balance. The magnetic pressure force accelerates the flow in the
lateral direction, which results in a significant increase of the jet
Lorentz factor, particularly in the outer layers of the jet. This may
alleviate the $\theta_{\rm jet}\Gamma\simeq 1$ problem of the magnetic
model. However, as soon as the rarefaction crosses the jet it is well
in the ballistic regime and the acceleration is over.\footnote{This is
in contrast with the highly robust mechanism of thermal acceleration,
where for an adiabatic index of $\gamma=4/3$ the jet Lorentz factor
grows linearly with the jet radius, $\Gamma\propto R$, even in the
ballistic regime.} Moreover, it does not ensure full conversion
of electromagnetic to kinetic energy. Should, it happen a bit too soon
and the jet remains Poynting-dominated.  Even under the best of
circumstances the resultant jet magnetization is still close to
$\sigma\simeq 1$, which is too high for effective shock
dissipation \citep{Leis05,MGA09,MA10}.

Given the problems with this basic case, other ideas have been put
forward. The most radical idea is to assume that relativistic
astrophysical jets do not become kinetic energy dominated but remain
Poynting dominated on all scales and that the observed emission comes
not from shocks but from magnetic dissipation cites \citep{B02,Lt06}.
In the context of the present work this may potentially serve as an
alternative to internal shocks in cases where for some reason the
magnetization remains high at large distances from the source.  Others
propose various ways of increasing the efficiency of magnetic
acceleration compared to the basic model, e.g., via allowing
non-axisymmetric instabilities and randomization of magnetic field
\citep{HB00}. In fact, the magnetic dissipation may also help the
transition from Poynting dominated to kinetic energy dominated states
\citep{Dre02,DS02}.

In this work we focus on the acceleration of an impulsive (strongly
time-dependent) highly magnetized relativistic outflow, which has
received relatively little attention so far. \citet{C95} was first to consider
the non-relativistic case of impulsive magnetic acceleration and
dubbed it an ``astrophysical plasma gun''. The relativistic version
presents a number of qualitatively different properties.  
In \S~\ref{sec:test_case} we present a detailed study of a
simplified test case featuring a cold and initially highly magnetized
($\sigma_0\gg 1$) one dimensional finite shell (of initial width
$l_0$) initially at rest (at $t = 0$), whose back end leans against a
``wall'' and with vacuum in front of it. The initial evolution
(\S~\ref{sec:ss-phase} and Appendix~\ref{app:self-sim}) is described
by a self-similar rarefaction wave traveling toward the wall and
accelerating the Poynting-dominated flow away from the
wall. At the end of this phase, at time $t_0\approx l_0/c$ when the
rarefaction wave reaches the wall, the mean Lorentz factor of the flow
is $\mean{\Gamma} \sim \sigma_0^{1/3}$.  Soon after $t_0$ the shell
separates from the wall and moves away from it
(\S~\ref{sec:ref-phase}). The shell continues to accelerate and keeps
an almost constant width of $\sim 2l_0$. Using both numerical
(\S~\ref{sec:num}) and analytical
(\S~\ref{sec:ref-phase},~\S~\ref{sec:after_separation} and
Appendixes~\ref{app:an-int}, \ref{sec:acc2}) methods, we find that
during the second phase the mean Lorentz factor grows as
$\mean{\Gamma} \sim (\sigma_0 t/t_0)^{1/3}\propto t^{1/3}$.  This
phase ends at time $t_c=t_0\sigma_0^2$, when the acceleration slows
down and the shell starts coasting. At this point $\mean{\Gamma}
\sim \sigma_0$ and $\sigma \sim 1$.  In \S~\ref{sec:back_envelope} we
present crude but simple derivations of the main results of
\S~\ref{sec:test_case} that allow us to understand the underlying
physics and show that the results are robust -- not very sensitive
to the exact initial configuration. The analysis of the coasting phase
(\S~\ref{sec:coast+summary}) shows that at $t>t_c$ the shell width
increases as $\Delta \sim 2l_0t/t_c \propto t$ while its magnetization
decreases as $\sigma \sim t_c/t \propto t^{-1}$, resulting in a
kinetic energy-dominated flow. 

In \S~\ref{sec:BM-effect} we address the apparent paradox of
self-acceleration -- how can the shell keep accelerating after it
separates from the wall? We analyze a variation of our simple test
case in which the wall is removed when the rarefaction wave reaches it
(at $t_0$). At subsequent times there are no external forces on the
system, implying that the center of momentum (CM) velocity or Lorentz
factor ($\Gamma_{\rm CM}$) remain constant and there is no global
acceleration at $t>t_0$ in this strict sense. Nevertheless, even
though we find that $\Gamma_{\rm CM}\sim\sigma_0^{1/2}$ remains
constant, the more relevant astrophysical quantity is the mean value
of $\Gamma$ weighted over the energy in the lab frame,
$\langle\Gamma\rangle_E$, and it indeed increases as
$\langle\Gamma\rangle_E \sim (\sigma_0t/t_0)^{1/3}$ at $t_0<t<t_c$.
In \S~\ref{sec:dis} we discuss the connection between our test case
and relativistic astrophysical flows and study the possible
implications of our impulsive acceleration mechanism for the dynamics
of GRB and AGN jets. We also briefly address the interaction of the
magnetized flow with the external medium for GRBs. Our main results
and conclusions are presented in \S~\ref{sec:conclusions}.

Soon after the first version of our paper had appeared on the
electronic archive (http://arxiv.org/archive/astro-ph/), an
independent study of impulsive magnetic acceleration was published
there as well \citep{Lt10a,Lt10b,LL10}, indicating growing interest in
this mechanism.  Where the covered topics overlap, the results of both
studies agree very well. As to the differences, their study focuses on
the initial phase of fast acceleration (at $t<t_0$) and shock
formation (when instead of pure vacuum the shell expands into a
rarefied plasma), whereas the main subject of our paper is the
operation of the impulsive acceleration mechanism after
the shell separates from the ``wall'' (at $t>t_0$).

\section{Test Case: Expansion of a Magnetized Shell into Vacuum}
\label{sec:test_case}

A good way of demonstrating the basic dynamics of the acceleration of
a highly magnetized impulsive flow is to start with a simple example
that can be analyzed analytically or using simple one-dimensional
simulations. To this aim we consider for our initial conditions a
uniform shell of width $l_0$ with high initial magnetization,
$\sigma_0 = B_0^2/4\pi\rho_0 c^2 \gg 1$, where $B_0$ is the initial
magnetic field and $\rho_0$ is the initial rest mass density. We
choose Cartesian coordinates in which the plane of the shell is
perpendicular to the $x$-axis and the magnetic field is aligned with
the $y$-axis. The right boundary of the shell is at $x=0$ and the left
one is at $x=-l_0$.  To the left of the shell is a solid conducting
wall and to the right is vacuum.

\subsection{Self-similar rarefaction phase}
\label{sec:ss-phase}

At time $t=0$ we let the shell expand into vacuum. This is a well
known problem that describes a simple rarefaction wave propagating
into the shell towards the wall. The self-similar simple wave solution
to the general case with non-vanishing thermal pressure is described
in Appendix~\ref{app:self-sim}. Here we focus only on the cold limit
(with no thermal pressure; the equations describing this case reduce
to those of the pure gas case with an adiabatic index $\gamma=2$).

Using units where $c=1$, the local wave speed is
\begin{equation}\label{lambda}\hspace{3.0cm}
\lambda = \frac{v-c_{\rm ms}}{1-vc_{\rm ms}}\ ,
\end{equation} 
where $ c_{\rm ms} $ is the fast magnetosonic speed as measured in the
fluid frame.\footnote{This is simply the Lorentz transformation of a
velocity component parallel to the relative velocity of two inertial
frames.}  In our (cold) limit
\begin{equation}\label{mspeed}\hspace{3.0cm}
  c^2_{\rm ms} = \frac{\sigma}{1+\sigma}\ ,
\end{equation}
where 
$$
   \sigma=\frac{B'^{\,2}}{4\pi\rho} = \frac{(B/\Gamma)^2}{4\pi\rho}
$$
is the local magnetization parameter, while $B'=B/\Gamma$ and $\rho$
are the magnetic field and the rest mass density, respectively, as
measured in the fluid rest frame. The equations of one-dimensional
motion yield
\begin{equation}\hspace{3.0cm}
\label{B-comove}
   \frac{B'}{\rho}=\frac{B}{\Gamma\rho}=\mbox{const}\ ,
\end{equation}
(see Appendix~\ref{app:self-sim}) and thus 
\begin{equation}\hspace{3.4cm}
\label{sigma}
   \sigma=\sigma_0\frac{\rho}{\rho_0}\ .   
\end{equation}

The backward characteristics of the simple wave (where the wave moves
in the direction opposite to that of the flow) are straight lines
described by
\begin{equation}\hspace{3.0cm}
\label{sv}
\xi = \lambda = \frac{v-c_{\rm ms}}{1-vc_{\rm ms}}, 
\end{equation}
where $\xi=x/t$ is the self-similar variable. 
Integration of the self-similar flow equation gives (see
Eq.~[\ref{eq:hydro_const}] for $\gamma = 2$ or
Eqs.~[\ref{eq:mag_const}] and [\ref{eq:c_ms}] for $a_0 = 0$),
\begin{equation}\hspace{2.7cm}
\frac{1+v}{1-v}\left(\frac{1+c_{\rm ms}}{1-c_{\rm ms}}\right)^2 = \mathcal{J}_+\ ,
\label{int}
\end{equation}
where
$$
\mathcal{J}_+
= \left(\frac{1+c_{\rm ms,0}}{1-c_{\rm ms,0}}\right)^2 =
\left(\sqrt{\sigma_0}+\sqrt{\sigma_0+1}\right)^4
\approx 16\sigma_0^2\ ,
$$
where the last equality holds for $\sigma_0\gg 1$. This equation, in
combination with Eqs.~(\ref{mspeed}) and (\ref{sigma}), allows to find
$\rho=\rho(v)$ and then Eq.~(\ref{B-comove}) gives $B=B(v)$. Finally,
Eq.~(\ref{sv}) allows us to find the dependence of all flow variables
on $\xi$.

\begin{figure}
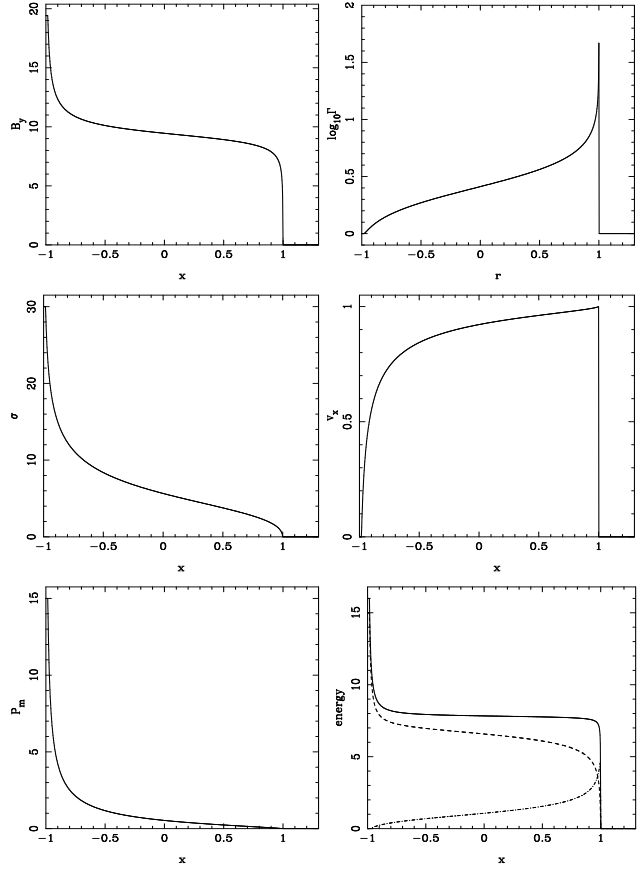

\includegraphics[width=37mm,height=41mm,angle=-90]{byb.eps}
\includegraphics[width=37mm,height=41mm,angle=-90]{lorb.eps}
\vspace{0.15cm}\\
\includegraphics[width=37mm,height=41mm,angle=-90]{sigb.eps}
\includegraphics[width=37mm,height=41mm,angle=-90]{vb.eps}
\vspace{0.15cm}\\
\hspace{-0.04cm}
\includegraphics[width=37mm,height=41mm,angle=-90]{pb.eps}
\hspace{0.02cm}
\includegraphics[width=37mm,height=40mm,angle=-90]{t00b.eps}
\caption{The self-similar rarefaction wave solution at $t=1$, using units of
$l_0 = 1$, $\rho_0 = 1$ and $c=1$. The initial conditions are a
uniform state with parameters $\sigma_0=30$ and $v_0=0$ at $-1<x<0$
and vacuum for $x>0$. Shown are the magnetic field $B_y$ ({\it top
left panel}), the Lorentz factor $\Gamma$ (as measured in the wall
frame; {\it top right panel}), the local magnetization parameter,
$\sigma=(B')^2/4\pi\rho$ ({\it middle left panel}), the flow velocity
$v_x$ ({\it middle right panel}), the magnetic pressure $p_m=(B')^2/8\pi$
({\it bottom left panel}) and the density of total energy ({\it solid
line}), magnetic energy ({\it dashed line}), and kinetic energy ({\it
dash-dotted line}) as measured in the wall frame ({\it bottom right
panel}).}
\label{fig:self-similar}
\end{figure}

Figure~\ref{fig:self-similar} shows the self-similar solution for
$\sigma_0=30$, in units where $\rho_0 = l_0 = c =1$, at time $t=1$
(when the left front of the rarefaction wave is about to reach the
wall). One can see that both the left and the right fronts of the wave
propagate at very close to the speed of light. The magnetic field and
the total energy density distributions in the expanding shell are
almost uniform (except for the thin boundary layers). This is expected
as the plasma inertia is very low and the electromagnetic part of the
solution must be close to the corresponding solution of the Maxwell
equations. Near the right front the distributions of most flow
parameters exhibit large gradients. In the plots of the Lorentz factor
and total kinetic energy density we see narrow spikes. The maximum
value of the Lorentz factor can be found from Eq.~(\ref{int}) by
setting $c_{\rm ms}=0$.  For $\sigma_0\gg 1$ we find
\begin{equation}\hspace{3.2cm}
  \Gamma_{\rm max} \approx 2\sigma_0\ . 
\end{equation}      
This is already a very high speed. However, only a very small fraction
of the flow energy is associated with this spike and the mean Lorentz
factor is much lower.  Figure~\ref{fig:self-similar} suggests that the
mean Lorentz factor must be close to that of the sonic point, $\xi=0$,
for which Eq.~(\ref{int}) gives (for $\sigma_0\gg 1$)
\begin{equation}\hspace{3.0cm}
  \Gamma(\xi=0) \approx \fracb{\sigma_0}{2}^{1/3}\ .
\end{equation}
More sophisticated averaging procedures (such as the weighted averages
over the energy or rest mass) described in
Appendix~\ref{app:average_Gamma} give values which are only slightly
higher (see right panel of Fig.~\ref{fig:gamma_av_t0}) and show that
\begin{equation}\hspace{3.2cm}
  \mean{\Gamma} \simeq \sigma_0^{1/3}
\end{equation}
is a very good estimate.

\subsection{Evolution after separation from the wall}
\label{sec:ref-phase}

At the time $t = t_0 = l_0/c_{\rm ms}(v=0)\approx l_0$ (where we still
use units of $c=1$) the left front of the rarefaction wave reaches the
wall, and then the evolution of the shell changes. A secondary
rarefaction wave is launched from the wall and propagates to the
right, trying to catch up with right front of the original
wave. However, both fronts propagate with speeds very close to the
speed of light, and the spatial separations separation between them
changes only very slowly -- to the first approximation it is $\approx
2l_0$. At $t < t_0$ the original rarefaction wave does not ``know''
about the existence of the wall, and therefore behaves according to
the self-similar solution for a semi-infinite shell. At $t > t_0$,
however, this is true only ahead of the reflected rarefaction wave, at
$x > x_*(t)$ or $\xi > \xi_*(t) = x_*(t)/t$, where $x_*(t)$ is the
location of the front of the secondary rarefaction, ($x_*(t_0) =
-l_0$). At $x > x_*(t)$ the fluid continues to be accelerated by the
pressure gradient created during the initial expansion.

\begin{figure}
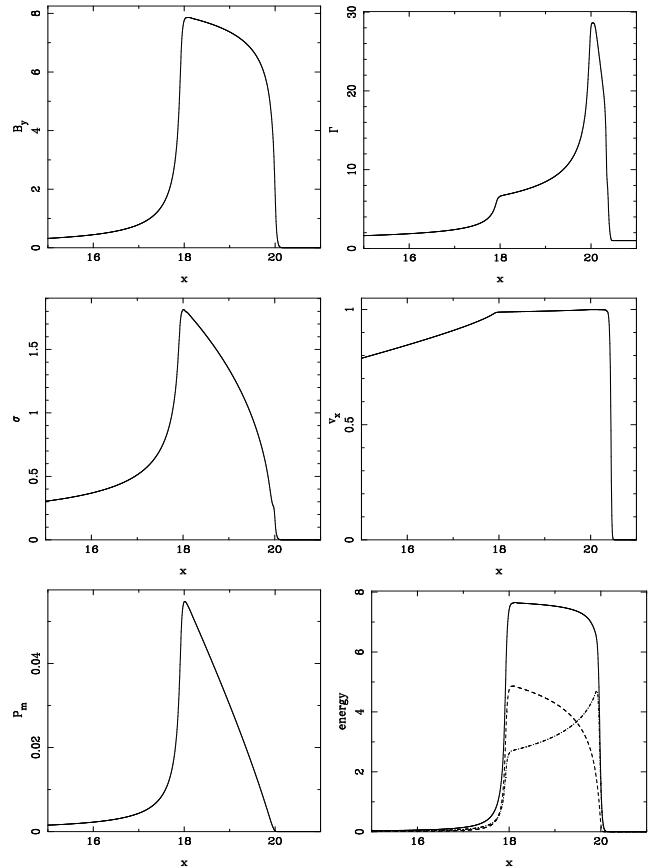

\includegraphics[width=37mm,height=41mm,angle=-90]{bya.eps}
\includegraphics[width=37mm,height=41mm,angle=-90]{lora.eps}
\vspace{0.15cm}\\
\includegraphics[width=37mm,height=41mm,angle=-90]{siga.eps}
\includegraphics[width=37mm,height=41mm,angle=-90]{va.eps}
\vspace{0.15cm}\\
\includegraphics[width=37mm,height=41mm,angle=-90]{pa.eps}
\hspace{0.7mm}
\includegraphics[width=37mm,height=41mm,angle=-90]{t00a.eps}
\caption{Propagation of a highly magnetized cold shell of plasma.
The plots describe the numerical solution at time $t=20$ for the same
initial data as in Fig.~\ref{fig:self-similar} and use the same units.
The {\it top panels} show the magnetic field $B_y$ ({\it top left
panel}), the Lorentz factor $\Gamma$, (as measured in the wall frame;
{\it top right panel}), the local magnetization parameter,
$\sigma=(B')^2/4\pi\rho$ ({\it middle left panel}), the flow velocity
$v_x$ ({\it middle right panel}), the magnetic pressure $p_m=(B')^2/8\pi$
({\it bottom left panel}), and the density of total energy ({\it solid
line}), magnetic energy ({\it dashed line}), and kinetic energy ({\it
dash-dotted line}) as measured in the wall frame ({\it bottom right
panel}). The front of secondary rarefaction is located at $x\simeq
18$.}
\label{fig:numeric}
\end{figure}

At $x < x_*(t)$, however, inside the secondary wave, the density and
pressure drop very rapidly and the fluid is decelerated by the strong
magnetic pressure gradient that develops just behind the head of this
wave. Moreover, the total rest mass in this region is very small and
one can describe the shell evolution as a separation from the wall.
This is in contrast with the non-relativistic version of this problem
considered by \citet{C95}, where there is no such separation and the
flow pressure and density peak at the wall.\footnote{Thus, the
relativistic dynamics of magnetized shell is even closer to the
``plasma gun'' action and also brings to mind the hypothetical
``phasers'', all too familiar to the fans of the science-fiction
series ``Star Trek''.} The Lorentz factor and kinetic energy, as
measured in the lab frame, drop strongly behind the front of this wave
and only the part of the initial flow that is not yet affected by the
right rarefaction wave significantly contributes to the total
energetics (see Fig.~\ref{fig:numeric}).  Therefore, the ``typical''
or mean (as averaged over the energy in the lab frame) Lorentz factor
of the shell should behave as the fluid Lorentz factor $\Gamma(\xi_*)$
at the front of the right rarefaction wave (or the back boundary of
the shell).

In the fluid frame the front of secondary rarefaction moves with the
local magnetosonic speed. In the lab frame this corresponds to
\begin{equation}\label{xs*}\hspace{1.8cm}
\beta_* \equiv \frac{dx_*}{dt} = 
\frac{v(\xi_*)+c_{\rm ms}(\xi_*)}{1+v(\xi_*)c_{\rm ms}(\xi_*)}\ .
\end{equation}
In the ultra-relativistic accelerating regime, where $v\simeq 1$ and
$c_{\rm ms}\simeq 1$ (where the latter requirement insures that there
is still plenty of magnetic energy to drive the acceleration:
$\langle\sigma\rangle \sim
\sigma(\xi_*) \gg 1$), it is more convenient to work with the
corresponding Lorentz factors, $\Gamma = (1-v^2)^{-1/2}$ and
$\Gamma_{\rm ms} = (1-c_{\rm ms}^2)^{-1/2}$, using the approximation
\begin{equation}\label{vLor}\hspace{1.7cm}
   v \approx 1-\frac{1}{2\Gamma^2}\ ,\quad 
   c_{\rm ms} \approx 1-\frac{1}{2\Gamma_{\rm ms}^2}\ .  
\end{equation}
Substituting these into Eq.~(\ref{int})  and Eq.~(\ref{sv})  yields 
\begin{equation}\label{WaW}\hspace{3.2cm}
\Gamma_{\rm ms}^2 \approx \frac{\sigma_0}{2\Gamma}\ .
\end{equation}
and
\begin{equation}\label{eq:xi_2}\hspace{2.7cm}
 \xi \approx \frac{(\Gamma/\Gamma_{\rm ms})^2-1}{(\Gamma/\Gamma_{\rm ms})^2+1}\ . 
\label{sWW}
\end{equation}
Combining  Eq.~(\ref{WaW}) with Eq.~(\ref{eq:xi_2}) we then obtain
\begin{equation}\label{Ws}\hspace{2.7cm}
\Gamma^3 \approx \frac{\sigma_0}{2}\left(\frac{1+\xi}{1-\xi}\right)\ . 
\end{equation}
The final step is to find $\xi_*=\xi_*(t)$ and substitute the result
into Eq.~(\ref{Ws}). In fact, in the ultra-relativistic
regime Eq.~(\ref{xs*}) yields  
\begin{equation}\hspace{0.5cm}
\label{dxdt}
\beta_* = \frac{dx_*}{dt}\approx  1-\frac{1}{8\Gamma^2(\xi_*)\Gamma_{\rm ms}^2(\xi_*)} 
\approx 1-\frac{1}{4\sigma_0\Gamma(\xi_*)}\ .
\end{equation}
When $\Gamma \ll \sigma_0$ this can be simply approximated as $\beta_*
= dx_*/dt \approx 1$, which gives us
\begin{equation}\label{xi_*}\hspace{1.2cm}
x_* \approx t - 2l_0\ ,\quad  \xi_* = \frac{x_*}{t} \approx 1 - \frac{2t_0}{t}\ ,
\end{equation}
(see Appendix~\ref{app:raref-front}).
Substituting this result in Eq.~(\ref{Ws})
we finally obtain 
\begin{equation}\hspace{1.0cm}
\label{final}
 \Gamma(\xi_*)=\left(\frac{\sqrt{\mathcal{J}_+}\,t}{8t_0}\right)^{1/3}
\approx \left(\frac{\sigma_0\,t}{2t_0}\right)^{1/3} \propto t^{1/3}\ .
\end{equation}
As a self-consistency check we note that since $\Gamma_* =
(1-\beta_*^2)^{-1/2} \gg 1$, then $\beta_* \approx 1-1/2\Gamma_*^2$
and equations (\ref{dxdt}) and (\ref{final}) imply
\begin{equation}\hspace{0.8cm}
\label{Gamma_*}
\beta_* \approx 1-\fracb{32\sigma_0^4t}{t_0}^{-1/3}\ ,\quad 
\Gamma_* \approx \fracb{4\sigma_0^4 t}{t_0}^{1/6}\ ,
\end{equation}
which  upon integration of $\beta_*$ yields 
\begin{equation}\hspace{0.6cm}
\label{xi*3}
\xi_* \approx 1-\frac{2t_0}{t}
\left[1+\frac{3}{2^{11/3}}\fracb{t}{\sigma_0^2t_0}^{2/3}\right] 
\approx 1-\frac{2t_0}{t}\ ,
\end{equation}
thus confirming the validity of Eq.~(\ref{xi_*}) for $t \ll
\sigma_0^2t_0$.

\begin{figure}
\begin{center}
\includegraphics[width=74mm,height=50mm,angle=0]{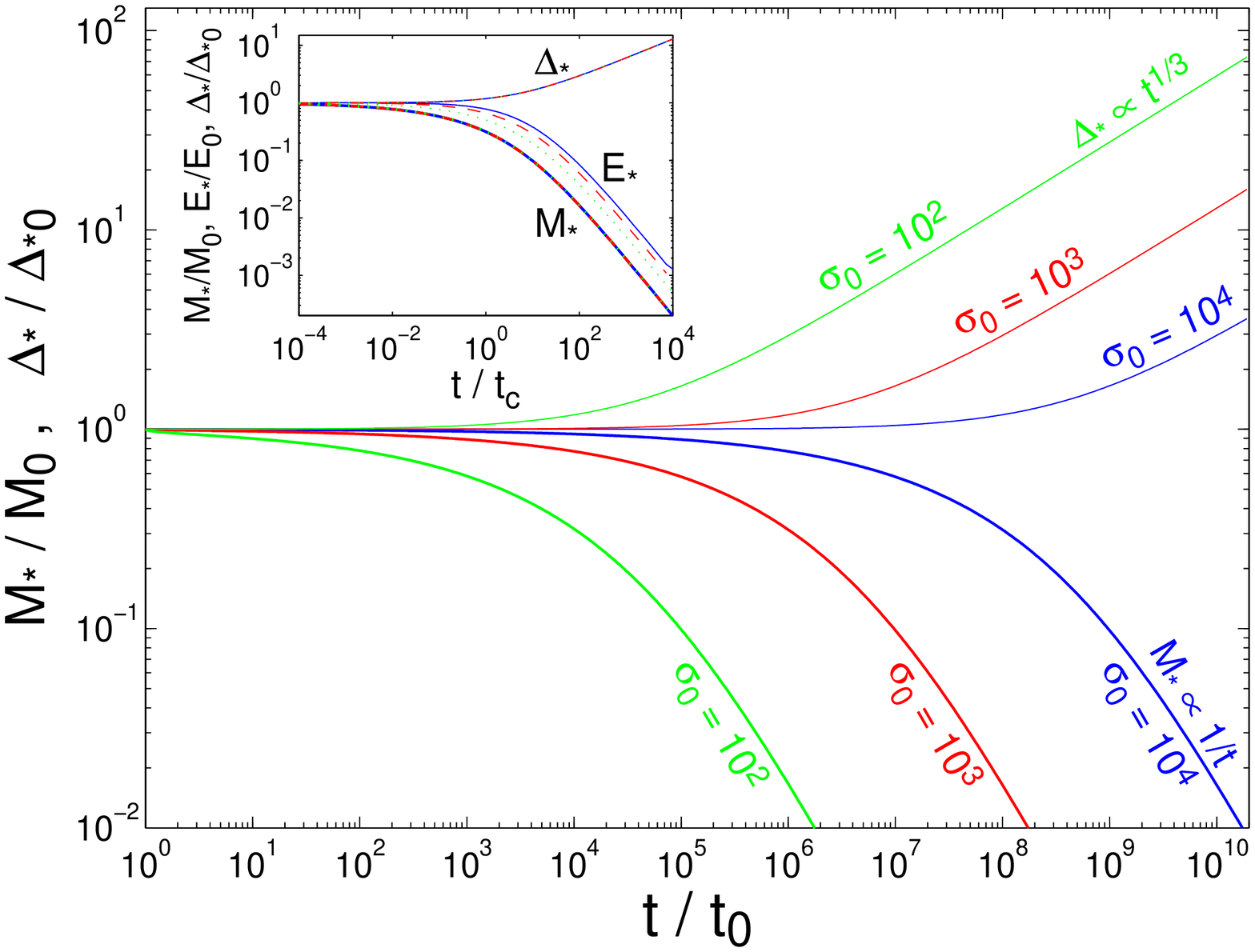}
\vspace{0.2cm}
\\
\includegraphics[width=74mm,height=50mm,angle=0]{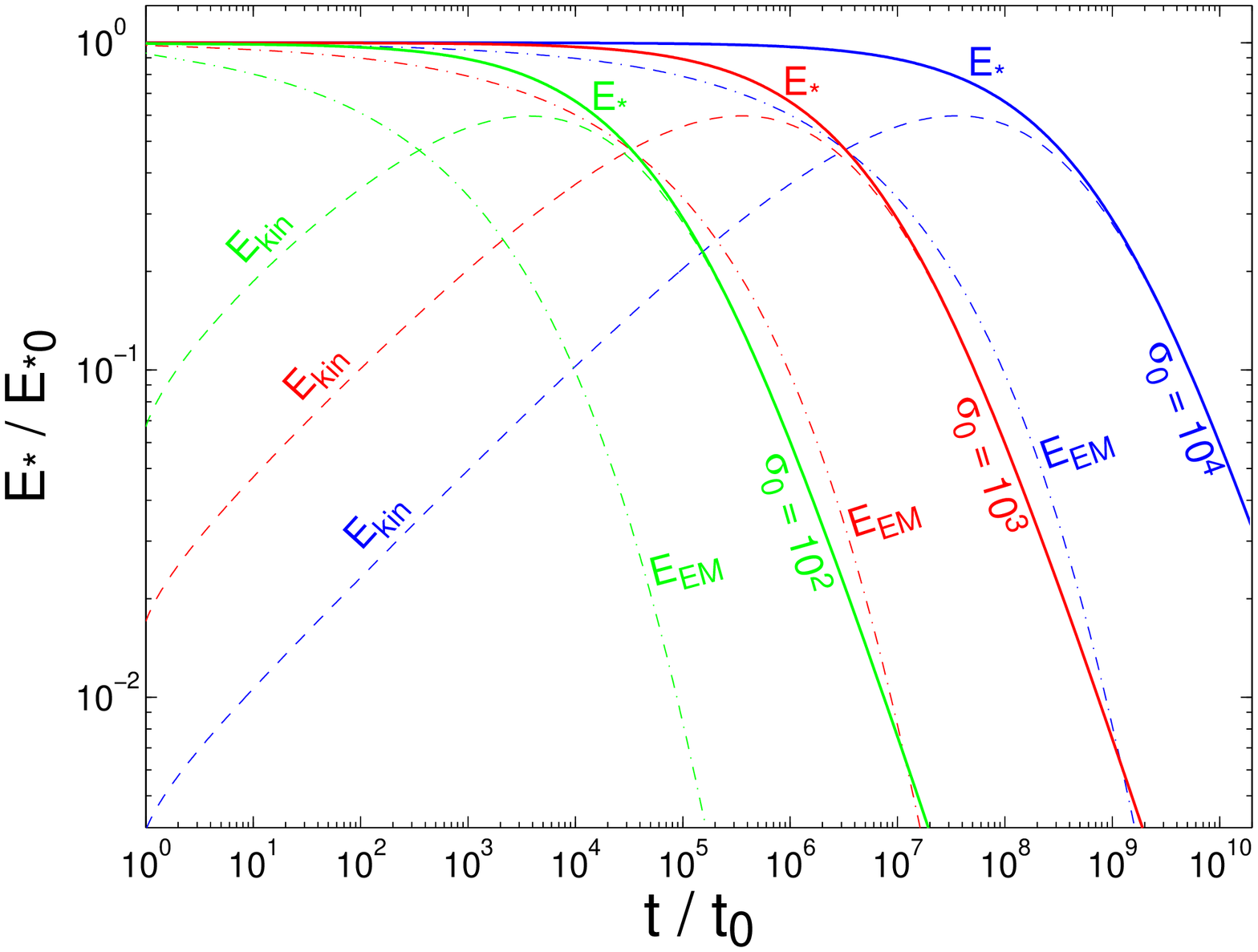}
\vspace{0.2cm}
\\
\includegraphics[width=74mm,height=50mm,angle=0]{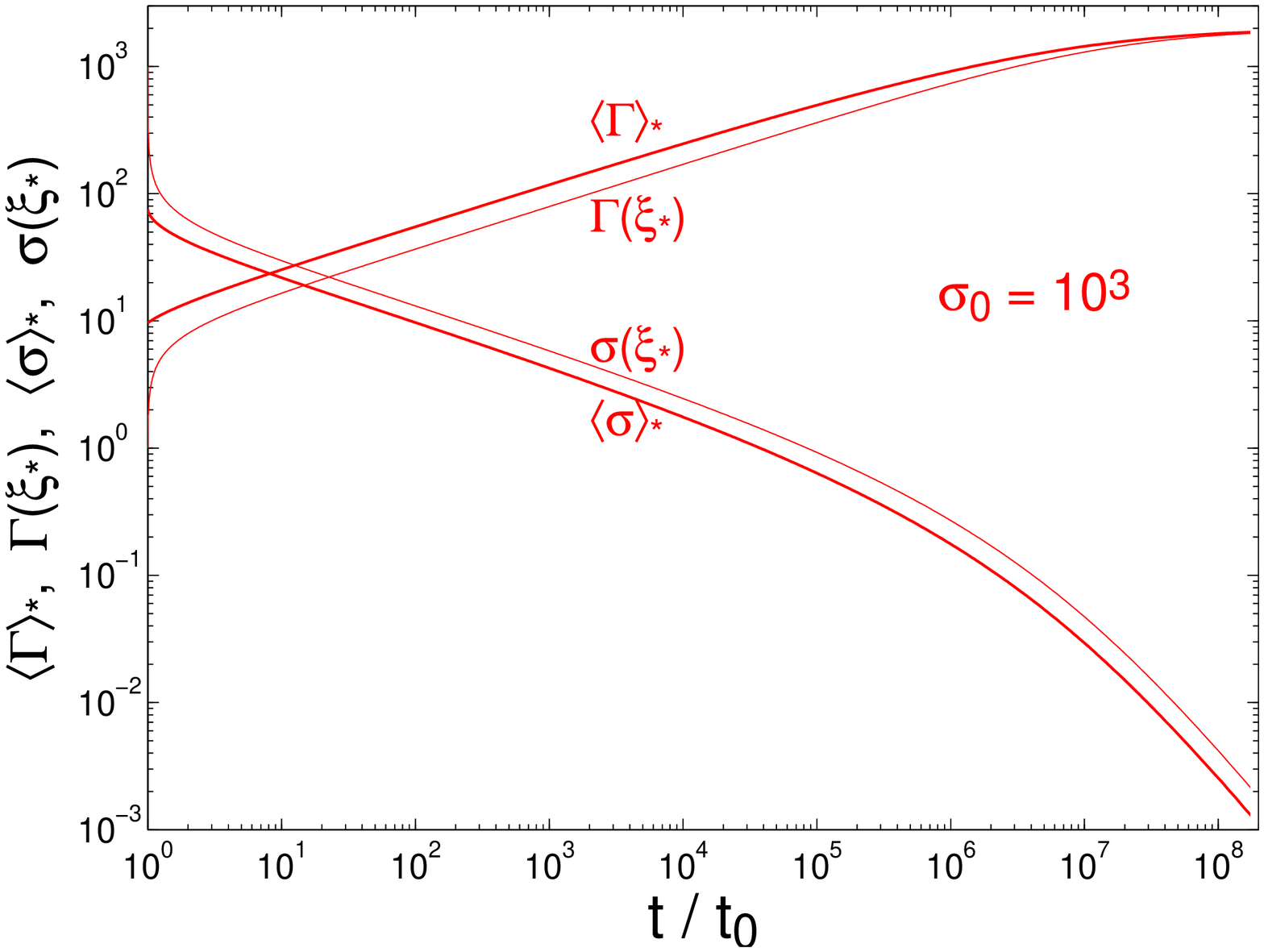}
\end{center}
\caption{Evolution of the shell, corresponding to the region between 
the front of secondary rarefaction wave ($\xi_*$) and vacuum interface
($\xi_h$), for three different values of the initial magnetization:
$\sigma_0 = 10^2$ ({\it green lines}), $\sigma_0 = 10^3$ ({\it red
lines}), and $\sigma_0=10^4$ ({\it blue lines}). The {\it top panel}
shows the width of this region, $\Delta_*$, and the rest mass $M_*$
within it, both normalized by their initial values at $t=t_0$ (when
the original rarefaction wave is secondary by the wall), as a function
of $t/t_0$. The magnetic flux, $\int Bdx$, has exactly the same
evolution as the total rest mass. The inset shows these quantities as
well as the total energy (kinetic+magnetic) $E_*$ within this region,
normalized by its initial value at $t=t_0$, $E_{*0} = E_{\rm EM,0} =
(\sigma_0/2)M_0c^2$, as a function of $t/t_c$ (where $t_c =
\sigma_0^2 t_0$); $\sigma_0 = 10^2,\,10^3,\,10^4$ are plotted with
dotted, dashed, and solid lines, respectively. The curves for
$M_*/M_0$ and $\Delta_*/\Delta_{*0}$ are practically on top of each
other, while those for $E_*/E_{*0}$ are slightly offset,
indicating a slower convergence in the limit $\sigma_0\to\infty$. The
{\it middle panel} shows the evolution of $E_*/E_{*0}$ ({\it
thick solid lines}), and its decomposition into kinetic ({\it dashed
lines}) and electromagnetic ({\it dashed-dotted lines}) energies. The
{\it bottom panel} shows for $\sigma_0 = 10^3$ the evolution of the
average values (weighted by energy -Eq.~[\ref{Gamma_av1}]) of $\Gamma$
($\langle\Gamma\rangle$) and $\sigma$ ($\langle\sigma\rangle$) within
this region ({\it thick lines}), as well as their values at the head
of the secondary rarefaction wave ($\xi_*$; {\it thin lines}).}
\label{fig:shell-int-an}
\end{figure}

Therefore, in this regime the mean Lorentz factor of the shell follows
the law $\mean{\Gamma} \propto t^{1/3}$.  Moreover, for $t = t_0$
Eq.~(\ref{final}) gives $\Gamma(\xi_*)
\sim \sigma_0^{1/3}$ in agreement with the results obtained in
\S~\ref{sec:ss-phase}. Thus, we may conclude that
\begin{equation}\hspace{3.0cm}
\label{final-1}
 \mean{\Gamma} \approx \left(\frac{\sigma_0\,t}{t_0}\right)^{1/3}.
\end{equation}
This regime continues until the magnetic and kinetic energies become
comparable (and $\Gamma_{\rm ms}\simeq 1$), which implies
$\mean{\sigma}\simeq 1$ and $\mean{\Gamma} \simeq \sigma_0$.  This
occurs at the time
\begin{equation}\hspace{3.4cm}
\label{eq:coasting-time}
t_c=t_0 \sigma_0^2\ ,  
\end{equation}
after which the shell starts coasting at a constant Lorentz factor
$\mean{\Gamma} \simeq \sigma_0$ (as described in \S~\ref{sec:coast+summary}). 

In Appendix~\ref{app:raref-front} we provide an alternative derivation
of Eq.~(\ref{final}), based on the explicit solution of the
self-similar rarefaction wave. Furthermore, analytic expressions are
derived for the rest mass $M_*$, kinetic energy $E_{\rm kin}$,
electromagnetic energy $E_{\rm EM}$, and total energy (excluding rest
energy) $E_*$, in the region between the head of the secondary
rarefaction wave and the vacuum interface: $\xi_*(t) <
\xi < \xi_h = 2[\sigma_0(1+\sigma_0)]^{1/2}/(1+2\sigma_0)$, as a
function of $\xi_*(t)$. Together with equation (\ref{t(xi_*)}) for
$t(\xi_*)$ these quantities can be parametrically expressed as a
function of the time $t$, and are presented in
Fig.~\ref{fig:shell-int-an} for $\sigma_0 = 10^2,\, 10^3,\, 10^4$
(presumably covering the range of values most relevant for GRBs).
Similarly, we also derive the average values (weighted over the
energy, according to Eq.~[\ref{Gamma_av1}]) of $\Gamma$
($\langle\Gamma\rangle_*$) and $\sigma$ ($\langle\sigma\rangle_*$)
within this region, which are shown in the bottom panel of
Fig.~\ref{fig:shell-int-an} for $\sigma_0=10^3$, along with
$\Gamma(\xi_*)$ and $\sigma(\xi_*)$.

\begin{figure}
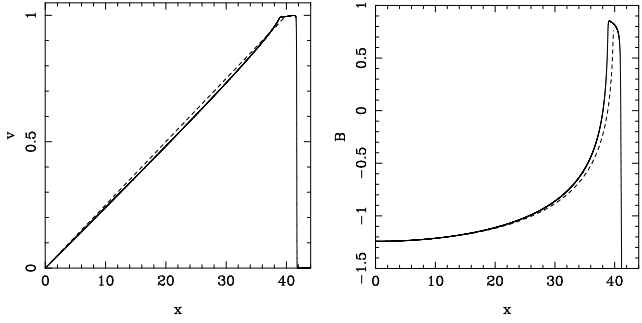

\includegraphics[width=41.5mm,angle=-90]{ss-v.eps}
\hspace{0.1cm}
\includegraphics[width=41.5mm,angle=-90]{ss-b.eps}
\caption{Numerical and self-similar solutions for shell's ``tail''.
The left panel shows the velocity and the right panel the magnetic
field $B$.  The numerical solution is represented by solid lines and
the self-similar solution (Eq.~\ref{similarity}) by dashed
lines. The problem parameters and units are the same as in
Fig.~\ref{fig:self-similar}.  The time is $t=40$ and the numerical
solution is shifted along the x axis so that the wall is now located
at $x=0$.  }
\label{fig:tail}
\end{figure}

One can see that up to the time $t\sim t_c = \sigma_0^2t_0$ the shell
width, $\Delta_*$, and total energy $E_*$ remain almost unchanged,
with $\Delta_*\approx 2l_0$ and $E_*$ being equal to the total
(excluding rest energy) energy in the initial solution, $E_{*0} =
E_{\rm EM,0} = (\sigma_0/2)M_0c^2$. At the same time, the shell's
total mass (and magnetic flux) slowly decrease due to the gradual
advance of the secondary rarefaction into the shell. At $t \ll t_c$,
$\langle\Gamma\rangle_*$ is slightly larger than $\Gamma(\xi_*)$ (with
the same scaling $\propto t^{1/3}$), as expected, while
$\langle\sigma\rangle_*$ is slightly lower than $\sigma(\xi_*)$ even
for $t \gg t_c$.

The shell's magnetic energy is gradually converted into its kinetic
energy: $E_{\rm kin} \sim \langle\Gamma\rangle M_0c^2 \sim
E_0(t/t_c)^{1/3}$ at $t\ll t_c = \sigma_0^2t_0$), as
$\mean{\Gamma}\approx (\sigma_0t/t_0)^{1/3}$ in this regime.  At
$t\simeq 0.03t_c$, when these energies become comparable, while
$\sigma(\xi_*)$ and $\langle\sigma\rangle_*$ drop below unity at $t/t_c
\approx 0.086$ and 0.0037, respectively. At $t>t_c$ the shell begins
to experience significant spreading ($\Delta_*/\Delta_{*0} \approx
2^{-7/3}3(t/t_c)^{1/3}$ at $t \gg t_c$).  Its total mass and
energy significantly decrease, indicating that the region between
vacuum and the secondary rarefaction no longer represents the shell
evolution.

The numerical solution presented in Fig.~\ref{fig:numeric} suggests
self-similar evolution with characteristic linear profile for the flow
velocity, $v \simeq x/t$, for the region between the wall and the
shell (we will refer to this region as the shell's tail).  
This is expected in the limit where the separation between the shell
and the wall becomes much larger compared to $l_0$, the only
characteristic length scale of the problem.  As shown in
Appendix~\ref{app:self-sim:tail}, such similarity solution does
exist, 
\begin{equation}\hspace{0.8cm}
\label{similarity}
v = \xi\ ,\quad \rho = \frac{1}{t}\frac{C_1}{\sqrt{1-\xi^2}}\ ,\quad   
B = \frac{1}{t}\frac{C_2}{1-\xi^2}\ ,
\end{equation}   
where $\xi=x/t<1$, and $C_i$ are constants. Figure~\ref{fig:tail}
compares the similarity solution with the numerical solution at $t=40$
($\sigma_0=10^3$). One can see that there is a reasonably good
agreement between them. The first equality in Eq.~(\ref{similarity})
shows that each fluid element moves with constant speed.  This implies
that the kinetic energy for any section $[\xi_1,\xi_2]$ of the
solution is conserved. However, the magnetic energy of such a section
decreases as $\propto t^{-1}$. This indicates that the magnetic energy
is transferred along the solution towards $\xi=1$, where this solution
is no longer applicable (as $\xi=1$ implies $\Gamma=\infty$). In order
to confirm this conclusion consider a conserved $Q$ that satisfies
equation
$$ 
\frac{\partial Q}{\partial
t} + \frac{\partial F}{\partial x} = 0.  
$$ 
Next consider a fluid element bounded by $x_1=\xi_1 t$ and $x_2 =\xi_2
t$.  The amount of $Q$ held by this element, 
$$
Q(\xi_1,\xi_2,t)=\int\limits_{\xi_1 t}^{\xi_2 t} Q(x,t)dx, 
$$
satisfies the equation 
$$ 
\frac{d}{dt} Q(\xi_1,\xi_2,t) = \bar{F}(\xi_1,t)-\bar{F}(\xi_2,t) 
$$
where 
$$
\bar{F} = F-\xi Q 
$$ 
is the flux of $Q$ through the boundary moving with speed $\xi$.  For
the energy, 
$$ 
\bar{F}_e = \frac{b^2}{2}v = \frac{(B/\Gamma)^2}{8\pi}v
= \frac{C_2^2}{8\pi t^2} \frac{\xi}{1-\xi^2}\ ,  
$$ 
which represents the work per unit area and time done by the
fluid behind $\xi$ on the fluid ahead of $\xi$ (the force per unit
area is simply the magnetic pressure, $f = b^2/2$, and thus $dW = fdx
= (b^2/2)vdt$). This is positive and monotonically increasing
function of $\xi$, which implies transport of energy through the tail
towards the shell ($\xi=1$), in the direction of motion of the flow.
Clearly this is due to the work done by the magnetic pressure during
the tail's spreading. In the tail's head this energy is presumably
converted into the kinetic energy.

At late times after most of the magnetic energy is transformed
into kinetic energy, this solution may still reasonably describe the
tail of the flow, corresponding to ballistic motion at
$\Gamma\ll\sigma_0$. It implies that in the tail there is
approximately equal rest mass per decade in $\Gamma$: $dM/d\ln\Gamma =
C_1/\beta$, $dM/d\ln u = C_1\beta$ and
$$ 
M(<\beta) = \frac{C_1}{2}\ln\left(\frac{1+\beta}{1-\beta}\right) =
C_1\ln\left[\Gamma(1+\beta)\right] \ , 
$$
and equal energy per unit 4-velocity, $dE/du = C_1$ or $E(<u) =
dE/d\ln u = C_1 u \propto u$, so that most of the energy is carried by
the fastest material. That is, deep in the tail there is a good part
of the total rest mass but a very small fraction of the total energy.
%

\subsection{Numerical simulations}
\label{sec:num}

In order to test the validity of our conclusions we have carried out
numerical simulations for the evolution of a cold finite shell,
initially highly magnetized and at rest, as it expands into vacuum. We
numerically integrate the relativistic magneto-hydro-dynamic (RMHD)
Eqs.~(\ref{eq:RMHD1}-\ref{eq:RMHD2}) in the cold limit, where the gas
pressure is set to zero. As shown in Appendix~\ref{app:spherical}, the
equations in spherical coordinates can be reduced to the planar case,
so it suffices to find the solution in the Cartesian one-dimensional
geometry.

\begin{figure*}
\includegraphics[width=16cm,height=8.4cm]{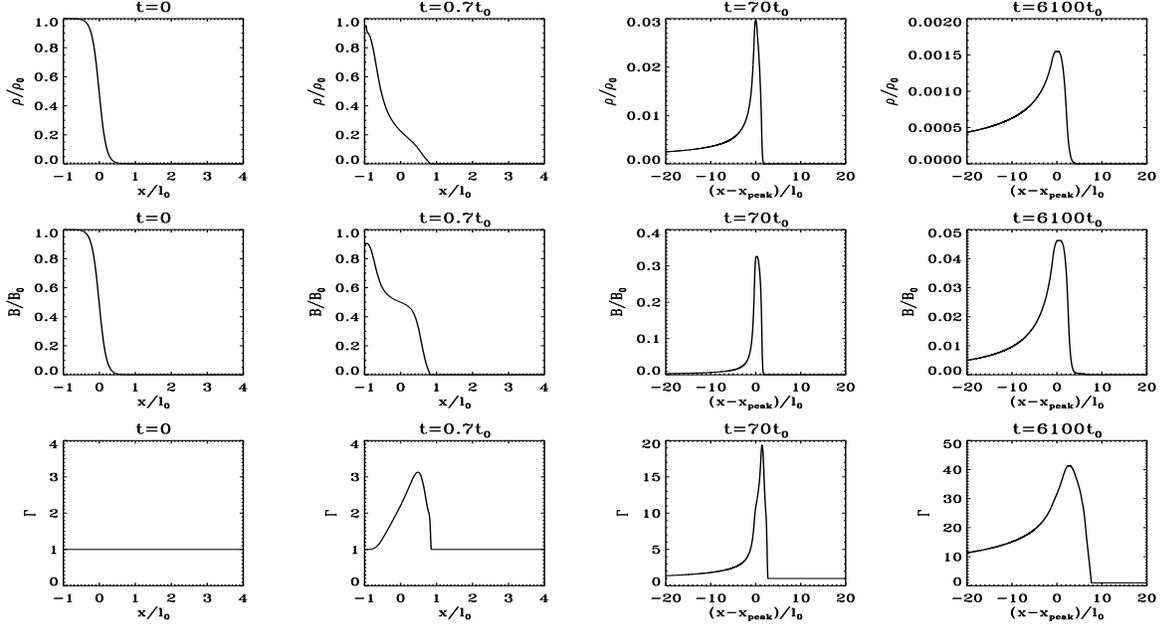}
\caption{Snapshots of physical quantities from the numerical
simulation of the evolution of a highly magnetized shell. Top row:
density; middle row: magnetic field; bottom row: Lorentz factor. Each
column corresponds to different times. Density and magnetic field are
normalized by $\rho_0$ and $B_0$ such that $B_0^2/4\pi\rho_0 c^2=\sigma_0$. In
the third and fourth columns, in order to follow the moving shell, the
x-coordinate is centered on the location $x_{\rm peak}$ of the peak of the
density of the shell.
}
\label{fig:sim1}
\end{figure*}
                                                                                         
\noindent{\bf Short term evolution:}
first, in order to validate our analytic treatment in
\S~\ref{sec:ref-phase} we used the exact same initial conditions as in our
analytic test case, namely a perfectly uniform, cold and highly
magnetized shell at rest.  At $t=0$ the shell occupies the region
$-l_0<x<0$, where at $x=-l_0$ it is bounded by a solid wall and in the
region $x>0$ there is vacuum.  The magnetic field is aligned with the
$y$-direction. We have used the initial magnetization of
$\sigma_0=30$.

In agreement with our analytic analysis the shell separates from the
wall at dimensionless time $t \approx t_0$ when its thickness in the
lab frame is $\Delta \approx 2l_0$. After this time the solution can
be described as a shell of constant thickness $\Delta \approx 2l_0$
followed by a low energy tail (see Fig.~\ref{fig:numeric}, which shows
the solution at $t=20t_0$).  In the tail of the flow, the velocity
$v_x$ grows linearly with $x$ as predicted in the self-similar
solution (see Appendix~\ref{app:self-sim:tail}).

\noindent{\bf Long term evolution:}
next we set out to test the long term evolution.  We used slightly
modified initial conditions: a shell of width $l_0$ with roughly
constant density and magnetic field, corresponding to a constant
magnetization of $\sigma_0=30$, whose back end touches a reflecting
wall on the left (at $x=-l_0$) and is tapered off to vacuum with a
hyperbolic tangent profile on the right over a thickness
$l_0/10$. That is, at $t=0$ and $x>-l_0$ we have $\rho/\rho_0 =
(B/B_0)^2 = [1-\tanh(10x/l_0)]/2$ and $\sigma = B^2/4\pi\rho c^2 =
B_0^2/4\pi\rho_0c^2 = \sigma_0$. We use a simple second-order accurate
Harten, Lax, and van Leer (HLL) scheme with Runge-Kutta third-order
time integration for the numerical algorithm. The resolution is $100$
cells per $l_0$, and the Courant number is $0.25$.  To follow the
evolution of the relativistically moving shell for long times without
enlarging the grid, we implemented a ``moving window" algorithm, where
all quantities are shifted to the left by $c \Delta t_{\rm shift}$
cells every $\Delta t_{\rm shift}=200$ time steps. Thus, the
simulation frame effectively flies to the right at the speed of light,
and the left wall becomes causally disconnected from the main domain.
The moving window algorithm turns on after the shell moves away from
the reflecting wall by about $70 l_0$. The size of the moving window
domain is $10^4$ cells corresponding to $100l_0$.

\begin{figure}
\hspace{-0.2cm}
\includegraphics[width=90mm]{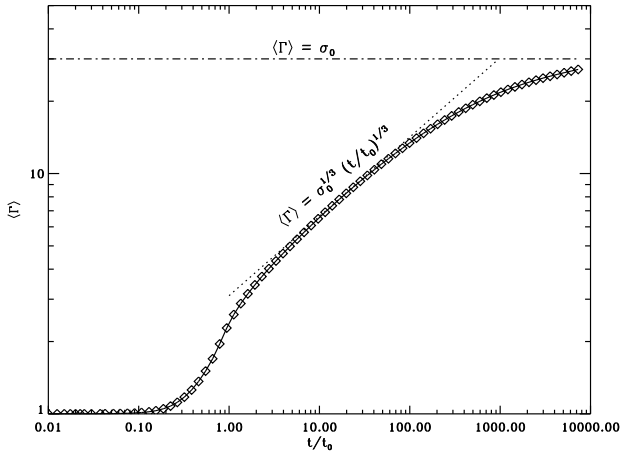}
\vspace{-0.4cm}
\caption{Time evolution of the energy-weighted average Lorentz factor
of the shell, showing the rarefaction wave, magnetic acceleration and
saturation stages.
}
\label{fig:sim2}
\end{figure}

Fig.~\ref{fig:sim1} shows the profiles of density, magnetic field and
Lorentz factor at several times during the simulation, while
Fig.~\ref{fig:sim2} shows the evolution of the energy-weighted average
Lorentz factor (defined in Eq.~\ref{Gamma_av1}),
$\langle\Gamma\rangle$, with time. We measure time and space in units
of the shell crossing time $t_0=l_0/c$ and initial shell widths $l_0$,
respectively. As expected, the evolution has several distinct
phases. First, the rarefaction wave propagates towards the reflecting
wall, as seen in the first two columns of Fig.~\ref{fig:sim1}. The
right end of the shell accelerates, and $\langle\Gamma\rangle$ reaches
$\sigma_0^{1/3}$ when the rarefaction wave crosses the shell at
$t=t_0$. At this point the shell decouples from the wall. As seen in
Fig.~\ref{fig:sim2}, at $t=t_0$ the evolution of
$\langle\Gamma\rangle$ changes to the accelerating stage which takes
it beyond $\sigma_0^{1/3}$, increasing as $t^{1/3}$. In this regime,
the shell remains thin ($\sim 2l_0$, third column in
Fig. \ref{fig:sim1}), leaving a low-density tail behind. This is the
``impulsive'' stage, where the right part of the shell
accelerates at the expense of the magnetic ``exhaust'' on the left. The
{\it dotted line} in Fig.~\ref{fig:sim2} shows the analytical
expectation during this stage, $\langle\Gamma\rangle=\sigma_0^{1/3}
(t/t_0)^{1/3}$, for the parameters of the simulation. The agreement
during the accelerating stage is very good. 

In the saturation (or coasting) stage, which starts around
$t/t_0\ga\sigma_0^2$, the shell starts to spread significantly in
the lab frame (last column in Fig.~\ref{fig:sim1}). The evolution of
$\langle\Gamma\rangle$ deviates from the earlier $t^{1/3}$ power-law
and begins to approach the asymptotic value $\langle\Gamma\rangle =
\sigma_0$ ({\it dash-dotted line} in Fig.~\ref{fig:sim2}), corresponding to
the complete conversion of magnetic to kinetic energy in the
shell.\footnote{In this asymptotic limit, $\langle\Gamma\rangle_M \to
1+\sigma_0/2$ for the mass-weighted average defined in
Eq.~(\ref{Gamma_av2}), while the exact asymptotic value for the
energy-weighted average $\langle\Gamma\rangle_E$ defined in
Eq.~(\ref{Gamma_av1}) depends on the asymptotic distribution of
$dM/d\Gamma$, which in turn depends on the exact initial
conditions. It is nonetheless always $\sim\sigma_0$.} In the far
asymptotic regime, the moving window of the simulation which flies at
the speed of light begins to outrun the shell, which moves with finite
Lorentz factor.  Thus, the last points in the evolution in
Fig.~\ref{fig:sim2} can be affected by the fact that a significant
fraction of the shell material is left outside the moving
window. However, the trend for saturation is clear. Overall, our
simulations support very well the analytical arguments about the
rarefaction wave, impulsive acceleration and the saturation (or
coasting) stages of the evolution of an impulsive flow. We have also
experimented with larger values of $\sigma_0=100,\, 1000$ of the
shell. We find that the $t^{1/3}$ evolution is robust and is seen in
both of these cases; however, we did not run the simulations long
enough to see the ultimate saturation, as the saturation time is much
longer, scaling as $\sigma_0^2$. We also checked that the evolution is
not sensitive to the exact shape of the initial shell.

\section{``Back of the envelope'' derivations}
\label{sec:back_envelope}
                                                                                         
In this section we re-derive the key results of previous sections
using crude but simple calculations which help to clarify its physics.
They also show that this phenomenon is rather generic and not very
sensitive to exact initial configuration.

\subsection{Initial acceleration}
\label{sec:initial_acc}

As the shell expands and the flow develops, the electromagnetic part 
of the solution closely follows that of vacuum electrodynamics. 
The shell (electromagnetic pulse) thickness increases by a factor of two, 
from $l_0$ to $l_1 \approx 2l_0$ and the magnetic field decreases by a factor of two, 
from $B_0$ to $B \approx B_0/2$. At the same time, an electric field $E\approx B$ is 
generated. Since the flow is still highly magnetically dominated the energy 
conservation implies 
$$
B_0^2l_0 \sim (E^2+B^2)l_1 \sim 2 B^2 l_1\ . 
$$
On the other hand the mass conservation reads 
$$
   \rho_0 l_0 \sim \rho \Gamma l_1\ , 
$$
where $\rho$ and $\Gamma$ are the characteristic (``mean'') density and Lorentz 
factor. From this we find that 
$$
\frac{B^2}{4\pi\rho\Gamma} \sim \frac{\sigma_0}{2}\ .
$$
Since the fluid frame magnetic field $B'=B/\Gamma$ this gives
$$
\Gamma \sigma \sim \frac{\sigma_0}{2}\ . 
$$ 
From the MHD viewpoint the shell separates from the wall (i.e. looses
causal contact with it) when its Lorentz factor just exceeds that of
the fast magnetosonic speed, given by equation~(\ref{mspeed}), which
corresponds to a 4-velocity $u_{\rm ms} = \sigma^{1/2}$.  For
$\sigma\gg 1$ this reads
$$
 \Gamma \approx \sigma^{1/2}\ .
$$
Combining the last two equations we find the anticipated results that
at $t_0$ when the shell separates from the wall,
$$
\mean{\Gamma} \sim \sigma_0^{1/3} \quad\mbox{and}\quad 
\mean{\sigma} \sim \sigma_0^{2/3}\ . 
$$
It is easy to see that these calculations are not sensitive to geometry 
and apply equally well to planar, spherical and cylindrical shells with 
a tangential magnetic field.

\subsection{Acceleration after the separation}
\label{sec:after_separation}

After the separation from the wall the total momentum of the shell no
longer increases and it is mainly in electromagnetic form.  However,
the shell plasma (corresponding to the front of the flow) continues to
be accelerated by the magnetic pressure gradient that has developed
during the first phase. (Although, in the laboratory frame the
magnetic field is almost uniform the magnetic pressure is given by the
strength of magnetic field in the comoving frame $B'=B/\Gamma$, which
is non-uniform.) Similarly, the plasma at the back of the flow (inside
the secondary rarefaction wave that develops and propagates into the
back of the shell) in decelerated by the magnetic pressure gradient
there.

Magnetic flux conservation implies that $Bl \approx {\rm const}$,
where $l$ is the shell width. Therefore, the electromagnetic energy
scales as
$$
E_{\rm EM} \propto B^2l \propto l^{-1}\ ,
$$
and thus it decreases significantly when the $l$ increases
significantly, say doubles its initial value of $l(t_0) = l_1 \approx
2l_0$. Since the part of the shell carrying most of the energy has a
spread in the Lorentz factor of the order of $\Delta\Gamma(t)\sim\Gamma(t)$
around its typical value, $\Gamma(t)$, it spreads such that its width
grows as
\begin{equation}\label{eq:l_shell}\hspace{2.0cm}
l \sim l_1 + \frac{t-t_0}{\Gamma^2(t)} \sim l_1 +\frac{t}{\Gamma^2(t)}\ ,
\end{equation}
where we use units of $c=1$ and the last approximate equality holds
for $t \gg t_0$ (factors of order unity are dropped for
simplicity). Now, $l$ increases significantly at the time $t_c$ when
the two terms on the r.h.s of the above equation become comparable,
$$
\frac{t_c}{\Gamma^2(t_c)} \sim l_1 \sim t_0\ .
$$
Since $t_c$ is also the time when the electromagnetic energy decreases
significantly, we know that at $t_c$ we must have $\sigma(t_c) \sim 1$
and $\Gamma(t_c) \sim \sigma_0$, regardless of the value of $t_c$,
which we want to derive here. Thus, we find that
$$
t_c \sim t_0\Gamma^2(t_c) \sim t_0\sigma_0^2\ .
$$
We have already derived in the previous subsection that $\Gamma(t_0)
\sim\sigma_0^{1/3}$ and therefore if indeed $\Gamma$ increases as a
power-law with time $t$ between $t_0$ and $t_c$ (which is the only
viable option) then the power law index must be
$$
\frac{d\log\Gamma}{d\log t} = \frac{\log[\Gamma(t_c)/\Gamma(t_0)]}{\log(t_c/t_0)} 
= \frac{\log(\sigma_0^{2/3})}{\log(\sigma_0^2)} = \frac{1}{3}\ .
$$
Thus we obtain the anticipated scaling $\Gamma \propto t^{1/3}$ at
$t_0 < t < t_c$. Since $\Gamma(t_0) \sim \sigma_0^{1/3}$ this implies
$$
\Gamma(t_0<t<t_c) \sim \fracb{\sigma_0t}{t_0}^{1/3}\ .
$$
An alternative derivation is provided in
Appendix~\ref{sec:acc2}. Thus, the scalings obtained for the test case
of initially uniform shell are in fact rather generic.

\subsection{Coasting phase and summary of main results}
\label{sec:coast+summary}

At $t > t_c$ the flow essentially becomes unmagnetized (i.e. with a
low magnetization, $\sigma < 1$), its internal (magnetic) pressure
becomes unimportant dynamically, and each fluid element within the
shell coasts at a constant speed (ballistic motion). The shell coasts
at a typical Lorentz factor of $\Gamma \sim \sigma_0$, where the
expansion of the shell during its acceleration stage results in a
dispersion $\Delta\Gamma \sim \Gamma$ in its Lorentz factor (i.e. that
of the part of the shell carrying most of the energy) around this
value. This causes an increase in the shell width in the lab frame,
according to Eq.~(\ref{eq:l_shell}), where at $t>t_c$ the second term
on the r.h.s becomes dominant, resulting in
\begin{equation}\hspace{2.3cm}
\frac{l}{l_1} \approx \frac{l}{2l_0} \sim \left\{\matrix{1
\quad & \zeta_c < 1\ ,\cr\cr 
\zeta_c \quad & \zeta_c > 1\ ,}\right. 
\end{equation}
where $\zeta_c = t/t_c \approx R/R_c$, while $t_c = t_0\sigma_0^2$ and
$R_c \approx t_c$ are the coasting time and radius,
respectively. Since $E_{\rm EM} \propto l^{-1}$ and at $t>t_c$ $E_{\rm
kin}\approx E = {\rm const}$, then $\sigma = E_{\rm EM}/E_{\rm
kin}\propto l^{-1} \propto t^{-1}$.  One can summarize this result in
terms of the lab frame time or the distance $x \approx ct$ of the
shell from the wall (or source), either in terms of $\zeta_c$,
\begin{equation}\label{eq:summary1a}\hspace{1.5cm}
\langle\Gamma\rangle \sim \left\{\matrix{\sigma_0\zeta_c^{1/3} 
\quad & \sigma_0^{-2} < \zeta_c < 1\ ,\cr\cr 
\sigma_0 \quad & \zeta_c > 1\ ,}\right. 
\end{equation}
\begin{equation}\label{eq:summary1b}\hspace{1.5cm}
\langle\sigma\rangle \sim \left\{\matrix{\zeta_c^{-1/3} 
\quad & \sigma_0^{-2} < \zeta_c < 1\ ,\cr\cr 
\zeta_c^{-1} \quad & \zeta_c > 1\ ,}\right.
\end{equation}
or in terms of $\zeta_0 = t/t_0 \approx R/R_0$,
\begin{equation}\label{eq:summary2a}\hspace{1.5cm}
\langle\Gamma\rangle \sim \left\{\matrix{(\sigma_0\zeta_0)^{1/3} 
\quad & 1 < \zeta_0 < \sigma_0^2\ ,\cr\cr 
\sigma_0 \quad & \zeta_0 > \sigma_0^2\ ,}\right. 
\end{equation}
\begin{equation}\label{eq:summary2b}\hspace{1.5cm}
\langle\sigma\rangle \sim \left\{\matrix{\sigma_0^{2/3}\zeta_0^{-1/3} 
\quad & 1 < \zeta_0 < \sigma_0^2\ ,\cr\cr 
\sigma_0^2\,\zeta_0^{-1} \quad & \zeta_0 > \sigma_0^2\ .}\right.
\end{equation}
%

\section{Self-Acceleration: a Paradox?}
\label{sec:BM-effect}

The apparent self-acceleration of the plasma shell, which was
described in \S\S~\ref{sec:test_case} and \ref{sec:back_envelope}, is
rather unusual and even somewhat perplexing. This self-acceleration
reminds of the outrageous tall tales of Baron Munchausen, particularly
the one where he escapes from a swamp by pulling himself up by his own
hair (or bootstraps).  In this section we try to resolve this apparent
paradox, and clarify how the shell keeps significantly accelerating
after losing causal contact with the wall.

At the heart of the apparent paradox lies the well known fact that
for a closed system with no external forces the center of momentum
(CM) velocity, $\vec{\beta}_{\rm CM}$, remains constant. This is valid
not only in the Newtonian regime, but also in special relativity,
where $\vec{\beta}_{\rm CM}$ is the velocity of an inertial frame,
$S_{\rm CM}$, where the total momentum of the system vanishes, $P' =
0$, as measured simultaneously in that frame.\footnote{When viewed
from this frame, it is obvious that in the absence of any external force
the total momentum $P' = 0$ remains unchanged, so this frame remains the CM
frame, and its velocity $\vec{\beta}_{\rm CM}$ as measured in any
other inertial frame remains constant.} If we denote the energy and
momentum as measured in $S_{\rm CM}$ by $E'$ and $P' = 0$,
then in an inertial frame $S$ in which $S_{\rm CM}$ moves at a
velocity $\vec{\beta}_{\rm CM} = \beta_{\rm CM}\hat{x}$, and the total
energy and momentum are $E$ and $P$, a simple Lorentz transformation
implies $P_z = P'_z = 0$, $P_y = P'_y = 0$, and
\begin{eqnarray}\nonumber
P = P_x = \Gamma_{\rm CM}(P'_x+\beta_{\rm CM}E') 
= \Gamma_{\rm CM}\beta_{\rm CM}E'\ ,
\\ 
E = \Gamma_{\rm CM}(E'+\beta_{\rm CM}P'_x) = \Gamma_{\rm CM}E'\ ,\quad
\Longrightarrow \quad \beta_{\rm CM} = \frac{P}{E}\ .
\end{eqnarray}
Since in the absence of external forces $P$ and $E$ remain constant, as
measured in frame $S$, so does $\beta_{\rm CM}$. 

Now, for simplicity let us consider a slight variation on our simple
test case from \S~\ref{sec:test_case}, where at the moment the
original rarefaction wave reaches the wall (i.e., at $t = t_0$ in the
lab frame, which is identified with frame $S$ here), the wall is
removed (and replaced by vacuum).  This modification should not have
any effect on the propagation speed and location of the head of the
secondary rarefaction wave, $\beta_* = dx_*(t)/dt$ and $\xi_*(t) =
x_*(t)/t$, or on the flow ahead of it, at $\xi > \xi_*(t)$. It would
affect only the region behind the head of the secondary rarefaction
wave. Therefore, it should not affect the local dynamics of the
``shell'' (where the shell refers to $\xi_*(t) <
\xi < \xi_h = \beta_{\rm max}$). However, in this case, at $t>t_0$
there is immediately no external force exerted on the flow, and
therefore its total momentum and energy are fixed to their values at
$t = t_0$ (for the energy this was true also before $t_0$ since the
wall was static in the lab frame):
\begin{eqnarray}\nonumber
P(t\geq t_0) = P(t_0) = \int_0^{t_0}Fdt = \frac{B_0^2}{8\pi}t_0 = 
M_0\frac{\sqrt{\sigma_0(1+\sigma_0)}}{2}\ ,
\end{eqnarray}
\begin{equation}
E(t\geq 0) = M_0\frac{2+\sigma_0}{2}\ ,\quad\Longrightarrow\quad 
\beta_{\rm CM} = \frac{\sqrt{\sigma_0(1+\sigma_0)}}{2+\sigma_0}\ .
\end{equation}
where $M_0 = \rho_0 l_0$ is the total rest mass (which like $P$, $E$
and $F$, is measured per unit area, given the 1D planar geometry). In terms of
the Lorentz factor,
\begin{equation}\hspace{0.5cm}
\Gamma_{\rm CM} \equiv \left(1-\beta_{\rm CM}^2\right)^{-1/2} 
= \frac{2+\sigma_0}{\sqrt{4+3\sigma_0}} \approx 
\frac{\sigma_0^{1/2}}{\sqrt{3}} \ll \sigma_0\ ,
\end{equation}
(the last two approximations hold for $\sigma_0\gg 1$). In
Appendix \ref{app:CM-frame} we derive the same result for $\beta_{\rm
CM}$ and $\Gamma_{\rm CM}$ by calculating the total momentum in a
general rest frame (simultaneously in that frame), and then requiring
that it vanishes.

In the CM frame, $S_{\rm CM}$, the total energy is
\begin{equation}\hspace{1.0cm}
E' = \frac{E}{\Gamma_{\rm CM}} = \frac{\sqrt{4+3\sigma_0}}{2}M_0
\approx \frac{\sqrt{3}}{2}\sigma_0^{1/2}M_0\ ,
\end{equation}
i.e., a factor of $\sim\sigma_0^{1/2}\gg 1$ larger than the rest
energy. Therefore, at late times when all of the magnetic energy is
converted into kinetic energy, the typical Lorentz factor of fluid in
this frame must be $\sim\sigma_0^{1/2}$, and in particular 
\begin{equation}\hspace{1.5cm}\label{eq:Gamma'_M}
\langle\Gamma'\rangle_M = \frac{E'}{M_0} = \frac{\sqrt{4+3\sigma_0}}{2}
\approx \frac{\sqrt{3}}{2}\sigma_0^{1/2}\ .
\end{equation}
However, since $P' = 0$ this implies that comparable fractions (of the
order of one half) of the rest mass would be moving at $u' \sim
\sigma_0^{1/2}$ and at $u' \sim-\sigma_0^{1/2}$, corresponding to 
$\Gamma \sim \sigma_0$ and $\Gamma\sim 1$, respectively, in the lab
frame. This picture is supported by a direct calculations in the CM
frame (for details see Appendix~\ref{app:CM-frame} and in particular
the discussion around Fig.~\ref{fig:gamma_med}).

This bares a lot of resemblance to the simple mechanical analogy that
is described in Appendix~\ref{app:mechanical-analogy}, of two masses,
$m$, initially moving together with a Lorentz factor $\Gamma$ in the
lab frame, and connected by a compressed ideal massless spring with
potential energy $E'_{\rm pot}$ in their initial rest frame. The
spring is then released and fully converts its potential energy into
kinetic energy of the two masses. In our case we can take $m = M_0/2$
and $E'_{\rm pot} = E'-M_0$ so that the final Lorentz factor of each
mass is $\Gamma_* = E'/M_0$ in their original rest frame. Their
velocities are parallel and anti-parallel to their original direction of
motion relative to the lab frame, denoted by subscripts `$+$' and
`$-$', respectively. If we choose $\Gamma = \Gamma_{\rm CM}$ then
$\Gamma_* =\sqrt{4+3\sigma_0}/2$ is slightly larger than $\Gamma_{\rm
CM}$ resulting in $\Gamma_+\approx\sigma_0$ and $\beta_-\approx-5/13$.
Alternatively, we could choose $\Gamma = \Gamma_* \neq\Gamma_{\rm CM}$
so that the mass at the back would be at rest in the lab frame:
$\Gamma_- = 1$ and $\Gamma_+=\Gamma^2(1+\beta^2)\approx 2\Gamma_*^2
\approx (3/2)\sigma_0 \sim \sigma_0$ (in this case $E$ is somewhat
larger than in the original case since we fixed $E'$ and slightly
increased $\Gamma$).  In either case the mass at the front ends up
with $\Gamma_+\sim \sigma_0$ and carries all (or almost all) of the
momentum and kinetic energy in the lab frame, while the mass at the
back has $\Gamma_-\sim 1$ and carries a negligible fraction of the
total energy and momentum. In the CM frame, however, the two masses
have equal energy and momenta of equal magnitude in opposite
directions.

Thus, the ``Baron Munchausen paradox'' described at the beginning of
this section is resolved as follows. First, while in the lab frame the
typical Lorentz factor at the time $t_0$ when the original rarefaction
wave reaches the wall is
$\langle\Gamma(t_0)\rangle\sim\sigma_0^{1/3}$, the center of momentum
Lorentz factor is significantly higher, $\Gamma_{\rm
CM}\sim\sigma_0^{1/2}$. This difference may be attributed to a
simultaneity effect: the Lorentz factor of a rest frame where the
total momentum vanishes as measured simultaneously {\it in the lab
frame} at $t_0$ is indeed $\sim \sigma_0^{1/3}$. However, the more
physically meaningful definition of the CM frame\footnote{For example,
with the former hybrid definition the velocity of that frame changes
with time and approaches the constant velocity of the proper CM frame
only at asymptotically late times.} requires that the momentum be
calculated simultaneously in that frame, and this accounts for the
difference. Second, even though $\Gamma_{\rm CM}$ remains constant, in
accord with our Newtonian intuition, we argue that the more
astrophysically relevant quantity is $\langle\Gamma\rangle_E$ -- the
energy weighted mean value of $\Gamma$ (in the lab frame), and
$\langle\Gamma\rangle_E$ does increase with time, approaching
$\sim\sigma_0$ at late times ($t>t_c$). This is justified below
(and in the discussion around Fig.~\ref{fig:gamma_med}).

It is by now clear that $\beta_{\rm CM}$ and $\Gamma_{\rm CM}$ remain
constant at $t\geq t_0$, while $\langle\Gamma\rangle_E$ grows with
time and approaches $\sim \sigma_0$ at late times. At such late
times, $t \gg t_c = t_0\sigma_0^2$, when all of the magnetic energy is
converted into kinetic energy,
\begin{equation}\hspace{1.1cm}\label{eq:beta_E}
\langle\beta\rangle_E \equiv \frac{\int dE\beta}{\int dE} \longrightarrow 
\frac{\int dM\Gamma\beta}{dM\Gamma} = \frac{P}{E} = \beta_{\rm CM} \ . 
\end{equation}
One might therefore ask, why is it more relevant to take the energy
weighted average of $\Gamma$, $\langle\Gamma\rangle_E$, rather than
that of $\beta$, $\langle\beta\rangle_E$, and then derive from it the
corresponding value of $\Gamma$, $(1-\langle\beta\rangle_E^2)^{-1/2}$,
which approaches $\Gamma_{\rm CM}$ at late times. The answer is that
$\langle\Gamma\rangle_E$ is more representative of the Lorentz factor
of the material that carries most of the energy in the lab frame,
which is the frame where all of our observations are made and the
external medium is at rest. This can be seen by using the simple
mechanical analogy outlined above of two equal masses $m=M_0/2$ that
end up with $\Gamma_-=1$ and $\Gamma_+\sim\sigma_0$. In this case
$E_\pm=\Gamma_\pm m$, and
\begin{eqnarray}\nonumber
\quad\quad\langle\Gamma\rangle_E = \frac{\Gamma_+^2+1}{\Gamma_++1}
\approx\Gamma_+\sim\sigma_0\ ,\quad\quad
\langle\beta\rangle_E = \frac{\Gamma_+\beta_+}{\Gamma_++1}\ ,
\\ \label{eq:vel_av}
\left(1-\langle\beta\rangle_E^2\right)^{-1/2} = 
\sqrt{\frac{\Gamma_++1}{2}} = \Gamma_{\rm CM} \sim \sigma_0^{1/2}\ ,\quad\quad
\end{eqnarray}
so that using $\langle\beta\rangle_E$ results in $\Gamma_{\rm CM} \sim
(\Gamma_+\Gamma_-)^{1/2}$, which gives too much weight to the mass
that ends up at rest ($\Gamma_-=1$), even though it carries only a
very small fraction of the energy in the lab frame,
$(\Gamma_++1)^{-1}\sim\sigma_0^{-1}\ll 1$. On the other hand,
$\langle\Gamma\rangle_E$ is very close to $\Gamma_+$, the Lorentz
factor of the mass that carries almost all of the energy in the lab
frame.

The situation where part of a closed system with no external forces is
accelerated to large positive velocities at the expense of another
part of that system, which attains large negative velocities, is
analogous to a rocket. If the rocket+fuel start at rest with no
external forces, then the total momentum remains zero all along. The
body of the rocket is accelerated to positive velocities while the
burnt fuel is thrown back with large negative velocities.  That is why
we had originally dubbed the impulsive acceleration of a shell the
``magnetic rocket'' effect. The analogy is not perfect, however, as
rocket acceleration implies a causal connection between the body of
the rocket and the exhaust. In the case of the magnetized shell, the
decelerated material behind the secondary rarefaction wave is causally
disconnected from the forward material.  In the self-similar solution
each fluid element is accelerated by the magnetic pressure gradient
towards the asymptotic value of $\Gamma\simeq 2\sigma_0$. The
secondary rarefaction, however limits the duration of such forward
acceleration. As soon as it reaches this fluid element its forward
acceleration is terminated and replaced by the deceleration forced by
the opposite pressure gradient behind this rarefaction. Thus, while
the head of the shell continues to accelerate, the center of momentum
speed for the whole flow remains unchanged, apart from the slow
increase due to the wall effect.

\section{Discussion}
\label{sec:dis}

\subsection{General points}

\subsubsection{Impulsive versus steady-state acceleration} 

The main shortcoming of the steady-state magnetic acceleration which
can be successfully overcome in the impulsive regime is best
illustrated in the case of an unconfined spherical outflow. In the
steady state regime such a flow accelerates effectively only up to the
fast magnetosonic point, where $\Gamma \sim \sigma_0^{1/3}$ and
$\sigma\sim\sigma_0^{2/3}$. After this point the acceleration becomes
very slow, with $\Gamma$ increasing only logarithmically with distance
\citep{T94,BKR98}, resulting in Poynting-dominated flows on length 
scales of astrophysical interest.  In contrast, the impulsive magnetic
acceleration allows effective conversion of electromagnetic energy,
which leads relatively quickly to a kinetic energy-dominated flow.
During the main phase of acceleration, after the separation from the
wall in our test case, the magnetization parameter decreases with
distance as $\sigma\propto R^{-1/3}$ ($\Gamma\propto R^{1/3}$) and
then during the coasting phase as $\sigma\propto R^{-1}$ ($\Gamma
\approx$ constant).

The steady-state magnetic acceleration of collimated flows (jets) is
generally more effective, leading to higher asymptotic Lorentz factors
and lower magnetization compared to the case of unconfined
flow. However, it still leads to the asymptotic values of
magnetization parameter $\sigma\geq 1$ \citep{KVKB09,L09,L10}.  When
the pressure distribution of confining medium is a power law $p_{\rm
ext} \propto R^{-\alpha}$, with $\alpha>2$ the external confinement is
in fact still rather insufficient. In such conditions, jets quickly
develop conical streamlines and do not accelerate efficiently
afterwards as the magnetic hoop stress, magnetic pressure, and
electric force finely balance each other.  The asymptotic value of the
Lorentz factor is $\Gamma \approx
\min(\sigma_0^{1/3}\theta_j^{-2/3},\,\sigma_0/2)$, where $\theta_j$ is
the asymptotic half-opening angle of the jet, and the corresponding
magnetization parameter is $\sigma \approx
\max(\frac{1}{2}\sigma_0^{2/3}\theta_j^{2/3},\, 1)$.

When $\alpha<2$, the shape of a steady-state flow is parabolic, $r_j
\propto R^{\alpha/4}$ (where $r_j$ is the cylindrical radius), and its
Lorentz factor grows as $\Gamma \propto r_j
\propto R^{\alpha/4}$ until reaching $\Gamma\approx\sigma_0/2$
($\sigma\approx 1$) after which the acceleration becomes ineffective
again \citep{KVKB09,L09,L10}.  Additional acceleration mechanisms,
such as the impulsive acceleration mechanism discussed in this paper,
are needed to produce kinetic-energy dominated flows (As we have
already mentioned in \S~\ref{sec:introduction}, $\sigma\sim 1$ is
still too high for effective shock dissipation.)  On the other hand,
for $2> \alpha > 4/3$ the steady-state acceleration is faster compared
to the impulsive one. However, even if magnetic acceleration initially
occurs in a steady-state fashion and then continues in an impulsive
fashion, the kinetic energy-dominated regime would still be reached at
the same distance from the source.  As we shall see later, such a
cooperation of two mechanisms is natural in astrophysical context.

A related issue is the level of variability at which the impulsive
mechanism becomes significant. The best case scenario is when short
bursts of activity are separated by rather long quiet periods, so that
the length of almost empty space between shells exceeds by an order of
magnitude (or more) the shell width. Then one can expect that the
collisions between shells effectively occur only in the coasting phase
where practically all of the shell energy is in the kinetic form (see
\S~\ref{sec:coast+summary}). The issue of interaction between multiple
shells is best addressed numerically and this is left to a future
work. More generally, the maximum fraction of magnetic energy that can
eventually be dissipated at standard MHD shocks in a variable flow,
generated via the impulsive plasma acceleration mechanism, can be
estimated as $f_B = (\mean{B^2}-\mean{|B|}^2)/\mean{B^2}$.
Essentially, this accounts for the decrease in magnetic energy during
the transition to uniform magnetic field\footnote{A similar issue
arises in the theory of striped pulsar winds, where smooth fast
magnetosonic waves from an oblique rotator eventually steepen into
multiple fast shocks and the same estimate can be used to estimate
their efficiency \citep{L03a}.}  This shows that in weakly variable
flows the impulsive mechanism becomes insignificant.  However, the
observations of AGN and micro-quasar jets indicate that violent bursts
rather than smooth variability can be more characteristic of their
central engines and a similar conclusion can be made regarding the GRB
jets from the often violent variability of their gamma-ray emission.
An additional shock dissipation may occur if within a single shell
the magnetic field rapidly alternates, like in the striped winds of
pulsars. In such a case, the typical gyration radius downstream of the
shocks caused by collisions between shells may exceed the stripes
separation, leading to fast dissipation of the alternating component
of the magnetic field \citep{L03b}.

Finally, let us address the efficiency of dissipation in the internal
shocks. If the source activity duration is $t_v$ and the duration of
the quiet phase between successive shell ejections is $t_{\rm gap}
\gtrsim t_v$ then the maximum fraction of the initial magnetic energy
that can be converted to other forms (namely kinetic or internal) is
$f_B = (1+t_v/t_{\rm gap})^{-1}$. If we define the mean value of
$\sigma$ as the ratio of the total magnetic to non-magnetic energies
then this implies that $\mean{\sigma}\geq t_v/t_{\rm gap}$ when the
different sub-shells collide.\footnote{And before the overall radial
extent of the flow increases appreciably, so that the expression $f_B
= (\mean{B^2}-\mean{|B|}^2)/\mean{B^2}$ that is based on a constant
total volume still holds.} However, internal energy is needed in order
to power the observed variable emission in GRBs, AGN or
micro-quasars. The fraction of the kinetic energy that is converted
into internal energy at the internal shocks depends on the local value
of $\sigma$ at the shocks (decreasing with increasing $\sigma$, especially at low
Mach number shocks). Internal shocks between different sub-shells
occur at $R_{\rm IS} \sim R_ct_{\rm gap}/t_v
\gtrsim R_c$ where the mean magnetization of the shell is
$\mean{\sigma(R_{\rm IS})}\sim t_v/t_{\rm gap}$, i.e. close to the
above lower limit. This suggests that the efficiency of internal
energy generation in internal shocks may significantly increase with
$t_{\rm gap}/t_v$. However, it could already be quite reasonable even
for $t_{\rm gap} \sim t_v$ for which $f_B \sim 0.5$ and even though
$\mean{\sigma(R_{\rm IS})}\sim 1$, the magnetization at the head of
each sub-shell is below average (which may improve the efficiency).

\subsubsection{Effects of geometry}

While our test case problem deals with flows with planar symmetry, the
effects of geometry are relatively minor. It is easy to verify that
the results of ``back of the envelope'' calculations of
\S~\ref{sec:back_envelope} remain unchanged for flows with
cylindrical and spherical geometry. Appendix~\ref{app:spherical} shows
the mathematical reason for this -- a suitable variable substitution
reduces the equations with spherical symmetry to those with the planar
symmetry. From the physical point of view this relative lack of
sensitivity to geometry is based on the fact that the key factor in
the impulsive acceleration is the flow expansion in the direction
parallel to the direction of motion, whereas the symmetry of the flow 
mainly regulates the rate of expansion in the transverse
direction. Due to the transverse expansion of jets the transverse
magnetic field, which we assume to be dominating, decreases as
$B_\perp\propto r_j^{-1}$ whereas the specific volume increases as
$V\propto r_j^2$, where $r_j$ is the transverse length scale.  The
specific electromagnetic energy remains unchanged, $E_{\rm EM}\propto
B_\perp^2 V \propto r_j^0$, and hence the transverse expansion does
not lead to magnetic acceleration.

\subsubsection{Test case and astrophysical flows}
\label{sec:ts-af}

The initial configuration of our test case problem can be relevant for
eruptive astrophysical phenomena involving fast magnetic reconnection
and restructuring of magnetic field configuration, like the magnetar
bursts \citep{Lt03}.  In many other cases, an astrophysical
central engine may operate rather steadily on relatively long time
scales. These scales have to be compared with the time scale required
for the flow, which is powered by the central engine, to reach the
fast magnetosonic point of the steady-state solution. Once the jet
propagates beyond this point, its inner part becomes much less
effected by the waves which are generated at the jet head.  In
particular, if the jet expands into a relatively empty channel, then
the rarefaction wave, which propagates in the comoving jet frame only
with the fast magnetosonic speed, will be confined to the jet head and
unable to propagate upstream. Therefore the fraction of the jet
affected by this wave will rapidly decrease in time.  For such cases,
a shell moving with super-fast-magnetosonic speed will be a more
suitable initial configuration compared to the static shell next to a
wall of our test problem.

After the shell of our test problem had separated from the wall, the
plasma acceleration was driven by the magnetic pressure gradient, that
had been developed in the shell prior to its separation. Thus, it is
reasonable to investigate whether a similar pressure distribution can
develop in the case where there is no wall but the shell is initially
moving with a super-fast-magnetosonic speed. In this case two
rarefaction waves will be moving into the shell, one from its head and
another from its tail. However, due to the properties of relativistic
velocity addition the head rarefaction will be moving across the shell
much faster compared to the tail rarefaction which will almost
``freeze'' at the shell tail. Indeed, in the laboratory frame the tail
rarefaction propagates with the speed
\begin{equation}\hspace{2.0cm}
\beta_{\rm t} = \frac{\beta+\beta_{\rm ms}}{1+\beta\beta_{\rm ms}}
\simeq 1-\frac{1}{8\Gamma^2\Gamma_{\rm ms}^2}\ ,
\label{dis1}
\end{equation}
where the last equality holds for $\Gamma,\,\Gamma_{\rm ms} \gg 1$.
The length of the jet affected by this wave grows at the rate
\begin{equation}\label{eq:delta-beta_t}\hspace{2.5cm}
\Delta \beta_{\rm t} = \beta_{\rm t} - \beta \simeq \frac{1}{2\Gamma^2}\ .
\end{equation}
The head rarefaction propagates with the speed    
\begin{equation}\hspace{2.2cm}
\beta_{\rm h} = \frac{\beta-\beta_{\rm ms}}{1-\beta\beta_{\rm ms}}
\simeq \frac{\Gamma^2-\Gamma_{\rm ms}^2}{\Gamma^2+\Gamma_{\rm ms}^2}
\label{eq:beta_h}
\end{equation}
and the length of the jet affected by this wave grows at the rate
\begin{equation}\hspace{1.8cm}
\Delta \beta_{\rm h} = \beta - \beta_{\rm h}
   \simeq \frac{2\Gamma_{\rm ms}^2}{\Gamma^2+\Gamma_{\rm ms}^2}\gg \Delta \beta_{\rm t}\ .
\label{eq:delta-beta_h}
\end{equation}
Thus, the head rarefaction crosses the shell first and creates the
magnetic pressure gradient which accelerates the shell in the  
direction of the head, just like in our test case after separation from the wall. 

Moreover, in the rest frame of the shell the head rarefaction starts
propagating much earlier than the tail rarefaction (due to
simultaneity effects, since they start more or less simultaneously in
the lab frame), and this is the reason why the tail rarefaction covers
only a very small fraction of the shell even though in this frame the
two rarefaction waves propagate at the same speed ($\beta_{\rm ms}$)
in opposite directions. In this frame, at the time the two rarefaction
waves meet (very close to the back end of the flow), the configuration
is very close to that of our modified test case, where the wall is
removed when the head rarefaction wave reaches it. This fact is used
later on, in the derivation leading to Eqs.~(\ref{eq:R_cr,t}) and
(\ref{eq:Gamma_cr,t}).

Another, important issue is whether the space between different ejecta
in astrophysical jets can be considered as empty. Indeed, if the jet
production is completely interrupted from time to time then the
external gas may rush into the gaps between different ejecta.
The speed of such a lateral flow is obviously limited by the speed of
light, and this sets the lower limit on the length scale at which the
gaps can be considered as empty,
$$
R_{\rm min}=ct_v \theta_j^{-1} \ ,
$$
where $t_v$ is the variability time-scale of the central engine and
$\theta_j$ is the jet half-opening angle. (At this distance from the
central source the shell's cylindrical radius is comparable to the
length of the gaps between shells.) For GRB jets with $t_v\gtrsim
4\;$ms and $\theta_j\sim 0.1$ this gives us $R_{\rm min} \gtrsim
1.2\times 10^{9}$cm, and for AGN jets with $t_v\sim 10\;$days and
$\theta_j\sim 1^\circ$ this gives us $R_{\rm min} \sim 1.5\times
10^{18}\;$cm. The sound speed of the surrounding gas at such
distances can be much lower than the speed of light and one may expect
the empty gaps to appear at smaller distances than $R_{\rm min}$
\citep{LL10}.  However, the ejecta will most certainly drive shock
waves into the surrounding gas, heating it to higher temperatures near
the jet channel. On the other hand, the increased buoyancy of this gas
will result in an outflow, which may become a supersonic wind.  This
will effectively reduce the speed with which this gas expands into the
jet openings. In fact, if $\theta_j>1/\mathcal{M}_w$, where
$\mathcal{M}_w$ is the wind Mach number, then the wind gas will be
unable to reach the jet axis.

Moreover, the jet may have this kind of protection from the beginning
if the accretion disk produces its own supersonic wind (we assume here
that the relativistic jet is driven by a Kerr black hole). Close to
the source, where $\mathcal{M}_w\le 1$, the disk wind may still try to
fill the polar region, thus creating an obstruction for the re-born
jet. However, it could be pushed aside by the jet on the time scale
required for the jet to overtake the wind, $\sim t_v v_w/(c-v_w)$,
where $v_w$ is the wind speed. Using the cited variability scales and
$v_w\sim 0.1c$ we then find that empty gaps may appear already
beginning from the distance of $\sim 10^7\;$cm for GRB jets and $\sim
3\times10^{15}\;$cm for AGN jets.

An impulsive operation of the central engine may well results in
trapping of some amount of weakly magnetized and dense external gas in
the gaps of intermittent highly magnetized jet.  This gas will then be
accelerated by the jet, leading to development of Rayleigh-Taylor
instability, turbulence and mixing.  Clearly, this important issue
requires further investigation.

\subsection{Application to GRB jets}

We start by considering the propagation of a single shell produced
during an active phase of central engine of duration $t_v$. This time
may correspond to the whole duration of gamma ray burst or to the
duration of one of many shells produced during the active phase of its
central engine.  The exact nature of the jet variability is not
known. In the collapsar model for long GRBs and the binary merger
models for short GRBs, this may be related to advection of magnetic
field with different polarity onto the black hole, similar to what has
been seen in recent numerical simulations \citep{BB09}.  In any case,
the shortest variability time scale in GRBs is probably given by the
``viscous'' time of the inner disk. For neutrino cooled disks this is
\begin{equation}\hspace{1.6cm} 
t_{v,\rm min} \approx 4 
\left(\frac{\alpha}{0.1}\right)^{-6/5}
\left(\frac{M}{M_{\odot}}\right)^{6/5} \mbox{ms}\ ,
\label{tdn}
\end{equation}
where $M$ is the black hole mass and $\alpha$ is the parameter of the
$\alpha$-disk model \citep{PWF99}. In the alternative model of GRB
central engine, which replace a super-accreting black hole with a
millisecond magnetar, the nature of variability has to be
different. It could be driven by a violent restructuring of magnetar
magnetosphere, e.g. rising buoyant magnetic loops and magnetic
reconnection. A relatively mild case of such restructuring, with the
characteristic timescale of order of $\sim 20\;$ms, has been seen in
recent numerical simulations of magnetar driven GRB jets \citep{KB07}.
This time scale gives us one of the characteristic length scale of
this problem, the shell width 
$$ 
l=ct_v=3\times10^{8}\fracb{t_v}{10\,{\rm ms}}\;\mbox{cm}\ .  
$$
(We use the name shell rather loosely here to describe the ejecta,
which can be rather elongated and better described as a jet close to
the central engine.) There are many other important scales in this
problem.

As we have already commented on, the initial acceleration of the flow
can proceed in a steady state fashion. This brings into consideration
the radius of the light cylinder, $r_{\rm lc}$, the distance to the
fast magnetosonic surface, $R_{\rm ms}$, the distance up to which the
steady-state acceleration mechanism remains effective, $R_s$, the
distance at which the impulsive acceleration mechanism kicks in,
$R_{\rm cr,t}$, the coasting distance $R_c$, and finally the distance
where the shell begins to decelerate due to the interaction with the
interstellar medium or stellar wind gas, $R_{\rm dec}$.  There are
many unknowns in this problem. In particular, it is difficult to say
what is the exact nature of the collimating medium. The jet is
unlikely to be in direct contact with the collapsing star. The hot jet
cocoon and the wind from accretion disk are more likely
candidates. Let us suppose that the external pressure scales as
$p_{\rm ext} \propto R^{-2}$, the most favourable case for the
steady-state collimation acceleration mechanism. Then the steady-state
jet is parabolic, $R\propto r^2$ (where $r$ is the cylindrical and $R$
is the spherical radius), and beyond the light cylinder, $r_{\rm lc} =
c/\Omega$, its Lorentz factor increases as $\Gamma \sim (r/r_{\rm
lc})\approx (R/r_{\rm lc})^{1/2}$ \citep[e.g.][]{KVKB09}.  At the
fast-magnetosonic surface $\Gamma\approx \sigma_0^{1/3}$ and thus this
surface is located at
\begin{eqnarray}\hspace{3.1cm}
R_{\rm ms} \sim r_{\rm lc} \sigma_0^{2/3}\ .
\end{eqnarray}
If the jet is powered by a rapidly rotating black hole ($a=0.9$) then
$r_{\rm lc} \approx 4R_g$, where $R_g=GM/c^2$ is the gravitational
radius of the black hole.  For the typical parameters of GRBs this
gives us
\begin{eqnarray}\hspace{2.3cm}
r_{\rm lc} = 6\times 10^5 \fracb{M}{M_\odot}\ \mbox{cm}\ ,
\end{eqnarray}
and 
\begin{eqnarray}\hspace{1.5cm}
R_{\rm ms} = 6\times10^7 \fracb{\sigma_0}{10^3}^{2/3} \fracb{M}{M_\odot}\ \mbox{cm}\ .
\end{eqnarray}
Thus, for the time scale of the central source variability
\begin{eqnarray}\hspace{1.5cm}
t_v > \frac{R_{\rm ms}}{c} \approx 2 \fracb{\sigma_0}{10^3}^{2/3} \fracb{M}{M_\odot}\;\mbox{ms}\ , 
\label{t_v1}
\end{eqnarray}
the source will be able to produce a steady-state
super-fast-magnetosonic flow. Since $\sim 2\;$ms is about the shortest
timescale for the variability of the central engine (see the end of
\S~\ref{sec:ts-af}) this must be always the case and the effects of
steady-state collimation acceleration have to be taken into account.

As the jet propagates into an almost empty channel cleared by the
previous ejections, there will be a rarefaction wave in its heads,
making its way into the jet. However, it will occupying only a small
fraction of the jet length. Indeed, in the source frame the speed of
this rarefaction is given by Eq.~(\ref{eq:beta_h}), which for
$\Gamma^2\gg\Gamma_{\rm ms}^2$ implies $\beta_{\rm h}\simeq 1$, and
thus the jet length grows much faster than the width of the
rarefaction wave in its head.

The collimation acceleration becomes ineffective when the jet
half-opening angle, $\theta_j$, exceeds the Mach angle, $\theta_{\rm
M}$, which is given by
\begin{equation}\hspace{2.2cm}
\sin\theta_{\rm M} = \frac{1}{\mathcal{M}_{\rm ms}} 
= \frac{\Gamma_{\rm ms} \beta_{\rm ms}}{\Gamma\beta}\ ,
\end{equation}
where $\mathcal{M}_{\rm ms}$ is the relativistic fast-magnetosonic
Mach number. For $\Gamma \gg \Gamma_{\rm ms} \gg 1$ and thus
$\theta_{\rm M} \ll 1$ this reduces to $\theta_{\rm M} \approx
\Gamma_{\rm ms}/\Gamma$ so that the critical half-opening angle is
given by
\begin{equation}\hspace{2.8cm}
\theta_j \sim \theta_{\rm M} \sim \frac{\Gamma_{\rm ms}}{\Gamma}\ , 
\end{equation}
where $\Gamma_{\rm ms} \sim \sigma^{1/2} \sim
\sigma_0^{1/2}\Gamma^{-1/2}$.  At this point the Lorentz factor
and magnetization parameter of the jet are
\begin{equation}\label{eq:Gamma_s}\hspace{1.1cm}
\Gamma_s \approx \sigma_0^{1/3}\theta_j^{-2/3} \approx 
46\fracb{\sigma_0}{10^3}^{1/3}\fracb{\theta_j}{0.1}^{-2/3}\,\,,
\end{equation}
and 
\begin{equation}\hspace{1.2cm}
\sigma_s \approx (\sigma_0\theta_j)^{2/3} \approx 
22\fracb{\sigma_0}{10^3}^{2/3}\fracb{\theta_j}{0.1}^{2/3}\ ,
\end{equation}
respectively. (The analysis of flows collimated by an external medium
with a power-law pressure distribution by \citet{L09} leads to a
result that differs from this one only by a factor of order unity.)
In principle, both $\sigma_0$ and $\theta_j$ can be estimated from
observations of GRBs and their afterglows.  In particular, $\sigma_0$
can be determined using the measurements of Lorentz factor via $\Gamma
\le \sigma_0$, where the equality corresponds to full conversion of
the electromagnetic energy into the bulk motion kinetic energy.  The
actual location of the point where the jet Mach angle reaches its
critical value and the jet enters the freely expanding regime is less
certain as it depends on the exact pressure distribution of the
confining medium.  For $p_{\rm ext}\propto R^{-2}$ we have
\begin{equation}\label{Rs-basic}\hspace{1.1cm}
R_s \sim r_{\rm lc}\Gamma_s^2 \sim r_{\rm lc}\sigma_0^{2/3}
\theta_j^{-4/3} \sim R_{\rm ms}\theta_j^{-4/3}.
\end{equation}
For the parameters typical for GRBs this gives us
\begin{equation}\hspace{0.5cm}
\label{Rs}
R_s \approx 
1.3\times 10^9\left(\frac{M}{M_\odot}\right)
\left(\frac{\sigma_0}{10^3}\right)^{2/3}
\left(\frac{\theta_j}{0.1}\right)^{-4/3}\;{\rm cm}\ .
\end{equation}
This is significantly lower compared to the radius of long GRB
progenitors, which is believed to be of the order on the Solar radius,
$R_\odot\approx 7\times 10^{10}\;$cm.  Beyond $R_s$ the
collimation acceleration is no longer effective.

In order to find the scale at which the impulsive acceleration comes
into play we first need to estimate how long it takes for the head
rarefaction to cross the shell. The length of the section of the shell
which is affected by the rarefaction grows with time at the rate
$\Delta \beta_{\rm h}$ given by Eq.~(\ref{eq:delta-beta_h}),
corresponding to a crossing time
\begin{eqnarray}\hspace{0.7cm}
t_{\rm cr,h} \approx \frac{t_v}{\Delta \beta_{\rm h}}
\approx \frac{t_v}{2}\left[\left(\frac{\Gamma}{\Gamma_{\rm ms}}\right)^2+1\right]
\approx \frac{t_v}{2}\left(\frac{\Gamma}{\Gamma_{\rm ms}}\right)^2  \ ,
\end{eqnarray}
where the last equality assumes a super-fast-magnetosonic regime. By
this time the shell will propagate the distance
\begin{eqnarray}\hspace{0.8cm}
\label{add2}
R_{\rm cr,h} \sim c t_{\rm cr,h} \sim 
    \frac{ct_v}{2}\fracb{\Gamma_s}{\Gamma_{\rm ms}}^2 \sim 
    ct_v \frac{\Gamma_s^2}{\sigma_s} \sim ct_v \theta_j^{-2}\ .
\end{eqnarray}
For the typical GRB parameters this gives us 
\begin{eqnarray}\hspace{1.5cm}
R_{\rm cr,h} \sim 3\times10^{10} \fracb{t_v}{10\mbox{ms}} 
           \fracb{\theta_j}{0.1}^{-2}\ .
\end{eqnarray}
In the frame moving at a Lorentz factor $\Gamma_s$, at the time the
two rarefaction waves meet (very close to the back end of the shell),
the configuration is very close to that of our modified test case at
the time when the head rarefaction wave reaches the wall and the wall
is removed. The main difference is that the initial magnetization
parameter is $\sigma_s$ and the initial shell width is $\Gamma_sct_v$.
Thus, after the passage of the head rarefaction wave the typical shell
Lorentz factor in this frame is
\begin{equation}\hspace{2.5cm}
\Gamma_* \sim \sigma_s^{1/3} = (\sigma_0 \theta_j)^{2/9}\ ,
\end{equation}
and the typical value of the magnetization parameter is $\sigma_{\rm
cr,h} \sim \sigma_s^{2/3} \sim (\sigma_0\theta_j)^{4/9}$. Therefore,
the shell Lorentz factor in the lab frame is
\begin{equation}
\Gamma_{\rm cr,h} \sim \Gamma_* \Gamma_s \sim \sigma_0^{5/9} \theta_j^{-4/9} 
\approx 130\left(\frac{\sigma_0}{10^3}\right)^{5/9}
\left(\frac{\theta_j}{0.1}\right)^{-4/9}\ .
\end{equation}
This Lorentz factor is only one order of magnitude below the maximum
value given by $\sigma_0$. In fact, this may still be only a
conservative estimate as we have not taken into account the
acceleration related to the transverse expansion of the jet when it
crosses the stellar surface \citep{KVKB09,TNM09}.  This additional
acceleration may well increase the mean Lorentz factor by a factor of
few (in what follows we denote this factor of $\kappa$). This brings
the Lorentz factor up to $\Gamma_{\rm cr,t} =
\kappa\Gamma_{\rm cr,h}$ and the magnetization parameter down to $\sigma_{\rm cr,t} =
\sigma_{\rm cr,h}/\kappa$. However, even after this the jet magnetization is
still too high for effective shock dissipation.

For simplicity we assume that the stellar radius $R_*$ where $\Gamma$
increases by a factor of $\kappa$ is $R_{\rm cr,h} \leq R_*\ll R_{\rm
cr,t}$, where $R_{\rm cr,t}$ is the radius where the tail rarefaction
crosses half of the original shell. We now use the similarity between
our modified test case at $t_0$ and the shell in its rest frame prior
to the crossing of the head rarefaction (referred to as the shell's
``initial'' rest frame) at the time when the head and tail rarefaction
waves meet, as discussed below Eq.~(\ref{eq:delta-beta_h}). Here the
shell's ``initial'' rest frame would be moving with a Lorentz factor
$\Gamma_f\sim\kappa^{3/2}\Gamma_s$ relative to the lab frame, rather
than $\Gamma_s$. This can be understood from the fact that we require
that after the passage of the head rarefaction $\sigma \sim
\sigma_f^{2/3} \sim
\kappa^{-1}\sigma_{\rm cr,h} \sim \kappa^{-1}\sigma_s^{2/3}$ so that
$\sigma_f \sim \kappa^{-3/2}\sigma_s$ and $\Gamma_f \sim
\sigma_0/\sigma_f \sim \kappa^{3/2}(\sigma_0/\sigma_s) \sim
\kappa^{3/2}\Gamma_s$.

Now, recall that in our test case significant additional acceleration
after the shell separates from the wall (i.e. at $t>t_0$) starts only
when the head of the secondary rarefaction wave -- identified here
with the tail rarefaction wave after it meets the head rarefaction --
reaches fluid with $\Gamma \sim \sigma_0^{1/3}$,
i.e. $\Gamma(\xi_*)\sim\sigma_0^{1/3}$. This corresponds to $\xi_*\sim
0$ i.e. $\xi'_* \sim 0$ in the frame moving at Lorentz factor
$\Gamma_f$ (referred to as the comoving frame), corresponding to
the middle of the shell after the passage of the head rarefaction
wave. In our original test case this corresponded to a single
dynamical time ($\approx t_0$, i.e. between $t = t_0$ and $t\approx
2t_0$), so we did not pay attention to this. The comoving shell width
at the time when the two rarefaction waves meet is $2\Gamma_fct_v$ and
therefore in our present case it takes the tail rarefaction wave a
time $t'_{\rm cr,t}\sim\Gamma_ft_v$ to reach $\xi'_* \sim 0$ in the comoving frame, 
which corresponds to a time $t_{\rm cr,t} \sim \Gamma_f^2t_v$ in
the lab frame. This result can also be obtained using
Eq.~(\ref{eq:delta-beta_t}) with $\Gamma = \Gamma_f$, $t_{\rm cr,t}
\approx t_v/\Delta\beta_{\rm t}\sim\Gamma^2t_v\to\Gamma_f^2t_v$. 
This corresponds to a radius
\begin{eqnarray}\label{eq:R_cr,t}
R_{\rm cr,t} &\sim& \Gamma_f^2ct_v 
\sim \kappa^3\sigma_0^{2/3}\theta_j^{-4/3}ct_v
\\ \nonumber
&\approx& 6.5\times 10^{11}\kappa^3\fracb{\sigma_0}{10^3}^{2/3}
\fracb{\theta_j}{0.1}^{-4/3}\fracb{t_v}{10\mbox{ms}}\;{\rm cm}\ ,
\end{eqnarray}
where the impulsive acceleration with $\Gamma\propto R^{1/3}$ begins,
\begin{equation}\hspace{1.0cm}\label{eq:Gamma_cr,t}
\Gamma \sim \frac{\sigma_0}{\sigma} \sim \sigma_0\fracb{R}{R_c}^{1/3}
\sim \Gamma_{\rm cr,t}\fracb{R}{R_{\rm cr,t}}^{1/3}\ ,
\end{equation}
for $R_{\rm cr,t}<R<R_c$. As a consistency check we verify that this
gives $R_c \sim R_{\rm cr,t}(\sigma_0/\Gamma_{\rm cr,t})^3 \sim R_{\rm
cr,t}\sigma_{\rm cr,t}^3 \sim \sigma_0^2 ct_v$, as it should from the
general considerations outlined in \S~\ref{sec:after_separation}.

We find that $R_{\rm cr,t} \gg R_s$, which suggests that the
steady-state collimation acceleration and the impulsive acceleration
are scale separated. At $R=R_{\rm cr,t}$ the ratio of the shell's
cylindrical radius, $r_j=\theta_j R$, to its width, $l_j=c t_v$, is
\begin{equation}\hspace{0.5cm}
  \frac{r_j}{l_j} \approx \kappa^3\sigma_0^{2/3}\theta_j^{-1/3} = 
215\kappa^3\fracb{\sigma_0}{10^3}^{2/3}\fracb{\theta_j}{0.1}^{-1/3}\ .   
\end{equation}
Thus, ``shell'' is indeed a suitable name for the flow at the stage of
impulsive acceleration. The coasting radius is given by
For the typical parameters of GRBs this gives us 
\begin{equation}\hspace{0.8cm}
\label{R_c_j}
R_c \sim \sigma_0^2ct_v \approx 
3\times 10^{14} \fracb{\sigma_0}{10^3}^2
\left(\frac{t_v}{10\,{\rm ms}}\right)\;\mbox{cm}\ ,
\end{equation}
and at $R>R_c$ the shell coasts at $\Gamma\sim\sigma_0$ while its
magnetization rapidly decreases as $\sigma \sim R_c/R$.

Within this model, prompt gamma-ray emission due to dissipation in
internal shocks between different shells within the highly variable
outflow naturally occurs in the region $R\sim (1$--$10)R_c$. On the
one hand, the mean plasma magnetization is $\sigma\sim 1$ at $R=R_c$
and then it decreases linearly with the distance. Thus, one of the
conditions for effective shock dissipation, $\sigma \ll 1$, is
satisfied in this region.  On the other hand, the width of individual
shells begins to grow linearly with the distance, allowing their
collisions. (For $R < R_c$ the shells keep an almost constant width.)
Moreover, the variation of the flow Lorentz factor in the coasting
regime is rather large, $\Delta \Gamma \sim \Gamma$, which may
potentially help increase the efficiency of shock dissipation up to
$\sim10\%$ \citep{B00} or even higher \citep{KS01}.

In order to test the viability of our impulsive magnetic acceleration
mechanism, the coasting radius, $R_c$, has to be compared with the
deceleration radius, $R_{\rm dec}$, at which most of the energy is
transferred to the swept-up shocked external medium.  In the ``thin''
shell regime (see below) where $R_c<R_{\rm dec}$ and $\Gamma(R_{\rm
dec})\sim\sigma_0$, $R_{\rm dec}$ is given
by~\citep[e.g.,][]{Granot05},
\begin{eqnarray}\label{R_dec}
R_{\rm dec} = \left[\frac{(3-k)E_{\rm iso}}{4\pi
Ac^2\sigma_0^2}\right]^{1/(3-k)}
\quad\quad\quad\quad\quad\quad\quad\quad\quad\quad\quad
\\ \nonumber
\quad
=\left\{\matrix{2.5\times 10^{16}n_0^{-1/3}E_{\rm
  iso,53}^{1/3}\sigma_{0,3}^{-2/3}\;{\rm cm} & \quad & k=0\ ,\cr\cr
1.8\times 10^{13}A_*^{-1}E_{\rm iso,53}\sigma_{0,3}^{-2}\;{\rm cm} & \quad &
k=2\ ,}\right.
\end{eqnarray}
for a spherical external rest mass density profile $\rho_{\rm ext} =
AR^{-k}$, where $\sigma_{0,3}= \sigma_0/10^3$, $E_{\rm iso}
=10^{53}E_{\rm iso,53}$ is the isotropic equivalent energy in the
flow, $n = n_0\;{\rm cm^{-3}}$ is the external number density for a
uniform external medium ($k=0$) and $A=5\times 10^{11}A_*\;{\rm
g\;cm^{-1}}$ for a stellar wind environment ($k=2$). In some GRBs the
afterglow onset time is observed, which is identified with the
observed deceleration time, $t_{\rm dec} \sim R_{\rm
dec}/2c\Gamma^2(R_{\rm dec})$, and may be used to infer the values of
$\Gamma(R_{\rm dec})$ and $R_{\rm dec}$, typically giving values of
$\Gamma(R_{\rm dec})$ of a few hundred and $R_{\rm dec} \sim
10^{17}\;$cm
\citep{SP99,Liang10},
\begin{eqnarray}\label{R_dec2}
R_{\rm dec} = \left[\frac{(3-k)E_{\rm iso}t_{\rm dec}}{2\pi
Ac(1+z)}\right]^{1/(4-k)}
\quad\quad\quad\quad\quad\quad\quad\quad\quad
\\ \nonumber
\quad
=\left\{\matrix{1.0\times 10^{17}n_0^{-1/4}E_{\rm
  iso,53}^{1/4}t_{\rm dec,2}^{1/4}\;{\rm cm} & \quad & k=0\ ,\cr\cr
1.8\times 10^{16}A_*^{-1/2}E_{\rm iso,53}^{1/2}t_{\rm dec,2}^{1/2}\;{\rm cm} & \quad &
k=2\ ,}\right.
\end{eqnarray}
where $t_{\rm dec}/(1+z) = 100t_{\rm dec,2}\;$s. Note, however, that
this method has an observational bias towards low values of
$\Gamma(R_{\rm dec})$ and large values of $R_{\rm dec}$ that
correspond to large $t_{\rm dec}$ values, since small $t_{\rm dec}$
values are hard to measure as optical or X-ray follow-up observations
usually start at least tens of seconds after the start of the prompt
gamma-ray emission. Nevertheless, even though $\Gamma(R_{\rm dec})$ is
the Lorentz factor of the shocked external medium and it is close to
that of the original ejecta only for a Newtonian or mildly
relativistic reverse shock (the ``think shell'' case, where $t_{\rm
dec} > T_{\rm GRB}$, $T_{\rm GRB}$ being the observed duration of the
gamma-ray emission from the GRB), even for the ``thick shell'' case
(where $t_{\rm dec} \sim T_{\rm GRB}$) this method gives
$\Gamma(R_{\rm dec})
\sim \Gamma_{\rm cr}$ 
and $R_{\rm dec} \sim R_{\rm cr}$, which is the correct deceleration
radius in this regime (the critical values of the Lorentz factor,
$\Gamma_{\rm cr}$, and radius, $R_{\rm cr}$, are provided below). In
both regimes $t_{\rm dec} \gtrsim T_{\rm GRB}$, so using $T_{\rm GRB}$
instead of $t_{\rm dec}$ in Eq.~(\ref{R_dec2}) gives $R_{\rm cr}$,
which is a lower limit on the value of $R_{\rm dec}$.

Only when $R_c\lesssim 0.1 R_{\rm dec}$ the internal shock mechanism
can be sufficiently effective to explain the prompt gamma-ray
emission.  Equations (\ref{R_dec2}), (\ref{R_c_j}) and (\ref{tdn})
show that this is satisfied only when the characteristic variability
time scale of the central engine is not much longer than the viscous
time scale of the inner disk ($t_v \sim 10^{-2}\;$s), even though for
$R_{\rm dec} \sim 10^{17}\;$cm, $R_c \lesssim 10^{16}\;$cm requires
$t_v \lesssim 0.3\;$s or an observed variability time $\lesssim 1\;$s
for a typical redshift of $z\sim 2$. For long GRBs, with the mean
duration in the source frame of about $\sim 10\;$s, this implies
between a few tens to about one thousand of individual shells.  For
short GRBs, with the mean duration in the source frame of about $\sim
0.3\;$s, this number can be reduced down to between about a few to a
few tens.  Moreover, it is generally easier to obtain $R_c < R_{\rm
dec}$ for a uniform external medium than for a stellar wind
environment, since $R_{\rm dec}$ is typically much smaller for a
stellar wind.

Now we briefly discuss the interaction with the external medium, and
when it strongly affects the flow (a more detailed analysis will be
presented in a separate work). For simplicity, we shall consider a
single shell and discard factors of order unity. Let us consider a
spherical outflow of duration $t_0$, radial width $R_0\approx ct_0$,
energy $E$, and luminosity $L\approx E/t_0$, propagating into a
spherical external rest mass density profile $\rho_{\rm ext} =
AR^{-k}$ (with $k<10/3$).  The regime where the $R_c < R_{\rm cr} <
R_{\rm dec}$ (i.e., where at $R_c$ only a small fraction of the total
energy is transferred to the shocked swept-up external medium),
corresponds to the well-known ``thin shell'' (or initially Newtonian
reverse shock) case for the deceleration of a coasting unmagnetized
($\sigma < 1$) shell~\citep{SP95,Sari97}, which has been investigated
in the context of GRBs. Due to the spreading of the shell at $R > R_c$
(because of a spread $\Delta\Gamma\sim\Gamma$ in its Lorentz factor),
the reverse shock gradually strengthens and becomes mildly
relativistic at $R_{\rm dec}$ (which in this regime is given by
Eq.~[\ref{R_dec}]) where it finishes crossing the shell, and may
produce a bright emission that peaks at an observed time $t_{\rm
dec}\sim (1+z)R_{\rm dec}/c\sigma_0^2 \sim (R_{\rm dec}/R_c)T_{\rm
GRB} > T_{\rm GRB}$ (i.e., after the the end of the prompt GRB
emission).

For the other ordering of the critical radii, $R_{\rm dec} = R_{\rm
cr} < R_c$ (which may occur for large values of $t_0$ or a stellar
wind external medium), the outflow generally never reaches a coasting
phase (so $R_c$ loses its physical meaning as a coasting radius),
since the magnetized shell starts being significantly affected by the
swept-up shocked external medium when the latter still has only a
small fraction of the total energy. The impulsive acceleration,
$\Gamma \sim (\sigma_0R/R_0)^{1/3}$, proceeds from $R_0$ up to a
radius $R_u \sim R_{\rm cr}(\sigma_0/\Gamma_{\rm cr})^{-4/(10-3k)}$,
where $\Gamma_{\rm cr}\sim (ER_0/Ac^2)^{1/2(4-k)}$ and $R_{\rm cr}\sim
R_0\Gamma_{\rm cr}^2\sim (ER_0/Ac^2)^{1/(4-k)}$. Then, at $R_u < R <
R_{\rm cr}$ the typical Lorentz factor of the magnetized shell becomes
similar to that of the swept-up external medium, $\Gamma \sim
(L/Ac^3)^{1/4}R^{(k-2)/4}$, and is determined by the pressure balance
at the contact discontinuity that separates these two regions. This is
a phase of either a modest deceleration (for $k < 2$) or a reduced
acceleration (for $2 < k < 10/3$) that occurs while the outflow is
still highly magnetized ($\sigma \gg 1$ for $R_{\rm dec}\ll
R_c$). Therefore, there might not be a reverse shock going into the
original magnetized outflow, and even if such a shock develops, then
it would be very weak and could dissipate only a very small fraction
of the total energy.
Finally, at $R_{\rm cr}$ where $\Gamma \sim \Gamma_{\rm cr}$, most of
the energy is transferred to the shocked external medium (so that in
this regime $R_{\rm dec} = R_{\rm cr}$). Therefore, at $R > R_{\rm
cr}$ the flow approaches the \citet{BM76} self-similar solution for a
spherical constant energy relativistic blast-wave going into an
unmagnetized external medium.\footnote{The previous regime, $R_u < R <
R_{\rm cr}$ corresponds to another variant of that solution, with
energy injection into the shocked external medium by a relativistic
wind from the central source.} At $R_{\rm cr}$ the magnetization is
still high, $\sigma(R_{\rm cr})\sim\sigma_0/\Gamma_{\rm cr}$, where this
generalized ``thick shell'' regime ($R_{\rm dec} = R_{\rm cr} < R_c$)
corresponds to $\sigma_0 >\Gamma_{\rm cr}$, while the ``thick shell''
regime ($R_c < R_{\rm cr} < R_{\rm dec}$) corresponds to
$\sigma_0<\Gamma_{\rm cr}$.

Altogether, the acceleration of an initially highly magnetized
($\sigma_0 \gg 1$) impulsive outflow via the impulsive effect
and its deceleration due to the interaction with the external medium
are tightly coupled and cannot be fully treated in isolation. That is,
the magnetic acceleration naturally sets the initial conditions for
the interaction with the external medium, and realistically one cannot
simply assume any arbitrary initial configuration of the magnetized
outflow near the deceleration radius. Moreover, in the highly
magnetized ``thick shell'' regime there is an intermediate phase ($R_u
< R < R_{\rm cr} = R_{\rm dec}$), where the magnetic acceleration and
the deceleration because of the external medium balance each other,
resulting either in a reduced acceleration or in a relatively modest
deceleration (as outlined above). If the outflow starts highly
magnetized, then it can decelerate either in the unmagnetized ``thin
shell'' regime (with $\sigma(R_{\rm dec}) \sim R_c/R_{\rm dec} < 1$
for $\sigma_0 <\Gamma_{\rm cr}$) or in the highly magnetized analog of
the ``thick shell'' regime (with $\sigma(R_{\rm dec})\sim
\sigma_0/\Gamma_{\rm cr} > 1$ for $\sigma_0 >\Gamma_{\rm cr}$). 
There is no high magnetization ``thin shell'' regime, and in order to
be in the low magnetization ``thick shell'' regime, a single-shell
flow cannot start highly magnetized ($\sigma_0 \gg 1$).\footnote{This
might still be possible, under favourable conditions, for a highly
variable flow with a large number of high contrast sub-shells that
accelerate independently, and then quickly collide and merge into a
wider shell via internal shocks, soon after reaching their coasting
radius and $\sigma < 1$. This would amount to quasi-continuous energy
injection by subsequent sub-shells just after the deceleration radius
of the first sub-shell, thus increasing the deceleration radius of the
whole flow.}

\subsection{Application to AGN jets}

We can apply the results obtained in the previous section to AGN jets
simply via appropriate rescaling. First, the characteristic
masses of black holes are higher, $\sim 10^7\,$--$\,10^9M_\odot$. They are
radiation cooled and the corresponding shortest variability time scale
is
\begin{equation}\hspace{1.4cm}
t_{v,\rm min} \approx 10
\left(\frac{\alpha\,\delta^2}{10^{-3}}\right)^{-1}
\left(\frac{M}{10^8M_{\odot}}\right)\;\mbox{days}\ ,
\label{tdn_ss}
\end{equation}
where $\delta = H_d/R_d$ is the ratio of the disk height to its radius
\citep{SHS73}.

Second, the Lorentz factors of AGN jets can be measured directly via
observation of proper motion of their knots. Such observations (mainly
VLBI radio observations) indicate relatively low Lorentz factors, of
the order of a few for weak radio sources (FR-I type) and $\mean\Gamma
\sim 10$ with a tail extending up to $\Gamma \sim50$ for blazars
\citep{Lister09}. The rapid variability of gamma-ray emission from
some AGN suggests the possibility of even higher Lorentz factors
\citep[$\Gamma>50$,][]{Ahar07}.  Assuming, that the magnetic
acceleration is efficient in these sources, and hence the observed
$\Gamma$ is close to $\sigma_0$, we obtain a characteristic value of
$\sigma_0\sim10$, which is much lower compared to GRBs. In principle,
the value of $\sigma_0$ can be higher near the black hole and then
decrease downstream, e.g. as a result of some mass-loading
process. However, at present we have no concrete evidence for this.
Finally, the observed half-opening angles of blazar jets are smaller,
$\sim 1^\circ-3^\circ$ \citep{P09}.

The corresponding rescaling of the results for GRB jets yields the
light cylinder radius
\begin{eqnarray}\hspace{2.0cm}
r_{\rm lc} = 6\times 10^{13} \fracb{M}{10^8M_\odot}\ \mbox{cm}\ ,
\end{eqnarray}
the distance to the fast magnetosonic surface
\begin{eqnarray}\hspace{1.2cm}
R_{\rm ms} = 2.7\times10^{14} \fracb{\sigma_0}{10}^{2/3} \fracb{M}{10^8 M_\odot}\ \mbox{cm}\ ,
\end{eqnarray}
and the shortest variability time scale required for establishing a
steady-state super-fast-magnetosonic flow
\begin{eqnarray}\hspace{2.0cm}
t_v > 2.5 \fracb{\sigma_0}{10}^{2/3} \fracb{M}{10^8M_\odot}\,\mbox{hr}. 
\end{eqnarray}
Since the timescale of strong central engine variability is unlikely
to be shorter than the viscous timescale of inner accretion disk,
$$
t_{\rm vis}\sim 100 R_g/c^3 \sim 1.0 \fracb{M}{10^8 M_\odot}\,\mbox{days}\ ,  
$$
we conclude that, just like in the case of GRBs, the initial
acceleration of AGN jets up to super-magnetosonic speeds is provided
in a steady-state fashion.  Moreover, the recent observations of AGN
jets \citep{P09} clearly indicate that they satisfy the
$\Gamma\theta_j <1$ condition of effective steady-state collimation
acceleration \citep{KVKB09}. The observed decrease of half-opening
angle with distance in M87 jet also supports the theory of collimation
acceleration \citep{BJL02,GTB05}.

The distance at which one half of the electromagnetic energy is
converted in the energy of bulk motion is now
\begin{equation}\hspace{1.0cm}
R_s \approx
0.02 \left(\frac{M}{10^8M_\odot}\right)
\left(\frac{\sigma_0}{10}\right)^{2/3}
\left(\frac{\theta_j}{1^\circ}\right)^{-4/3}\;{\rm pc}
\end{equation}
\citep[see also ][]{KBVK07}. This scale is unresolved with modern VLBI systems. 
The recent numerical simulations show that the collimation
acceleration may continue a bit beyond this point, reducing the
magnetization down to $\sigma\sim 0.4$ within another decade of
distance from the source \citep{KBVK07}. This is still a relatively
high magnetization leading to relatively low efficiency of MHD shock
dissipation.  Additional impulsive accelerative can improve this.
This time, however, when the impulsive mechanism switches on this is
already the coasting regime.  Indeed, the fast magnetosonic speed
corresponding to $\sigma=0.4$ is $\beta_{\rm ms}\sim 0.5$. Then
instead of Eq.(\ref{eq:beta_h}) the speed of the head rarefaction is
given by
\begin{equation}\hspace{3.0cm}
\beta_{\rm h} \sim \frac{2\Gamma^2-2}{ 2\Gamma^2+1}
\end{equation}       
and the length of the section of the shell affected by the rarefaction
grows with time at the rate
\begin{equation}\hspace{2.2cm}
\Delta \beta_{\rm h} = \beta - \beta_{\rm h}
   \simeq \frac{2\Gamma_{\rm ms}^2}{\Gamma_{\rm ms}^2+\Gamma^2}\ .
\end{equation}
The scale of transition to impulsive regime is now  
\begin{equation}\hspace{2.5cm}
     R_{\rm cr,h} = \frac{t_v}{\Delta \beta_{\rm h}} \sim ct_v \Gamma^2.
\end{equation}
For the typical parameters of AGN jets this reads 
\begin{equation}\hspace{2.1cm}
     R_{\rm cr,h} \sim 1 \fracb{t_v}{10\mbox{d}} \fracb{\sigma_0}{10}^2 \mbox{pc}.  
\end{equation}
Basically, since $\beta_{\rm ms}$ is mildly relativistic we have
$R_{\rm cr,h} \sim R_{\rm cr,t}$, and since $\Gamma \sim \sigma_0$
there, the two distances are also of the order of $R_c \approx
\sigma_0^2ct_v$. Thus, the theory predicts effective dissipation at
internal shocks on the scales of $\sim 1\,$--$\,10\;$pc, exactly the
region where VLBI observations reveal bright super-luminal knots of
AGN jets.  This is the AGN counterpart of the prompt emission region
of GRBs.

\section{Summary and Conclusions}
\label{sec:conclusions}

In this paper we investigated the properties of magnetic acceleration
of relativistic impulsive flows. As a first step, we focused on a
relatively simple test case where a uniform cold and highly magnetized
($\sigma_0\gg 1$) shell of initial width $l_0$, whose back end leans
against a conducting ``wall'' and whose head faces vacuum.  The
evolution of the flow that develops in this test case splits into
three distinct phases.

The first phase can be described as a formation of a plasma pulse (or
a moving shell). During this phase, which lasts for the time $\sim
t_0\equiv l_0/c_{\rm ms,0} \approx l_0/c$, a self-similar rarefaction
wave develops at the interface with vacuum and travels towards the
wall.  At the end of this phase, the mean Lorentz factor of the
outflow is only $\mean{\Gamma}\sim\sigma_0^{1/3}$ and, apart from the
very thin layer at the vacuum interface, the shell of plasma is still
highly magnetized, with a mean magnetization parameter of
$\mean{\sigma} \sim \sigma_0^{2/3}$.

The first phase ends when the rarefaction wave reaches the wall. At
this point a secondary rarefaction wave forms that propagates from the
wall into the back of the shell and decelerates the material that
passes through it so that the shell quickly separates from the wall
and moves away from it.  During this second phase, the center of
momentum Lorentz factor of the shell remains fairly constant
($\Gamma_{\rm CM} \sim \sigma_0^{1/2}$). However, the leading part of
the plasma shell, ahead of the secondary rarefaction, continues to
accelerate at the same rate as in the self-similar solution.  It
contains most of the shell energy and its mean Lorentz factor grows as
$\mean{\Gamma}\propto t^{1/3}$.

At the end of the second phase, which lasts up to $\sim t_c \equiv
\sigma_0^2 t_0$, the magnetization of the shell drops down to $\sigma
\sim 1$, one half of the electromagnetic energy is converted into the
bulk motion kinetic energy of the plasma, and the growth of the mean
Lorentz factor begins to saturate at $\mean{\Gamma} \sim
\sigma_0$. Thus, the flow enters a phase of coasting.  During the
coasting phase the pulse width grows faster, approaching $l \propto
t$. The decrease of the magnetization parameter also accelerates,
approaching $\sigma\propto t^{-1}$, and the pulse soon becomes
kinetic-energy dominated. This property of impulsive magnetic
acceleration is most valuable in astrophysical context as the
efficiency of relativistic MHD shock dissipation decreases
dramatically with magnetization. In contrast to an impulsive flow, a
steady-state magnetized jet either remains highly magnetized ($\sigma
\gg 1$) all the way, or approaches $\sigma \approx 1$, depending on
the efficiency of external collimation.  This implies at best only
modest shock dissipation efficiency.

In some cases of truly explosive phenomena, such as magnetar flares,
our impulsive magnetic acceleration mechanism can be solely
responsible for the flow acceleration. In most other cases, such as
GRB and AGN jets, strong variability of their central engines is not
expected on time scales below the viscous time-scale of the inner
accretion disc around a black hole, which powers relativistic outflow.
This gives plenty of time to establish a quasi-steady
super-fast-magnetosonic flow near the source where it is accelerated
via the collimation mechanism. The observed strong collimation of
these jets supports our conclusion that the collimation mechanism
plays a part in their acceleration.  The impulsive acceleration
mechanism comes in force further out, where an individual ejecta
element starts being accelerated after the head rarefaction crosses it
and creates conditions similar to those of our test case flow in
phases two and three. The mean Lorentz factor of the shell, however,
starts increasing significantly above the value achieved by the
quasi-steady collimation acceleration only when the tail rarefaction
wave crosses about half of the shell. Provided the central engine
variability is sufficiently strong, so that the flow can be described
as individual ejecta shells separated by long gaps, the impulsive
acceleration mechanism can complete the acceleration process and
produce kinetic energy dominated relativistic flows on astrophysically
relevant distances from the central engine.  For short GRBs this may
still work well even if the ejecta effectively form a single uniform
shell.

Our analysis of GRBs show that a combination of the collimation and
impulsive mechanisms can accelerate GRB jets up to $\Gamma \gtrsim
10^3$, as has been inferred recently for several bright GRBs detected
by the Fermi Large Area Telescope, for both
long~\citep{080916C,090902B} and short~\citep{090510-phys} duration
GRBs.\footnote{We do note, however, that these lower limits on
$\Gamma$ from pair opacity are somewhat model dependent and a fully
self consistent calculation appropriate for an internal shock origin
of the gamma-ray emission gives limits that are a factor of $\sim 3$
lower \citep{Granot08,GRB090926A}, $\Gamma \gtrsim 10^{2.5}$, which
are significantly easier to satisfy.}  Moreover, their jets can become
kinetic energy dominated before the interaction with the interstellar
or stellar wind gas begin to decelerate the ejecta at $R_{\rm dec}\sim
10^{16}-10^{17}\;$cm.  The dissipation at internal shocks can become
efficient on scales $R\gtrsim R_c \approx
10^{13}(\sigma_0/300)^2(t_v/4\,{\rm ms})\;$cm.  The large variation of
Lorentz factor at the coasting phase, $\Delta\Gamma\sim\Gamma$,
insures that the internal shock will be strong and can dissipate and
radiate of the order of $\sim 10\%$ or so of the flow kinetic energy,
leading to a possibility of strong prompt emission.

The AGN jets are likely to be accelerated up to their observed Lorentz
factors already during the collimation acceleration phase. However,
the impulsive acceleration phase remains important, providing
effective conversion of remaining electromagnetic energy and producing
kinetic energy dominated flows. Our estimates show that efficient
shock dissipation region, analogous to the prompt emission region of
GRBs, is located around $\sim 1-10\;$pc, where VLBI observations
reveal the presence of super-luminal ``blobs''.

\section*{Acknowledgments}

We are grateful to Y.~Lyubarsky, J.~McKinney, M.~Lyutikov and A.~Levinson for 
constructive criticism and helpful discussions of the first draft of 
this paper.  
J.~G. gratefully acknowledges a Royal Society Wolfson Research Merit
Award. S.~S.~K. was supported by the STFC grant ``A Rolling Programme of
Astrophysical Research at Leeds''. A.~S. is supported by NSF grant AST-0807381.


\onecolumn

\appendix

\section{Self-Similar Rarefaction Wave in Planar Symmetry}
\label{app:self-sim}

The equations of relativistic MHD can be written as
\begin{equation}\label{eq:RMHD1}
\partial_\mu T^{\mu\nu} = 0\ ,\quad
\partial_\mu F^{*\mu\nu} = 0\ ,\quad
\partial_\mu (\rho u^\mu) = 0\ ,
\end{equation}
\citep[see][and references therein]{Komisarov99}, where 
\begin{equation}\label{eq:RMHD2}
T^{\mu\nu} = (\rho h_g + b^2)u^\mu u^\nu +
\left(p_g+\frac{b^2}{2}\right)g^{\mu\nu} - b^\mu b^\nu\ ,
\end{equation}
is the energy-momentum tensor. Here $\rho$, $w_g = \rho h_g$, $p_g$,
and $u^\mu = (u^0,\vec{u}) = \Gamma(1,\vec{v})$ are the fluid proper
rest mass density, enthalpy density, pressure, and four-velocity,
where $\Gamma = (1-v^2)^{-1/2}$ is its Lorentz factor, $g^{\mu\nu}$ is
the metric tensor, and we use units where $c = 1$ for
convenience. Furthermore,
\begin{equation}
F^{*\mu\nu} = b^\mu u^\nu - b^\nu u^\mu
\end{equation}
is the dual tensor of the electromagnetic field, and $b^\mu =
(b^0,\vec{b})$ where
\begin{equation}
b^0 = \vec{B}\cdot\vec{u} = \Gamma\vec{B}\cdot\vec{v}\quad ,\quad\quad
\vec{b} = \frac{\vec{B}+b^0\vec{u}}{u^0} =
\frac{\vec{B}}{\Gamma}+\Gamma(\vec{v}\cdot\vec{B})\vec{v}\ ,
\end{equation}
is the four-vector of the magnetic field, which is defined as
\begin{equation}
b_\alpha = \frac{1}{2}\eta_{\alpha\beta\gamma\delta}u^\beta
F^{\gamma\delta}\ ,
\end{equation}
where $F^{\gamma\delta}$ is the electromagnetic tensor and
$\eta_{\alpha\beta\gamma\delta}$ is the Levi-Civita alternating
tensor. In the fluid rest frame $b^\mu = (0,\vec{B})$ where $\vec{B}$
is the usual three-vector magnetic field, divided by $\sqrt{4\pi}$, so
that $w_m = 2p_m = b^2$. In general, $\vec{B}$ is measured in the lab
frame.  The three-vectors of the magnetic and electric fields in an
arbitrary frame are given by
\begin{equation}
\vec{B} = F^{*i0} = \vec{b}u^0 - \vec{u}b^0\ ,\quad
\vec{E} = \vec{b}\times\vec{u}\ .
\end{equation}
Similar to classical MHD, the electric current is given by the second
Maxwell equation,
\begin{equation}
J^\nu = \partial_\mu F^{\mu\nu}\ ,
\end{equation}
where it also includes the displacement current (time derivatives of
the electric field). Finally, $\nabla\cdot\vec{B} = \partial_i F^{*i0}
= 0$, $u_\mu b^\mu = 0$, $u_\mu u^\mu = -1$.

The RMHD equations simplify considerably under the assumption of a
flat space-time, $g^{\mu\nu} = \eta^{\mu\nu} = {\rm diag}(-1,1,1,1)$,
and planar symmetry, i.e. that all quantities depend only on $x$ and
$t$ in a Cartesian coordinate system \citep[see, e.g.,][]{GR06},
\begin{equation}
\frac{\partial}{\partial t}\left(
\begin{array}{c} 
\rho \Gamma                          \\ 
\rho h \Gamma^2 - p -\rho \Gamma - b^0b^0 \\ 
\rho h \Gamma^2 v^x - b^0b^x         \\ 
\rho h \Gamma^2 v^y - b^0b^y         \\ 
\rho h \Gamma^2 v^z - b^0b^z         \\ 
B^y                             \\ 
B^z  
\end{array}
\right)
+
\frac{\partial}{\partial x}\left( 
\begin{array}{c}
\rho \Gamma v^x                           \\ 
\rho h \Gamma^2 v^x - b^0b^x - \rho \Gamma v^x \\ 
\rho h \Gamma^2 v^x v^x + p - b^x b^x     \\ 
\rho h \Gamma^2 v^y v^x - b^x b^y         \\ 
\rho h \Gamma^2 v^y v^z - b^x b^z         \\ 
B^y v^x - B^x v^y                    \\ 
B^z v^x - B^x v^z 
\end{array}
\right) = 0\ .
\end{equation}
Here we consider the even simpler case where $v^y = v^z = 0$ and $B^x
= B^z = 0$ so that $\vec{v} = v\hat{x}$ and $\vec{B} = B\hat{y}$, i.e.
$u^\mu = \Gamma(1,v,0,0)$ and $b^\mu = (0,0,B/\Gamma,0)$. Under these
conditions, the RMHD equations further simplify to
\begin{equation}\label{1D_vx_By}
\frac{\partial}{\partial t}\left(
\begin{array}{c} 
\rho \Gamma                   \\ 
\rho h \Gamma^2 - p - \rho \Gamma  \\ 
\rho h \Gamma^2 v             \\ 
B
\end{array}
\right)
+
\frac{\partial}{\partial x}\left( 
\begin{array}{c}
\rho \Gamma v                \\ 
\rho h \Gamma^2 v - \rho \Gamma v \\ 
\rho h \Gamma^2 v^2 + p      \\ 
B v 
\end{array}
\right) = 0\ ,
\end{equation}
and the the magnetic field in the fluid rest frame is given by $B' =
B/\Gamma$, so that the equations for the evolution of $\rho\Gamma$ and
$B = B'\Gamma$ are the same and $B/\rho \Gamma = B'/\rho = {\rm
const}$.  Thus, we are left with three equations for three variables
(e.g., $\rho$, $\Gamma$, and $p_g$), where we also need to assume an
equation of state. In our notation
\begin{equation}
h = h_g + \frac{b^2}{\rho}\ ,\quad h_g = 1+\epsilon+\frac{p_g}{\rho}
= 1+\frac{\gamma}{\gamma-1}\frac{p_g}{\rho}\ ,
\end{equation}
where $\epsilon = e_{\rm int}/\rho$ and $e_{\rm int} = \epsilon\rho =
w_g - p_g -\rho$ is the proper internal energy density of the fluid,
while $\gamma$ is the adiabatic index of the fluid. The total pressure
is given by $p = p_g + p_m = p_g + b^2/2$.

We are looking for rarefaction wave solutions, which are self similar,
i.e. all quantities depend of $x$ and $t$ only through their ratio,
which is defined as the self-similar variable: $\xi \equiv x/t$. In
rarefaction waves the specific entropy, $s$, of every fluid element is
conserved, and therefore $0 = ds/dt = \partial s/\partial t +v\partial
s/\partial x$. Since $\partial/\partial x = (1/t)d/d\xi$ and
$\partial/\partial t = -(\xi/t)d/d\xi$, this implies $(v-\xi)s'= 0$
where a prime denotes a derivative with respect to $\xi$ ($s' \equiv
ds/d\xi$), and therefore $s' = 0$ and $s = {\rm const}$ (in general
$v \neq \xi$). Therefore, the flow is isentropic, and we may simply
relate the pressure to its value ahead of the rarefaction wave,
\begin{eqnarray}
b^2 &=& \left(\frac{B_0}{\rho_0}\right)^2 \rho^2 
= \rho_0\sigma_0\tilde{\rho}^{\,2}\ ,
\\
p_g &=& p_{g,0}\left(\frac{\rho}{\rho_0}\right)^\gamma
= \rho_0 a_0\tilde{\rho}^{\,\gamma}\ ,
\\
p &=& p_{g,0}\left(\frac{\rho}{\rho_0}\right)^\gamma 
    + \frac{1}{2}\left(\frac{B_0}{\rho_0}\right)^2 \rho^2
= \rho_0\left(a_0\tilde{\rho}^{\,\gamma} + 
\frac{\sigma_0}{2}\tilde{\rho}^{\,2}\right)\ ,
\\
\rho h &=& \rho 
+\frac{\gamma}{\gamma-1}p_{g,0}\left(\frac{\rho}{\rho_0}\right)^\gamma
+B_0^2\left(\frac{\rho}{\rho_0}\right)^2
= \rho_0\left(\tilde{\rho}+\frac{\gamma}{\gamma-1}a_0\tilde{\rho}^{\,\gamma}
+\sigma_0\tilde{\rho}^{\,2}\right)\ ,
\end{eqnarray}
where $\tilde{\rho} \equiv \rho/\rho_0$, $a_0 \equiv p_{g,0}/\rho_0$,
and $\sigma_0 \equiv B_0^2/\rho_0$ is the magnetization parameter of
the fluid ahead of the rarefaction wave (which is assumed to be at
rest in the lab frame: $\Gamma_0 = 1$).

equation \ref{1D_vx_By} can be expressed in terms of the self-similar
variable $\xi$ as
\begin{eqnarray}\label{eq1}
0 &=& (v-\xi)(\rho'+\rho \Gamma^2 vv') + \rho v'\ ,
\\ \label{eq2}
0 &=& (v-\xi)(\rho h \Gamma^2)' + \xi p' + \rho h \Gamma^2 v'\ ,
\\ \label{eq3}
0 &=& (1-v\xi)p' + (v-\xi)\rho h \Gamma^2 v'\ .
\end{eqnarray}
Let $c_s$, $c_A$, and $c_{\rm ms}$ denote the sound speed, the
Alf\'ven speed, and the fast magnetosonic speed, respectively.
We have $c_s^2 = (1/h_g)(\partial p_g/\partial
\rho)_s$ and $c_A^2 = b^2/\rho h$, so that $h_g/h = 1-c_A^2$ and
\begin{equation}
c_{ms}^2 = \frac{1}{h}\left(\frac{\partial p}{\partial\rho}\right)_s
= c_A^2 + c_s^2(1-c_A^2)\ ,
\end{equation}
which implies $p' = c_{\rm ms}^2 h\rho'$. Therefore, Eq.~(\ref{eq3}) can be
rewritten as
\begin{equation}\label{eq5}
0 = (1-v\xi)c_{ms}^2 \rho' + (v-\xi)\rho \Gamma^2 v'\ .
\end{equation}
equation. \ref{eq1} and \ref{eq5} imply $c_{\rm ms}^2 = [(v-\xi)/(1-v\xi)]^2$
and therefore
\begin{equation}\label{c_ms}
c_{\rm ms} = \pm\,\frac{v-\xi}{1-v\xi}\ ,
\end{equation}
where the plus and minus signs correspond to rarefaction waves
propagating to the left and right, respectively. This also implies
\begin{equation}\label{eq:xi_v}
\xi = \frac{v \mp c_{\rm ms}}{1 \mp vc_{\rm ms}}\ .
\end{equation}
The velocities of the tail (where $v = 0$) and of the head (where
$c_{\rm ms} = 0$) of the rarefaction wave are given by
\begin{equation}\label{xi_h_t}
\xi_t = \mp c_{\rm ms} \quad,\quad \xi_h = \pm v_{\rm max}\ ,
\end{equation}
where $v_{\rm max} = \max|v|$ is obtained at the head of the
rarefaction wave. As expected, the tail of the rarefaction wave
propagates into the fluid at rest at the fast magnetosonic speed.

equation. \ref{eq5} and \ref{c_ms} imply
\begin{equation}\label{J_pm}
\Gamma^2 dv \pm \frac{c_{ms}}{\rho}d\rho = 0 \ \Longrightarrow\ J_\pm =
\frac{1}{2}\ln\left(\frac{1+v}{1-v}\right) \pm
\int_0^{\tilde{\rho}}\frac{c_{ms}(\tilde{\rho}^{\,\prime})}
{\tilde{\rho}^{\,\prime}}d\tilde{\rho}^{\,\prime} = {\rm const}\ .
\end{equation}
Under our assumptions, 
\begin{equation}\label{eq:c_ms}
c_{ms}(\tilde{\rho}) = \sqrt{\frac{\gamma a_0\tilde{\rho}^{\,\gamma-1}+\sigma_0\tilde{\rho}}
{1+\frac{\gamma}{\gamma-1}a_0\tilde{\rho}^{\,\gamma-1}+\sigma_0\tilde{\rho}}}\ ,
\end{equation}
so that the integral in Eq.~(\ref{J_pm}) can be calculated analytically
in the simple cases where $\sigma_0 = 0$ ($B_0 = 0$), or $a_0 = 0$
($p_{g,0} = 0$). In the first limit ($\sigma_0 = 0$, i.e. no magnetic
field),
\begin{equation}
J_\pm = \frac{1}{2}\ln\left(\frac{1+v}{1-v}\right) \pm
\frac{1}{\sqrt{\gamma-1}}
\ln\left(\frac{\sqrt{\gamma-1}+c_s}{\sqrt{\gamma-1}-c_s}\right) 
= {\rm const}\ ,
\end{equation}
\citep{MM94} so that 
\begin{equation}\label{eq:hydro_const}
\left(\frac{1+v}{1-v}\right)\left(\frac{\sqrt{\gamma-1}+c_s}
{\sqrt{\gamma-1}-c_s}\right)^{\pm\frac{2}{\sqrt{\gamma-1}}} = {\rm const}\ ,
\end{equation}
and in the limit $a_0 \gg 1$ and $\gamma = 4/3$ this implies
$\Gamma_{\rm max} \gg 1$ which is approximately given by,
\begin{equation}\label{Gamma_max}
\Gamma_{\rm max} \approx \frac{1}{2}\left[\frac{4\gamma a_0}{(\gamma-1)}\right]^{(\gamma-1)^{-1/2}}
=
\left\{
\matrix{2^{4\sqrt{3}-1}a_0^{\sqrt{3}} & (\gamma = 4/3)\ ,\cr\cr
4a_0=4p_{g,0}/\rho_0 & (\gamma = 2)\ .
}
\right.
\end{equation}
In the second limit ($a_0 = 0$) we find
\begin{equation}
J_\pm = \frac{1}{2}\ln\left(\frac{1+v}{1-v}\right) \pm
2\rm{ArcSinh}\left(\sqrt{\sigma_0\tilde{\rho}}\right)
= \frac{1}{2}\ln\left(\frac{1+v}{1-v}\right) \pm
2\ln\left(\sqrt{\sigma_0\tilde{\rho}}+\sqrt{\sigma_0\tilde{\rho}+1}\right)
= {\rm const}\ ,
\end{equation}
so that
\begin{equation}\label{eq:mag_const}
\left(\frac{1+v}{1-v}\right)\left(\sqrt{\sigma_0\tilde{\rho}}
+\sqrt{\sigma_0\tilde{\rho}+1}\right)^{\pm 4} = {\rm const}\ ,
\end{equation}
and in the limit $\sigma_0 \gg 1$ we have
\begin{equation}
\Gamma_{\rm max} \approx 2\sigma_0\ .
\end{equation}
It can be seen that the purely magnetic case, $a_0 = 0$, is equivalent
to the pure hydrodynamic case, $\sigma_0 = 0$, for $\gamma = 2$ and
$a_0
\to \sigma_0/2$.  In the more general case,
\begin{equation}
J_\pm = \frac{1}{2}\ln\left(\frac{1+v}{1-v}\right) \pm
I(\tilde{\rho})
= \pm I(1) = {\rm const}\ ,
\end{equation}
where 
\begin{equation}
I(\tilde{\rho}) =
\int_0^{\tilde{\rho}}\frac{d\tilde{\rho}}{\tilde{\rho}}
\sqrt{\frac{\gamma a_0\tilde{\rho}^{\gamma-1}+\sigma_0\tilde{\rho}}
{1+\frac{\gamma}{\gamma-1}a_0\tilde{\rho}^{\gamma-1}+\sigma_0\tilde{\rho}}}\ ,
\end{equation}
so that 
\begin{equation}\label{v_max}
v_{\rm max} = \frac{\exp[2I(1)]-1}{\exp[2I(1)]+1}\quad,\quad 
\Gamma_{\rm max} = \frac{\exp[I(1)]}{1+v_{\rm max}} 
= \frac{\exp[2I(1)]+1}{2\exp[I(1)]}
\end{equation}
and
\begin{equation}
v = \pm\frac{\exp[2\tilde{I}(\tilde{\rho})]-1}{\exp[2\tilde{I}(\tilde{\rho})]+1}
\quad,\quad 
\Gamma = \frac{\exp[2\tilde{I}(\tilde{\rho})]+1}{2\exp[\tilde{I}(\tilde{\rho})]}\ ,
\end{equation}
where
\begin{equation}
\tilde{I}(\tilde{\rho}) = I(1) - I(\tilde{\rho}) =
\int_{\tilde{\rho}}^1\frac{d\tilde{\rho}}{\tilde{\rho}}
\sqrt{\frac{\gamma a_0\tilde{\rho}^{\gamma-1}+\sigma_0\tilde{\rho}}
{1+\frac{\gamma}{\gamma-1}a_0\tilde{\rho}^{\gamma-1}+\sigma_0\tilde{\rho}}}\ .
\end{equation}

\section{The Average Lorentz Factor}
\label{app:average_Gamma}

The maximal Lorentz factor $\Gamma_{\rm max}$ is only asymptotically
reached at the very head of the rarefaction wave, and only a small
amount of material which carries a small fraction of the total energy
has $\Gamma \sim \Gamma_{\rm max}$. Therefore, it makes sense to calculate some
average value of the Lorentz factor, which would reflect better the
Lorentz factor of the material that carries most of the energy.
A natural definition is the weighted average over the energy,
\begin{equation}\label{Gamma_av1}\hspace{6.0cm}
\langle \Gamma\rangle_{E} \equiv \frac{\int \Gamma dE}{\int dE} 
= \frac{\int dx\,T^{00}\,\Gamma}{\int dx\,T^{00}} =
\frac{\int_{\xi_t}^{\xi_h} d\xi\,T^{00}\,\Gamma}{\int_{\xi_t}^{\xi_h}d\xi\,T^{00}}\ ,
\end{equation}
where
\begin{equation}\hspace{5.0cm}
T^{00} = \rho_0\left[\Gamma^2\left(\tilde{\rho}
+\frac{\gamma}{\gamma-1}a_0\tilde{\rho}^\gamma +
\sigma_0\tilde{\rho}^2\right)-a_0\tilde{\rho}^\gamma
-\frac{\sigma_0}{2}\tilde{\rho}^2\right]\ .
\end{equation}
Another possible definition is the weighted average over the rest mass,
\begin{equation}\label{Gamma_av2}\hspace{6.0cm}
\langle \Gamma\rangle_{M} \equiv \frac{\int \Gamma dM}{\int dM} 
= \frac{\int dx\,\Gamma^2\rho}{\int dx\,\Gamma\rho} =
\frac{\int_{\xi_t}^{\xi_h} d\xi\,\Gamma^2\rho}{\int_{\xi_t}^{\xi_h}d\xi\,\Gamma\rho}\ .
\end{equation}
We note that for a cold magnetized shell, at late times when almost all
of the energy is in kinetic form and the magnetic energy becomes
negligible, the enumerator approaches $E/c^2$, and since the
denominator is simply the rest mass $M$, then $\langle
\Gamma\rangle_{M}$ approaches $E/Mc^2 = 1+\sigma_0/2$.  Alternative
options to define a ``typical'' Lorentz factor are its average over
space
\begin{equation}\label{Gamma_av3}\hspace{6.5cm}
\langle \Gamma\rangle_{x} \equiv \frac{\int dx\,\Gamma}{\int dx}
= \frac{\int_{\xi_t}^{\xi_h} d\xi\,\Gamma}{\int_{\xi_t}^{\xi_h}
d\xi}\ , 
\end{equation}
or its value at the point where there are equal energies on either
side within the rarefaction wave in the lab frame (i.e. the ``energy
median'' value),
\begin{equation}\label{Gamma_av4}\hspace{1.8cm}
\langle \Gamma\rangle_{E,{\rm med}} \equiv 
\left\{\Gamma(x_{1/2})\left|\,\,\int_{x_{\rm min}}^{x_{1/2}}
dx\,T^{00} = \int_{x_{1/2}}^{x_{\rm max}}dx\,T^{00} \right.\right\}  =
\left\{\Gamma(\xi_{1/2})\left|\,\, \int_{\xi_t}^{\xi_{1/2}} d\xi\,T^{00} =
\int_{\xi_{1/2}}^{\xi_h}d\xi\,T^{00}\right.\right\}\ ,
\end{equation}
where $x_{\rm min} = t \xi_t$ and $x_{\rm max} = t \xi_h$ (see
Eq.~\ref{xi_h_t}). 

\begin{figure}
\centerline{\includegraphics[width=7.9cm,height=5.4cm]{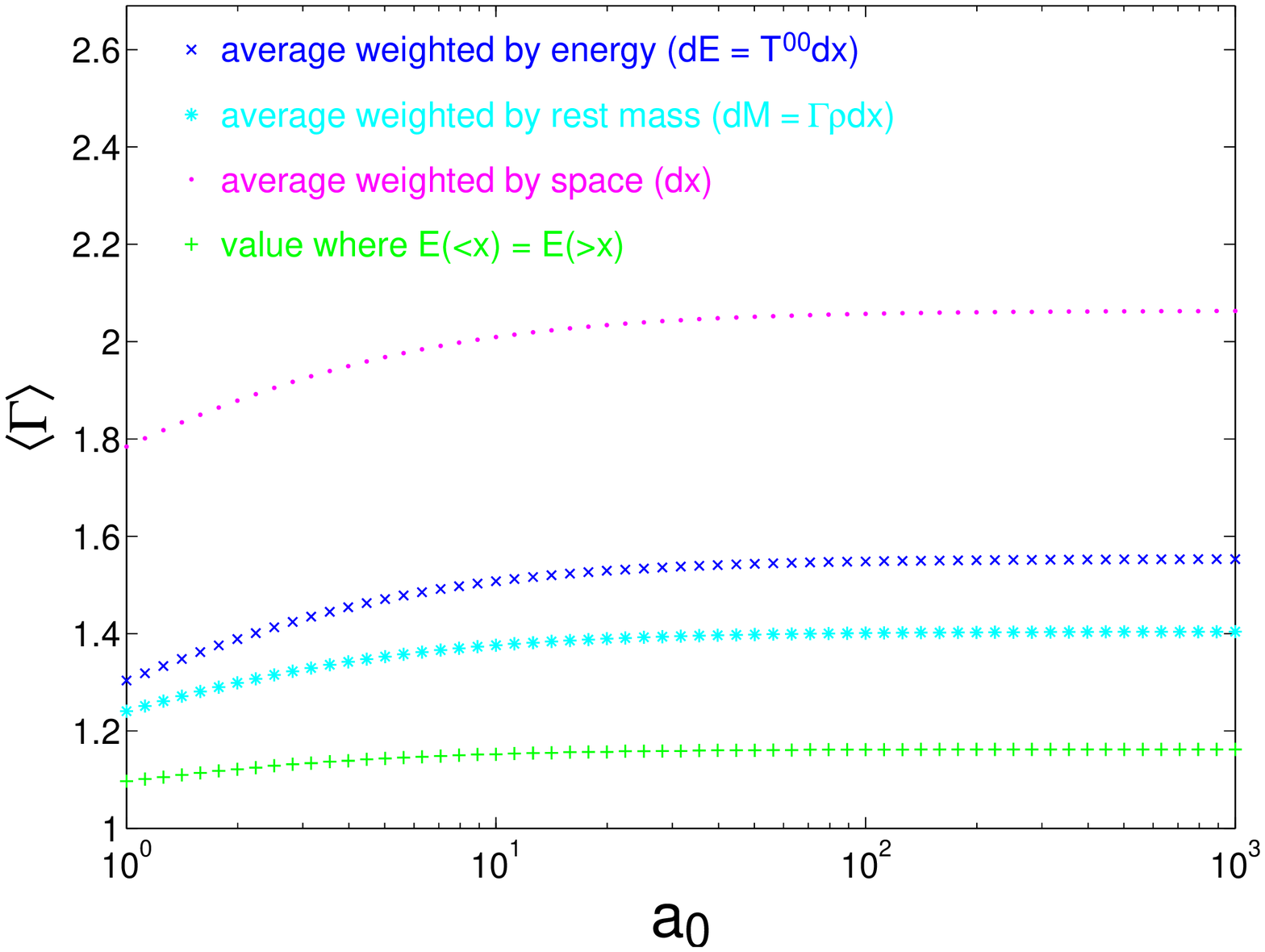}
\hspace{1.0cm}
\includegraphics[width=8.2cm,height=5.5cm]{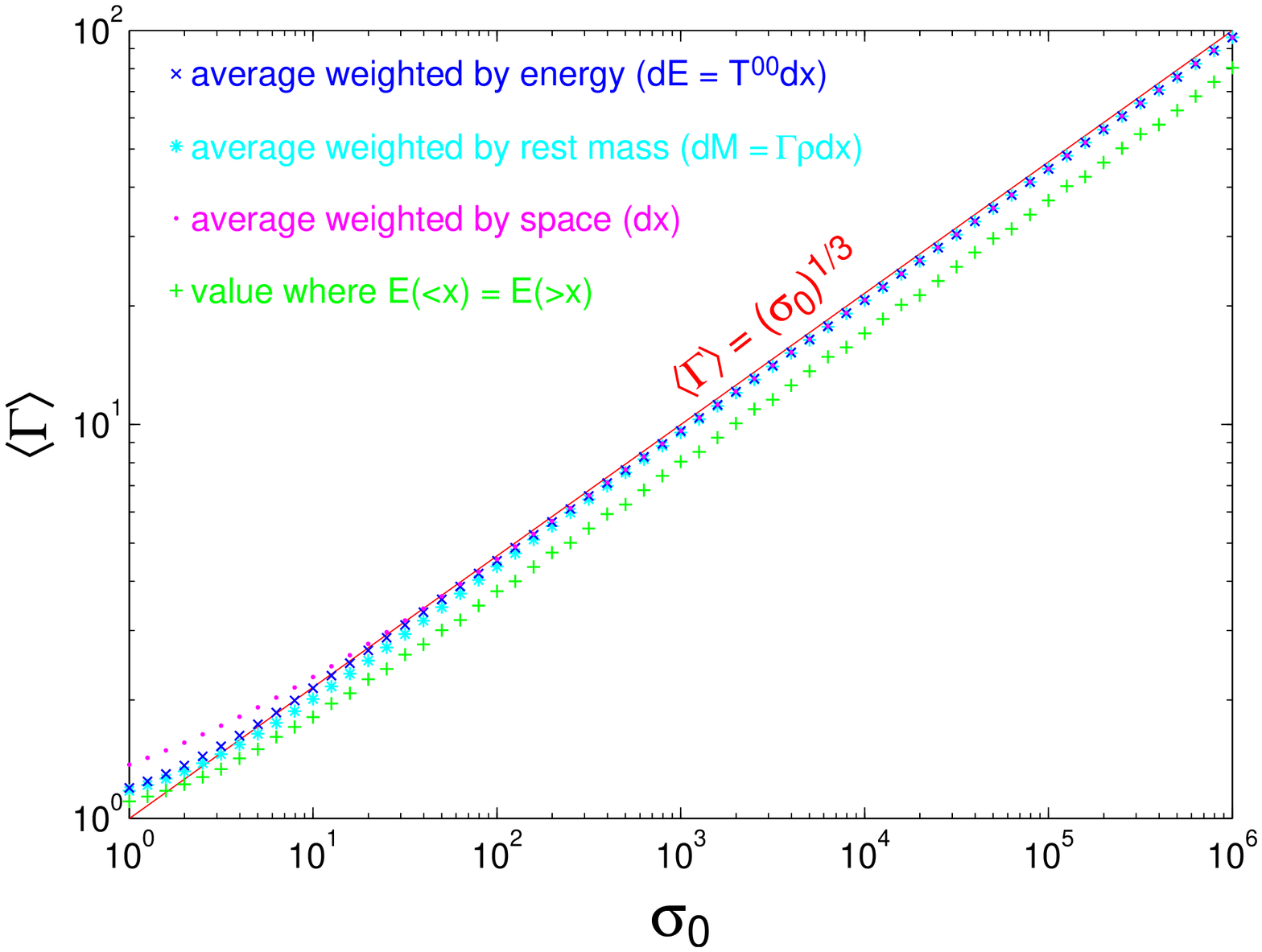}}
\caption{
{\bf Right panel:} four different estimates for the ``typical'' Lorenz
factor, $\langle \Gamma\rangle$, within a rarefaction wave for the
pure hydrodynamic case ($\sigma_0 = 0$) and for an adiabatic index of
$\gamma = 4/3$, as a function of $a_0 = p_0 /\rho_0 c^2$. Various
symbols show different weightings of $\Gamma$.  {\bf Left panel:} the
same four estimates of the ``typical'' Lorenz factor within the
rarefaction wave for the pure magnetic case ($a_0=0$), as a function
of $\sigma_0$.  The red solid line corresponding to $\langle
\Gamma\rangle = \sigma_0^{1/3}$ has been added for reference.
}
\label{fig:gamma_av_t0}
\end{figure}

Figure~\ref{fig:gamma_av_t0} shows these three estimates for the
typical Lorentz factor within the rarefaction wave, for the pure
hydrodynamic case ($\sigma_0 = 0$; {\it left panel}) and for the pure
magnetic case ($a_0 = 0$; {\it right panel}). In the pure
hydrodynamics case the typical Lorentz factor of the material in the
rarefaction wave is only mildly relativistic even in the limit of $a_0
\gg 1$, where it approaches a constant value, while $\Gamma_{\rm max}$
rapidly increases with $a_0$ (see Eq.~\ref{Gamma_max}). In the purely
magnetic case, we find that the typical value of the Lorentz factor
within the rarefaction wave is $\langle \Gamma \rangle \approx
\sigma_0^{1/3}$, while its maximal value at the head of the
rarefaction wave is $\Gamma_{\rm max}
\approx 2\sigma_0$.

\section{Analytic derivations for the rarefaction waves}
\label{app:raref-front}

\subsection{Explicit solution for the original self-similar rarefaction wave}

Equation (\ref{sv}) can be written as 
\begin{equation}\hspace{7.9cm}
\label{sv1}
\delta_\xi^2 = \delta_v^2\delta_{\rm c_{ms}}^{-2}\ ,
\end{equation}
and Eq.~(\ref{int}) as 
\begin{equation}\hspace{5.8cm}
\label{int1}
\delta_v^2\delta_{\rm c_{ms}}^4 = \delta_{\rm c_{ms,0}}^4 = \mathcal{J}_+ 
= \left(\sqrt{\sigma_0}+\sqrt{\sigma_0+1}\right)^4\ ,
\end{equation}
where $\delta_X$ is defined via
$$
  \delta_X^2=\frac{1+X}{1-X}\ , \quad\quad X = \frac{\delta_X^2-1}{\delta_X^2+1}\ .
$$
This allows us to find all flow variables as explicit functions of $\xi$: 
\begin{equation}\label{eq:Gamma}\hspace{1.0cm}
\label{deltav}
\delta_v = (\delta_{\rm c_{ms,0}}\delta_\xi)^{2/3}\ ,\quad\quad
v = \frac{(\delta_{\rm c_{ms,0}}\delta_\xi)^{4/3}-1}
{(\delta_{\rm c_{ms,0}}\delta_\xi)^{4/3}+1}\ ,\quad\quad
\Gamma = \frac{(\delta_{\rm c_{ms,0}}\delta_\xi)^{4/3}+1}
{2(\delta_{\rm c_{ms,0}}\delta_\xi)^{2/3}}\ ,\quad\quad
u = \Gamma v = \frac{(\delta_{\rm c_{ms,0}}\delta_\xi)^{4/3}-1}
{2(\delta_{\rm c_{ms,0}}\delta_\xi)^{2/3}}\ ,
\end{equation}
\begin{equation}\hspace{4.97cm}
\label{deltac}
\delta_{\rm c_{ms}} = \frac{\delta_{\rm c_{ms,0}}^{2/3}}{\delta_\xi^{1/3}}\ ,
\quad\quad
c_{\rm ms}^2 = \frac{\sigma}{1+\sigma} = 
\left[\frac{\delta_{\rm c_{ms,0}}^{4/3}\delta_\xi^{-2/3}-1}
{\delta_{\rm c_{ms,0}}^{4/3}\delta_\xi^{-2/3}+1}\right]^2\ ,
\end{equation}
\begin{equation}\hspace{5.5cm}
\label{B}
\frac{\sigma}{\sigma_0} = \frac{\rho}{\rho_0} = \frac{B'}{B_0} = \frac{B}{\Gamma B_0} =
\frac{1}{4\sigma_0}\left(\frac{\delta_{\rm c_{ms,0}}^{2/3}}{\delta_\xi^{1/3}}
-\frac{\delta_\xi^{1/3}}{\delta_{\rm c_{ms,0}}^{2/3}}\right)^2\ ,
\end{equation}
\citep[this result is due to][with a small correction in 
Eq.~(\ref{B})]{Lt10a}.  The extent of the rarefaction wave is given by
the conditions $\delta_{\rm c_{ms}}=1$ for the right front and
$\delta_v=1$ for the left front. They yield $\delta_{\rm
c_{ms,0}}^{-1} \le \delta_\xi \le
\delta_{\rm c_{ms,0}}^2$.

\subsection{Motion of the head of the secondary rarefaction wave}
\label{app:xi*}

The overall impression created by Fig.~\ref{fig:numeric} is that of an effective 
separation of the shell from the wall. The region between $x=18l_0$ and 
$x=20l_0$ contains most of the total mass and energy of the initial 
solution. This somewhat surprising result can be verified in a different way. 
In the fluid frame the front of secondary rarefaction moves with the local 
magnetosonic speed. In the lab frame this corresponds to 
\begin{equation}\hspace{5.8cm}
\beta_* \equiv \frac{dx_*}{dt} = 
\frac{v(\xi_*)+c_{\rm ms}(\xi_*)}{1+v(\xi_*)c_{\rm ms}(\xi_*)} = 
\frac{\delta_{\rm c_{ms,0}}^{8/3}\delta_{\xi_*}^{2/3}-1}
{\delta_{\rm c_{ms,0}}^{8/3}\delta_{\xi_*}^{2/3}+1}\ ,
\label{xs*2}
\end{equation}   
where $v(\xi)$ and $c_{\rm ms}(\xi)$ describe the self-similar
solution for the initial rarefaction (Eqs.~[\ref{deltav}],
[\ref{deltac}]). Noting that $d\xi_*/dt = [(dx_*/dt)-\xi_*]/t$, and
$d\delta_{\xi_*}^2/dt = (d\delta_{\xi_*}^2/d\xi_*)(d\xi_*/dt) =
(d\xi_*/dt)(\delta_{\xi_*}^2+1)^2/2$, equation (\ref{xs*2}) can be
rewritten as
\begin{equation}\hspace{6.3cm}
\frac{d\delta_{\xi_*}^2}{d\ln t} =
\delta_{\xi_*}^2+1 - \frac{(\delta_{\xi_*}^2+1)^2}
{\delta_{\rm c_{ms,0}}^{8/3}(\delta_{\xi_*}^2)^{1/3}+1}\ , 
\label{xs*3}
\end{equation}
which has the solution \citep{Lt10a}
\begin{equation}\label{t(xi_*)}\hspace{6.0cm}
\frac{t}{t_{0,*}} = (\delta_{\xi_*}^2+1)
\left[1-\fracb{\delta_{\xi_*}^2}{\delta_{\rm c_{ms,0}}^4}^{2/3}\right]^{-3/2}\ ,
\end{equation}   
where $\delta_{\xi_*}<\delta_{\rm c_{ms,0}}^2$ and
\begin{equation}\hspace{5.7cm}
\frac{t_{0,*}}{t_0} = \frac{(\delta_{\rm c_{ms,0}}^4-1)^{3/2}}
{\delta_{\rm c_{ms,0}}^4(\delta_{\rm c_{ms,0}}^2+1)} =
\frac{4\sigma_0^{3/4}(1+\sigma_0)^{1/4}}
{\left(\sqrt{1+\sigma_0}+\sqrt{\sigma_0}\right)^2}\ .
\label{eq:t_0*}
\end{equation}
For $\sigma_0 \gg 1$ we find that $t_{0,*}\approx t_0$ so that at $t
\gg t_0$ we have $\delta_{\xi_*}^2 \approx 2/(1-\xi_*) \gg 1$ and
\begin{equation}\hspace{0.0cm}
\frac{t}{t_0} \approx 
\frac{2}{1-\xi_*}\left[1-\frac{1}{4\sigma_0^{4/3}(1-\xi_*)^{2/3}}\right]^{-3/2}\ ,
\quad\quad\quad\quad
1-\xi_* \approx \frac{1}{8\sigma_0^2}\left[1+\fracb{t}{16\sigma_0^2t_0}^{-2/3}\right]^{3/2}
\approx \left\{\matrix{2t_0/t
\quad & t \ll 16\sigma_0^2t_0 \ ,\cr\cr 
1/8\sigma_0^2
\quad & t \gg 16\sigma_0^2t_0\ ,}\right. 
\label{xs*sol2}
\end{equation} 
For $\sigma_0 \gg 1$ we also have $\delta_{\rm c_{ms,0}}^2 \approx 4\sigma_0$,
so that for $t_0 \ll t \ll 16\sigma_0^2t_0$, where $\delta_{\xi_*}^2
\approx t/t_0$, equations (\ref{eq:Gamma}) and (\ref{xs*}) imply
\begin{equation}\hspace{4.4cm}
\Gamma(\xi_*) \approx \fracb{\sigma_0t}{2t_0}^{1/3}\ ,\quad\quad\quad\quad\quad
\Gamma_* = (1-\beta_*^2)^{-1/2} \approx \fracb{4\sigma_0^4t}{t_0}^{1/6}\ .
\end{equation}
Note that $\Gamma_*$ is the Lorentz factor of the motion of the head
of the secondary rarefaction wave, while $\Gamma(\xi_*)$ is the
Lorentz factor of the fluid at that location.

\subsection{Analytic calculation of physical quantities at $\xi > \xi_*(t)$}
\label{app:an-int}

It is instructive to calculate the values of relevant physical
quantities in the region between the head of the secondary rarefaction
wave and the vacuum interface, which corresponds to $\xi_*(t) < \xi
<\beta_{\rm max}$. In particular, it can help verify that this region
contains most of the energy and rest mass in the flow during the
magnetic rocket acceleration phase (at $t \ll t_c = \sigma_0^2 t_0$).
Using equations (\ref{eq:Gamma}) and (\ref{B}), at a given time $t$ we
have $dx = td\xi$ so that the rest mass per unit area at $\xi>\xi_*$
is given by
\begin{equation}\hspace{1.5cm}
M[>\xi_*(t)] = \int_{x_*(t)}^{x_{\rm vac}(t)}dx\Gamma(x)\rho(x) =
\frac{\rho_0 t}{8\sigma_0}\int_{\xi_*(t)}^{\beta_{\rm max}}d\xi
\left(\delta_{\rm c_{ms,0}}^{2/3}\delta_\xi^{2/3}+
\frac{1}{\delta_{\rm c_{ms,0}}^{2/3}\delta_\xi^{2/3}}\right)
\left(\frac{\delta_{\rm c_{ms,0}}^{2/3}}{\delta_\xi^{1/3}}-
\frac{{\delta_\xi^{1/3}}}{\delta_{\rm c_{ms,0}}^{2/3}}\right)^2\ ,
\end{equation}
where 
\begin{equation}\label{beta_max}\hspace{6.5cm}
\beta_{\rm max} = \frac{2\sqrt{\sigma_0(1+\sigma_0)}}{1+2\sigma_0}\ ,
\end{equation}
is the maximal fluid velocity, which is obtained at the vacuum
interface, while $\xi_*(t)$ is given implicitly by equation
(\ref{t(xi_*)}).  One can change variables of integration to
$\delta_\xi^2$ and then to $y = \delta_\xi^{2/3}$, using the relations
\begin{equation}\hspace{6.3cm}
d\xi = \frac{2d\delta_\xi^2}{(\delta_\xi^2+1)^2} = 
\frac{6y^2dy}{(y^3+1)^2}\ ,
\end{equation}
and use the simple expression for the initial mass, $M_0 = \rho_0l_0 =
\rho_0t_0[\sigma_0/(1+\sigma_0)]^{1/2}$, to obtain the following
expression for the fraction of the initial rest mass at
$\xi>\xi_*(t)$,
\begin{equation}\hspace{0.6cm}
\frac{M[>\xi_*(t)]}{M_0} = \frac{3\sqrt{1+\sigma_0}}{4\sigma_0^{3/2}}\fracb{t}{t_0}
\int_{y_{\rm min}(t)}^{a^2}dy\frac{y^2}{(y^3+1)^2}\left(ay+\frac{1}{ay}\right)
\left(\frac{a^2}{y}-2+\frac{y}{a^2}\right) =
\frac{\sqrt{1+\sigma_0}}{4\sigma_0^{3/2}}
\fracb{t}{t_0}\frac{(a^2-y_{\rm min})^3}{a^3(y_{\rm min}^3+1)}\ ,
\end{equation}
where $y_{\rm min}(t) = \delta_{\xi_*(t)}^{2/3}$ and $a = \delta_{\rm
c_{ms,0}}^{2/3}$.  This result can be written more explicitly and
simplified using Eqs.~(\ref{t(xi_*)}) and (\ref{eq:t_0*}),
\begin{equation}\hspace{1.2cm}
\frac{M[>\xi_*(t)]}{M_0} = 
\frac{4(1+\sigma_0)}{\left(\sqrt{1+\sigma_0}+\sqrt{\sigma_0}\right)^2}
\left[1+\fracb{\delta_{\xi_*(t)}^2}{\delta_{\rm c_{ms,0}}^4}^{1/3}\right]^{-3}
\left(\delta_{\xi_*(t)}^{2}+1\right)\fracb{t_0}{t}
\approx \left\{\matrix{1
\quad & t \ll \sigma_0^2t_0 \ ,\cr\cr 
2\sigma_0^2t_0/t
\quad & t \gg \sigma_0^2t_0\ ,}\right. 
\end{equation}
where the last asymptotic values are valid for $\sigma_0 \gg 1$.

Using Eq.~(\ref{xs*sol2}), one can calculate the fractional change in
the width of the region between the secondary rarefaction wave and the
vacuum rarefaction, $\Delta_*(t) = t[\xi_h-\xi_*(t)]$, where $\xi_h =
\beta_{\rm max}$,
\begin{equation}\hspace{1.2cm}
\frac{\Delta_*(t)}{\Delta_*(t_0)} = \frac{t[\xi_h-\xi_*(t)]}{t_0[\xi_h-\xi_t]} 
\approx \left[1+\fracb{t}{16\sigma_0^2t_0}^{2/3}\right]^{3/2}-\frac{t}{16\sigma_0^2t_0}
\approx \left\{\matrix{1+(3/2)(t/16\sigma_0^2t_0)^{2/3} \quad & t \ll 16\sigma_0^2t_0 \ ,\cr\cr 
(3/2)(t/16\sigma_0^2t_0)^{1/3}
\quad & t \gg 16\sigma_0^2t_0\ ,}\right. 
\end{equation}
where $\xi_t = -c_{\rm ms,0} = -[\sigma_0/(1+\sigma_0)]^{1/2}$,
$\Delta_*(t_0) =
t_0c_{\rm ms,0}\frac{3+4\sigma_0}{1+2\sigma_0} \approx
2t_0$, and the asymptotic values are valid for $\sigma_0 \gg 1$.

The kinetic energy, $E_{\rm kin} = \int dx\Gamma(\Gamma-1)\rho$, is
given by
\begin{eqnarray}\nonumber\hspace{0.7cm}
\frac{E_{\rm kin}[>\xi_*(t)]}{M_0} &=& 
\frac{3\sqrt{1+\sigma_0}}{8\sigma_0^{3/2}}\fracb{t}{t_0}
\int_{y_{\rm min}(t)}^{a^2}dy\frac{y^2}{(y^3+1)^2}\left(ay+\frac{1}{ay}\right)
\left(ay+\frac{1}{ay}-2\right)
\left(\frac{a^2}{y}-2+\frac{y}{a^2}\right) 
\\
&=&
\frac{\sqrt{1+\sigma_0}}{8\sigma_0^{3/2}}
\fracb{t}{t_0}\left[f(a^2)-f(y_{\rm min}(t))\right]\ ,
\\ \nonumber
f(y) &=& \frac{2a+6a^4+2a^7-(4a^2+6a^5+a^8)y+(1+6a^{3}+4a^6)y^2}{a^4(1+y^3)}
\\
& &+\frac{1-2a^2-2a^6+a^8}{\sqrt{3}\,a^4}\arctan\left(\frac{2y-1}{\sqrt{3}}\right) 
+3\ln(y)+\frac{(a^2+1)^3(a^2-1)}{3a^4}\ln\left(\frac{1+y}{\sqrt{1+y^2-y}}\right)\ .
\end{eqnarray}
The electromagnetic energy, $E_{\rm EM} = \int
dx\rho_0\sigma_0(\Gamma^2-\frac{1}{2})(\rho/\rho_0)^2$, can be
calculated in a similar way,
\begin{eqnarray}\nonumber\hspace{0.5cm}
\frac{E_{\rm EM}[>\xi_*(t)]}{E_{\rm EM,0}} &=& 
\frac{2}{\sigma_0}\frac{E_{\rm EM}[>\xi_*(t)]}{M_0} =
\frac{3\sqrt{1+\sigma_0}}{16\sigma_0^{5/2}}\fracb{t}{t_0}
\int_{y_{\rm min}(t)}^{a^2}dy\frac{y^2}{(y^3+1)^2}\left(a^2y^2+\frac{1}{a^2y^2}\right)
\left(\frac{a^2}{y}-2+\frac{y}{a^2}\right)^2
\\
&=&
\frac{3\sqrt{1+\sigma_0}}{16\sigma_0^{5/2}}
\fracb{t}{t_0}\left[g(a^2)-g(y_{\rm min}(t))\right]\ ,
\\ \nonumber
g(y) &=& \frac{y}{a^2}-\frac{a^2}{y}
-\frac{1+8a^6+a^{12}-7a^4y-4a^{10}y+4a^2y^2+7a^8y^2}{3a^6(1+y^3)}
\\
& &-\frac{4(1-2a^2-2a^6+a^8)}{3^{3/2}\,a^4}\arctan\left(\frac{2y-1}{\sqrt{3}}\right) 
-4\ln(y)-\frac{4(a^2+1)^3(a^2-1)}{9a^4}\ln\left(\frac{1+y}{\sqrt{1+y^2-y}}\right)\ .
\end{eqnarray}
The total energy (including rest energy) is simply $\int T^{00}dx =
M+E_{\rm kin}+E_{\rm EM}$. This can be used for the normalization when
calculating the average values of quantities weighed by 
\begin{equation}\hspace{2.5cm}
T^{00}(y,a) = \frac{\rho_0}{64\sigma_0}
\left[4\left(ay+\frac{1}{ay}\right)^2\left(\frac{a^2}{y}-2+\frac{y}{a^2}\right)
+\left(a^2y^2+\frac{1}{a^2y^2}\right)\left(\frac{a^2}{y}-2+\frac{y}{a^2}\right)^2\right]\ .
\end{equation}
We find that
\begin{equation}\hspace{2.5cm}
\frac{E[>\xi_*(t)]}{E_0} = \frac{2E[>\xi_*(t)]}{(2+\sigma_0)M_0} = 
\frac{\sqrt{1+\sigma_0}}{16(2+\sigma_0)\sigma_0^{3/2}}\fracb{t}{t_0}
\frac{(a^2-y_{\rm min})^3\left[y_{\rm min}(1+a^6)+3a^2(1+a^2y_{\rm min}^2)\right]}
{a^6y_{\rm min}(y_{\rm min}^3+1)}\ ,
\end{equation}
and the same holds for the energy above some $\xi>\xi_*(t)$ where
$y_{\rm min}(t) = \delta_{\xi_*(t)}^{2/3}$ is replaced by $y(t) =
\delta_{\xi}^{2/3}$.  In order to calculate the mean
Lorentz $\Gamma$ factor or magnetization $\sigma$ at $\xi>\xi_*$
(denoted by $\langle\Gamma\rangle_*$ and $\langle\sigma\rangle_*$,
respectively) one needs to calculate the following integrals:
\begin{equation}
\frac{1}{M_0}\int_{x_*(t)}^{x_{\rm vac}(t)}dx\,T^{00}\,\Gamma = 
\frac{\sqrt{1+\sigma_0}}{\sigma_0^{3/2}}\fracb{t}{t_0}
\left[f_\Gamma(a^2)-f_\Gamma(y_{\rm min}(t))\right]\ ,
\end{equation}
\begin{eqnarray}\nonumber
f_\Gamma(y) &=& -\frac{3(a^2-y^4)}{128ay^2} 
-\frac{9a^4(1+a^6)+a^2(1-25a^6+a^{12})y-(1-25a^6+a^{12})y^2}{64a^7(1+y^3)}
-\frac{3(a^6-1)}{64a^3}\ln(1+y^3)+\frac{3(a^6-2)}{64a^3}\ln(y)
\\
& &+\frac{(1+a^2)(1-7a^6+a^{12})}{64\sqrt{3}\,a^7}\arctan\left(\frac{2y-1}{\sqrt{3}}\right)
+\frac{(a^2-1)(1-7a^6+a^{12})}{192a^7}
 \ln\left(\frac{1+y}{\sqrt{1+y^2-y}}\right)\ .
\end{eqnarray}
\begin{equation}
\frac{1}{M_0}\int_{x_*(t)}^{x_{\rm vac}(t)}dxT^{00}\sigma = 
\frac{\sqrt{1+\sigma_0}}{\sigma_0^{3/2}}\fracb{t}{t_0}
\left[f_\sigma(a^2)-f_\sigma(y_{\rm min}(t))\right]\ ,
\end{equation}
\begin{eqnarray}\nonumber
f_\sigma(y) = -\frac{3(a^8-4a^6y+4a^2y^3-y^4)}{256a^4y^2}
+\frac{2a^2(1-25a^6+a^{12})-(1-34a^6-8a^{12})y-a^4(8+34a^6-a^{12})y^2}{128a^8(1+y^3)}
-\frac{3}{128}\ln(y)
\\
+\frac{1+10a^4-16a^6-16a^{10}+10a^{12}+a^{16}}{128\sqrt{3}\,a^8}
\arctan\left(\frac{2y-1}{\sqrt{3}}\right)
+\frac{(a^2+1)^5(1-5a^2+5a^4-a^6)}{384a^8}
 \ln\left(\frac{1+y}{\sqrt{1+y^2-y}}\right)\ .
\end{eqnarray}

\section{Generalizing to a Spherical flow}
\label{app:spherical}

Here we consider the case of cold ($p_g=0$) radial flow.  We assume
that the flow is one-dimensional and the magnetic field is
perpendicular to the radial direction. Obviously, this is not fully
self-consistent, but this is a reasonable approximation for an
equatorial wedge. More accurate two dimensional treatments are saved
for future works.

The basic equations for a one dimensional flow in spherical symmetry are
\begin{eqnarray}\label{a1}
\partial_{t}(\rho\Gamma) + \frac{1}{r^2} \partial_{r} (r^2\rho\Gamma v ) &=& 0 
\quad\quad {\rm (continuity)}\ ,
\\ \label{a2}
\partial_{t} (b\Gamma) + \frac{1}{r} \partial_{r} (r b \Gamma v) &=& 0
\quad\quad {\rm (magnetic\ field)}\ ,
\\ \label{a3}
\partial_{t} \left[(\rho+b^2)\Gamma^2 - \frac{b^2}{2} \right] + 
\frac{1}{r^2} \partial_{r} \left[r^2(\rho+b^2)\Gamma^2v\right] &=& 0 
\quad\quad {\rm (energy)}\ ,
\\ \label{a4}
\partial_{t} \left[(\rho+b^2)\Gamma^2 v\right] + 
\frac{1}{r^2} \partial_{r} \left[r^2 \left((\rho+b^2)\Gamma^2v^2  
+\frac{b^2}{2}\right)\right] &=& 0 \quad\quad {\rm (momentum)}\ ,
\end{eqnarray}
where $v$ is the velocity, $b=B/\sqrt{4\pi}\,\Gamma$, and $B$ is the
(azimuthal) magnetic field as measured in the source frame.

One can introduce new variables, $\bar{b}$, $\bar{\rho}$ and $x$ as follows
\begin{equation}
  \rho= r^{-2}\bar{\rho}\ , \quad b=r^{-1}\bar{b}\ , \quad x=r\ ,
\label{c1}
\end{equation}
which upon substitution into Eqs.~(\ref{a1}-\ref{a3}) yield
\begin{eqnarray}\label{c2}
\partial_{t} (\bar{\rho}\Gamma) + \partial_{x}(\bar{\rho}\Gamma v) &=& 0\ ,
\\ \label{c3}
\partial_{t} (\bar{b}\Gamma) + \partial_{x}(\bar{b}\Gamma v) &=& 0\ ,
\\ \label{c4}
\partial_{t} \left[(\bar{\rho}+\bar{b}^2)\Gamma^2 -\frac{\bar{b}^2}{2}\right] 
+ \partial_{x}\left[(\bar{\rho}+\bar{b}^2)\Gamma^2 v\right] &=& 0\ ,
\\ \label{c5}
\partial_{t} \left[(\bar{\rho}+\bar{b}^2)\Gamma^2v\right] 
+ \partial_{x}\left[\left((\bar{\rho}+\bar{b}^2)\Gamma^2 v^2  
+\frac{\bar{b}^2}{2}\right)\right] &=& 0\ . 
\end{eqnarray}
These equations are identical to those of plane cold ($p_g = 0$)
flow. Therefore, all the results obtained for the planar case, including the
self-similar solution, can be utilized in the spherical case.

After the substitution $ \bar{p}\to\bar{b}^2/2$, Eqs.~(\ref{c2})
,(\ref{c4}), and (\ref{c5}) also become identical to those of
unmagnetized plasma.  From Eqs.~(\ref{c2}) and (\ref{c3}) it
follows that $\bar{b}/\bar{\rho} = {\rm const}$, and $\bar{p}
\propto\bar{\rho}^2$. Thus, the ratio of specific heats for this
plasma is $\gamma=2$.

\section{Self-similar solution for the shell's tail.} 
\label{app:self-sim:tail}

Our starting point is Eqs.~(\ref{c2})-(\ref{c5}), which are valid
for both the planar and spherical (after the substitution (\ref{c1}))
cases.  Equations (\ref{c2}) and (\ref{c3}) imply
\begin{equation}
\label{eq:db_drho}
\frac{d}{dt}\left(\frac{\bar{b}}{\bar{\rho}}\right) = 0\ ,
\end{equation}
that is $\bar{b}/\bar{\rho}$ remains constant for each fluid
element, and is determined by the initial conditions.

Just like in Appendix~\ref{app:self-sim} we introduce the self-similar variable 
$ \xi = x/t$, but this time we seek solutions of the form 
\begin{equation} 
v=V(\xi)\ ,\quad \bar{\rho} = t^\alpha F(\xi)\ ,\quad \bar{b} = t^\alpha G(\xi)\      
\label{bb4}
\end{equation}
(Since $\bar{b}/\bar{\rho}$ remains constant for each fluid element,
$\bar{b}$ and $\bar{\rho}$ must have the same temporal
scaling.). Using Eqs.~(\ref{c2}) and (\ref{c3}),
Eq.~(\ref{c5}) can be reduced to
\begin{equation} 
\bar{\rho}\Gamma \left( \frac{\partial \Gamma v}{\partial t} + 
                        v\frac{\partial \Gamma v}{\partial x}  \right) + 
 \bar{b}^2\Gamma^2 \frac{\partial v}{\partial t} + 
 \frac{1}{2}\frac{\partial b^2}{\partial x} =0. 
\label{bb5}
\end{equation}
Substituting the expressions (\ref{bb4}) into this equation we obtain 
\begin{equation} 
   t^\alpha F\Gamma (V-\xi) (\Gamma V)'+t^{2\alpha}(GG'-G^2\Gamma^2\xi V')=0.
\label{bb6}
\end{equation}
This equation is satisfied for any $t$ only in the following two cases. 
First, if $\alpha=0$ --  
this is the case of simple rarefaction wave analyzed in Appendix~\ref{app:self-sim}. 
Second, if
\begin{equation}
   V = \xi\ , 
\label{bb7}
\end{equation}
and
\begin{equation}
   GG'-G^2\Gamma^2\xi V'=0\ .
\label{bb8}
\end{equation}
Integrating the last equation (after substitution of equation
[\ref{bb7}] into it) we find that
\begin{equation}
\label{bb9}
G(\xi) = \frac{A}{\sqrt{1-\xi^2}} = A\Gamma(\xi)\ .
\end{equation}
Substitution of  expressions (\ref{bb4}) into Eqs.~(\ref{c2}) and (\ref{c3}) leads to 
\begin{equation}
(1+\alpha)F = 0\ ,\quad (1+\alpha)G = 0\ ,
\label{bb10}
\end{equation}
which are satisfied when $\alpha=-1$. Function $F(\xi)$, however, remains 
undefined. 

If the initial solution is uniform, as in the planar case considered in 
\S~\ref{sec:test_case}, 
with $\rho_0$ and $b_0$ being the initial rest mass density and magnetic field 
respectively, then from Eq.~(\ref{eq:db_drho}) we obtain  
\begin{equation}
   \bar{\rho} = \frac{\bar{\rho}_0}{\bar{b}_0}\bar{b}\ ,
\end{equation}
and thus, 
\begin{equation}
F(\xi) = \frac{\bar{\rho}_0}{\bar{b}_0} G(\xi) 
= \frac{\bar{\rho}_0}{\bar{b}_0} \frac{A}{\sqrt{1-\xi^2}}\ .
\end{equation}

Thus, a self-similar solution of the required form does exist. 
In planar geometry, this is 
\begin{equation}
v = \xi\ ,
\quad b = \frac{1}{t} \frac{A}{ (1-\xi^2)^{1/2} } \ ,\quad 
\rho = \frac{F(\xi)}{t} \ ,
\end{equation}
and in the spherical geometry
\begin{equation}
v = \xi \ ,\quad 
b = \frac{1}{t^2} \frac{A}{ \xi(1-\xi^2)^{1/2} } \ ,\quad
\rho = \frac{1}{t^3} \frac{F(\xi)}{\xi^2} \ .
\end{equation}
In both cases the magnetization parameter of fluid elements 
decreases linearly with time:
\begin{equation}
\sigma = \frac{b^2}{\rho} = \frac{\bar{b}^2}{\bar{\rho}} \propto t^{-1}\ .
\end{equation}

Clearly, this solution is only applicable for $\xi<1$.  Moreover, it
cannot be simply truncated at some large $\xi$ and continued with
vacuum. Instead, it should terminate at a shock or smoothly transform
into a non-self-similar flow. For our simple test problem, this
solution cannot become asymptotically valid up to $\xi = \beta_{\rm
max}$ given by Eq.~(\ref{beta_max}) at very late times ($t \gg t_c$),
since it cannot simultaneously satisfy the global conservation of
energy and rest mass for an initially uniform shell. This implies a
limited region of applicability, in the tail of the flow but not at
its head.\footnote{Likely, once $\sigma(\xi_*)$ drops below unity the
flow can no longer efficiently rearrange itself from the original
self-similar structure (described in Appendix~\ref{app:self-sim}) to
the new one (described here), as the secondary rarefaction wave
becomes weak).} General considerations, however, show that some
similar scalings still apply at the head of the flow at late times:
$\sigma \propto t^{-1}$ while $\rho \propto t^{-1}$ ($t^{-3}$) in
planar (spherical) geometry. Moreover, $v = \xi$ at asymptotic late
times when the magnetic pressure becomes dynamically unimportant and
each fluid element moves ballistically.

Finally, we calculate the scalings of the total kinetic 
$E_{\rm kin}(t,\xi_1,\xi_2) $ and magnetic $E_B(t,\xi_1,\xi_2)$ energy of a 
fluid element bounded  by $x_1=\xi_1 t$ and $x_2=\xi_2 t$.  Both in planar 
geometry, where $dV \propto dx = td\xi$, and in
spherical geometry, where $dV \propto r^2 dr = t^3\xi^2d\xi$, 
we obtain
\begin{eqnarray}
E_{\rm kin}(t,\xi_1,\xi_2) &=&  \int dV\rho\Gamma(\Gamma-1) \propto t^0\ ,
\\ \nonumber
\\
E_B(t,\xi_1,\xi_2) &=& \int dV\, b^2\left(\Gamma^2-\frac{1}{2}\right) \propto t^{-1}\ 
\end{eqnarray}

\section{Acceleration after the separation from the wall} 
\label{app:temp}

\subsection{Mechanical analogy: two masses and a spring}
\label{app:mechanical-analogy}

At $t < t_0$ the acceleration of the shell as a whole occurs mainly
because the back end of the shell is pushing against the wall, and
therefore this mode of acceleration can remain effective only as long
as the shell remains in causal contact with the wall. Therefore, the
initial shell crossing by the rarefaction wave accelerates the shell
up to $\langle\Gamma\rangle\sim \sigma_0^{1/3}$, and soon after $t_0$
the shell can no longer effectively push against the wall, as the
magnetic pressure at the wall drops dramatically, and the subsequent
change in the total momentum $P$ of the shell due to the force $F$
exerted on it by the wall ($dP = Fdt$) becomes negligible.

It is a somewhat surprising result that after the shell separates from
the wall its mean Lorentz factor continues to increase with time
despite the apparent lack of any external force.  This can be
understood as follows. The total energy and momentum of the shell are
indeed conserved in the lack of an external force (or energy losses or
gains). However, the shell expands under its own pressure, and
develops a considerable relative velocity between its leading and
trailing edges. In its center of mass frame the energy and momentum of
the front and back ends of the shell are comparable. However, if the
expansion is relativistic in the center of mass (or comoving) frame
then in the lab frame the energy and momentum of the leading part are
much larger than those of the trailing part, and thus the leading part
dominates the total energy and the Lorentz factor when averaged over
the energy in the lab frame.

This may be illustrated by the following simple example. Consider two
identical masses $m$ moving together with a compressed ideal massless
spring between them, with potential energy $E_{\rm pot}$ in its own
rest frame ($S_*$, which is also the rest frame of the two masses,
hereafter the comoving frame). The energy of the system in its own
(comoving) rest frame is $E' = 2mc^2 + E_{\rm pot}$, and in a frame
where this system is moving at a Lorentz factor $\Gamma =
(1-\beta^2)^{-1/2}$ in the positive x-direction (hereafter, the lab
frame), its energy is $E = \Gamma E' = \Gamma(2mc^2+E_{\rm pot})$ and
its momentum is $P_x = \Gamma\beta E'/c = \Gamma\beta(2mc^2+E_{\rm
pot})/c = \beta E/c$ in the x-direction (while $P_y = P_z = 0$). Then
the spring is released and all of its potential energy is converted to
kinetic energy of the two masses, which in the comoving frame now move
at a Lorentz factor $\Gamma_* = (1-\beta_*^2)^{-1/2}$ such that
$E_{\rm pot} = 2(\Gamma_*-1)mc^2$ and $\Gamma_*=E'/2mc^2$, in the
positive and negative x'-directions, respectively (the two masses are
thus denoted by subscripts `$+$' and `$-$' accordingly). In the
comoving frame their energy-momentum 4-vectors read
$u^{\prime\;\mu}_\pm = \Gamma_*(1,\pm\beta_*,0,0)$, and a simple
Lorentz transformation shows that in the lab frame
\begin{equation}\hspace{5.7cm}
u_\pm^\mu =
[\Gamma\Gamma_*(1\pm\beta\beta_*),\Gamma\Gamma_*(\beta\pm\beta_*),0,0]\ ,
\end{equation}
which indeed satisfies $E = E_+ + E_- = mc^2(u_+^0 + u_-^0) =
2\Gamma\Gamma_* mc^2 = \Gamma E'$ and $P_x = mc(u_+^1 + u_-^1) = \beta
E/c$ (and $P_y = P_z = 0$), as it should, while $\Gamma_\pm =
E_\pm/mc^2 = \Gamma\Gamma_*(1\pm\beta\beta_*)$. Thus, the ratios of
the energy and momentum of the two masses, and the fractions of the
total energy and momentum that each mass holds are given by
\begin{equation}\hspace{1.6cm}
\frac{E_+}{E_-} = \frac{\Gamma_+}{\Gamma_-} =
\frac{1+\beta\beta_*}{1-\beta\beta_*}\ ,\quad\quad
\frac{P_{x+}}{P_{x-}} =
\frac{\Gamma_+\beta_+}{\Gamma_-\beta_-} =
\frac{\beta+\beta_*}{\beta-\beta_*}\ ,\quad\quad
\frac{E_\pm}{E} = \frac{1\pm\beta\beta_*}{2}\ ,\quad\quad 
\frac{P_{x\pm}}{P_x} = \frac{\beta \pm \beta_*}{2\beta}\ .
\end{equation}
For $\Gamma,\,\Gamma_* \gg 1$ we have 
\begin{equation}\hspace{5.2cm}
\frac{E_-}{E_+}   
\approx \frac{1}{4}\left(\frac{1}{\Gamma^2}+\frac{1}{\Gamma_*^2}\right)\ll 1\ ,
\quad
\frac{\Gamma_+}{\Gamma}\approx 2\Gamma_*\gg 1\ .
\end{equation}
Therefore, almost all of the energy in the lab frame is in the leading
mass (or leading part of the shell), which has greatly increased its
Lorentz factor. For $\Gamma_* = \Gamma$ the only energy left in the
trailing mass (or trailing part of the shell) is its rest mass energy
and all of the potential energy is converted into the kinetic energy
of the leading mass, which in this case also carries all of the
momentum.
Thus we can see that the leading mass, which constitutes one half of
original rest mass, ends up with almost all of the energy and with a
much higher Lorentz factor than what it started with.  Going back to
our magnetized shell, the potential energy in a ``spring'' is the
analog of the magnetic energy in the shell, and similarly to the
mechanical analog eventually most of the energy ends up in a good
fraction of the original rest mass, that can reach a very high Lorentz
factor (much larger than the initial Lorentz factor of the shell).

\subsection{Evolution of $\mean{\Gamma}$ after the separation: an alternative derivation of the scaling $\mean{\Gamma}\propto t^{1/3}$}
\label{sec:acc2}

Here we follow the mean shell parameters but drop `$\mean{}$' in the
notation, for simplicity. Let us consider a planar\footnote{The same
same reasoning essentially also holds for a spherical geometry, as
demonstrated in Appendix~\ref{app:spherical}.}  shell initially (at
lab frame time $t = 0$) at rest in some rest frame $S_0$, which we
refer to as the lab frame, in which it has a width $l_0$, magnetic
field $B_0$, rest-mass density $\rho_0$, magnetization $\sigma_0 =
B_0^2/4\pi\rho_0c^2 \gg 1$, energy $E_0$, and no (or negligible)
thermal pressure (for simplicity). This shell can either be leaning
against a wall to one end (at $x = -l_0$), or be half of an unbounded
shell (initially occupying $-2l_0
\leq x \leq 0$).  We have shown that initially the shell expands due
to the passage of a self-similar rarefaction wave, which crosses the
shell over a time $t_0 \approx l_0/c$, and is accelerated to a typical
Lorentz factor of $\Gamma_1 \sim
\sigma_0^{1/3}$.

Now, even though the shell is no longer perfectly uniform as in our
initial configuration, we consider the part of the shell that carries
most of its energy (as measured in $S_0$), which is expected to
be roughly uniform (the relevant physical quantities not changing by
more than factors of order unity within that region), and make the
analogy between it in its own rest frame, $S_1$ (which moves at a
Lorentz factor $\sim \Gamma_{1,0} \sim
\sigma_0^{1/3}$ relative to $S_0$), and our original configuration
(that was quantified in $S_0$). Even though this analogy is not
perfect, we still expect a similar qualitative behavior, and a similar
quantitative behavior up to factors of order unity (which we discard
here, as we are interested only in the relevant scaling laws).  

One difference, however, is that for reasonably smooth initial
conditions we no longer have a strong rarefaction wave crossing the
shell, which eventually splits it in two (as for a perfectly uniform
shell with sharp edges surrounded by vacuum on both sides, where the
two rarefaction waves from both sides meet and are secondary -- our
``wall'' for a one-sided shell). Thus, the shell is basically the
smooth peak of the lab frame energy density (which scales as $B^2$
when $\sigma \gg 1$; see the profile of $B$ at $t=70t_0$ in
Fig.~[\ref{fig:sim1}]). The shell still significantly expands in its
own rest frame, reaching speeds of order of its fast magnetosonic
speed on its fast magnetosonic (or light) crossing time. However, this
spreading is smooth and continuous, and the shell does not split in
two, but instead it remains a smooth peak of the lab frame energy
density. Material ahead of the peak in the pressure (at the front of
the shell) is accelerated, while material behind this peak (at the
back of the shell) is decelerated (by the pressure gradient, in both
cases).

Nonetheless, it is instructive to divide this process into discrete
steps or phases. The approximate initial conditions of the ``second
phase'' in the evolution of the shell, as expressed in frame $S_1$
(using a subscript ``1'' for all the relevant quantities, when
measured in this frame) are 
$$
l_1 \sim \sigma_0^{1/3}l_0\ ,\quad\quad B_1 \sim \frac{B_0}{\sigma_0^{1/3}}\ ,\quad\quad
\rho_1 \sim \frac{\rho_0}{\sigma_0^{1/3}}\ ,\quad\quad
\sigma_1 \sim \sigma_0^{2/3}\ ,\quad\quad
E_1 \sim \frac{E_0}{\sigma_0^{1/3}}\ .
$$
In frame $S_1$, the two sides of the shell are expected to accelerate
in opposite directions, and develop velocities of the order of the
shell's magnetosonic speed, $\Gamma_{2,1} \sim \sigma_1^{1/3} \sim
\sigma_0^{2/9}$, on the shell's magnetosonic (or light) crossing time,
$t_1 \sim l_1/c \sim \sigma_0^{1/3}t_0$ (as measured in frame
$S_1$). The same scalings as before should approximately hold here as
well,
$$
l_2 \sim \sigma_1^{1/3}l_1 \sim \sigma_0^{5/9}l_0\ ,\quad\quad
B_2 \sim \frac{B_1}{\sigma_1^{1/3}} \sim \frac{B_0}{\sigma_0^{5/9}}\ ,\quad\quad
\rho_2 \sim \frac{\rho_1}{\sigma_1^{1/3}} \sim \frac{\rho_0}{\sigma_0^{5/9}}\ ,\quad\quad 
\sigma_2 \sim \sigma_1^{2/3} \sim \sigma_0^{4/9}\ ,\quad\quad
E_2 \sim \frac{E_1}{\sigma_1^{1/3}} \sim \frac{E_0}{\sigma_0^{5/9}}\ .
\quad\quad\quad\quad
$$
We note that in frame $S_0$ (i.e. in the lab frame) almost all of the
energy ends up in the front part of the shell (which was accelerated
in the direction of motion of $S_1$ relative to $S_0$), whose rest
frame $S_2$ moves at a Lorentz factor $\Gamma_{2,0}
\sim \Gamma_{1,0}\Gamma_{2,1} \sim \sigma_0^{5/9}$ relative to $S_0$, 
and has an energy of $E_{2,0} \sim \Gamma_{2,0} E_2 \sim E_0$ as
measured in $S_0$. The back part of the shell has a Lorentz factor of
$\sim \Gamma_{1,0}/\Gamma_{2,1} \sim \sigma_0^{1/9}$ in frame $S_0$,
while its energy in frame $S_0$ is $\sim \sigma_0^{1/9}E_2 \sim
E_0/\sigma_0^{4/9}$, i.e. a factor of $\sim \sigma_0^{4/9} \gg 1$
smaller than that of the forward half of the shell, so it can be
safely discarded as we are interested in the part of the shell that
carries most of the energy in frame $S_0$.
In frame $S_0$, the the second phase of acceleration takes a time
$t_{1,0} \sim \Gamma_{1,0}t_1 \sim \sigma_0^{2/3}t_0$. During this
time $\Gamma$ increased from $\Gamma_{1,0} \sim \sigma_0^{1/3}$ to
$\Gamma_{2,0} \sim \sigma_0^{5/9}$, i.e. by a factor of $\sim
\sigma_0^{2/9}$, implying
\begin{equation}\hspace{5.2cm}
\frac{\Gamma_{2,0}}{\Gamma_{1,0}} \sim \sigma_0^{2/9} \sim 
\left(\frac{t_{1,0}}{t_0}\right)^{1/3} \quad \Longrightarrow \quad
\Gamma \propto t^{1/3}\ .
\end{equation}

Similarly, recursively repeating the same procedure $n$ times, it can
be shown that
$$
\vec{U}_n \sim \sigma_{n-1}^{1/3}\vec{U}_{n-1} 
\sim \sigma_0^{1-(2/3)^n}\vec{U}_0, 
$$ where $\vec{U}_n =
(l_n,\,t_n,\,\Gamma_{n,0},\,B_n^{-1},\,\rho_n^{-1},\,\sigma_n^{-1},\,E_n^{-1})$
and $\Gamma_{0,0}=1$.
Noting that $t_{n,0} \sim \Gamma_{n,0}t_n \sim t_n^2/t_0$, this
implies that
$$
\frac{\log\sigma_n}{\log\sigma_0} \sim \left(\frac{2}{3}\right)^n\ ,\quad\quad
\frac{\log\Gamma_{n,0}}{\log\sigma_0} \sim \left[1-\left(\frac{2}{3}\right)^n\right]\ ,\quad\quad
\frac{\log\left(\frac{t_{n-1,0}}{t_0}\right)}{\log\sigma_0}
\sim 2\left[1-\left(\frac{2}{3}\right)^{n-1}\right]\ ,\quad\quad
\frac{\log\left(\frac{\Gamma_{n,0}}{\Gamma_{1,0}}\right)}{\log\sigma_0}
\sim \frac{2}{3}\left[1-\left(\frac{2}{3}\right)^{n-1}\right]\ ,\quad\quad
$$
and
\begin{equation}\hspace{5.5cm}
\frac{\log(\Gamma_{n,0}/\Gamma_{1,0})}{\log(t_{n-1,0}/t_0)} = \frac{1}{3}
\quad \Longrightarrow \quad \Gamma \propto t^{1/3}\ .
\end{equation}
It can also be seen that in the limit\footnote{In practice the
approximation that $\sigma \gg 1$ breaks down at $\sigma = \sigma_{\rm
min} \sim {\rm a\ few}$, after $\approx \log(\log\sigma_{\rm
min}/\log\sigma_0)/\log(2/3)$ steps, or $\sim 4-6$ for $\sigma_0\sim
10^3$ and $\sigma_{\rm min}\sim 1.8-3.9$, so even for $\sigma_0\gg 1$
the shell becomes kinetic-energy dominated within a rather small
number of steps.} of $n \gg 1$, $\Gamma_{n,0}
\to
\sigma_0$, i.e. the Lorentz factor approaches its asymptotic value that 
is achieved when $\sigma \sim 1$ at $t_c \sim \sigma_0^2 t_0$. This
implies a transition radius to the coasting phase of $R_c
\sim\sigma_0^2 l_0 \sim \Gamma_c^2 l_0$, where $\Gamma_c = \Gamma(R_c)
\sim\sigma_0$. Up until this time the shell width in the lab frame
($S_0$) remains approximately constant, $l_{n,0} \sim l_n/\Gamma_{n,0}
\sim l_0$.

This derivation relies on the fact that the shell, which represents
the leading part of the flow, carries most of the total energy and
rest mass during the impulsive acceleration phase, so that its energy
and rest mass are practically constant. As we have seen in
\S~\ref{sec:ref-phase} this condition is satisfied for at least as
long as the shell remains highly magnetized, $\sigma \ga 1$.

\section{The center of momentum frame and calculations in different frames of reference} 
\label{app:CM-frame}

Let us consider a frame of reference $S_1$ moving at a dimensionless
velocity $\beta_1$ in the positive $x$-direction relative to the lab
frame , and let us denote quantities measured in this frame with a
prime. Also, let $(t_a,\,x_a)=(t'_a,\,x'_a)=(0,\,0)$ correspond the
event `$a$' of exposing the original front of the magnetized shell at
rest to the vacuum, i.e. the onset of motion of the shell material.
It is easy to show that the event `$b$' of the original rarefaction
wave hitting the wall corresponds to $(t_b,\,x_b) = (t_0,-l_0)$, where
$t_0 = l_0/c_{\rm ms,0} = l_0[(1+\sigma_0)/\sigma_0]^{1/2}$, and
$(t'_b,\,x'_b) = t_0\Gamma_1(1+\beta_1 c_{\rm ms,0},\,-\beta_1 -c_{\rm
ms,0},)$. Initially (at $t'=0$) the shell width, density and rest-mass
are $l'_0 = l_0/\Gamma_1$, $\rho'_0 =\Gamma_1\rho_0$ and $M_0 =
\rho_0l_0 = \rho'_0 l'_0$, respectively, so that
\begin{equation}\hspace{4.35cm}
T^{\prime,0x} = -\rho_0(1+\sigma_0)\Gamma_1u_1\ ,
\quad\quad
P'_0 = l'_0T^{\prime,0x} = -M_0(1+\sigma_0)u_1\ .
\end{equation}
Now, between $t'=0$ and $t'=t'_b = \Gamma_1(1+\beta_1 c_{\rm
ms,0})l_0[(1+\sigma_0)/\sigma_0]^{1/2}$ the momentum increases due to
the external force exerted by the wall, $dP'/dt' = \rho_0\sigma_0/2$
so that at $t'=t'_b$ when the rarefaction wave reaches the wall and
the wall is removed (so that from that point on there are no external
forces and the total momentum $P'$ and energy $E'$ remain constant) we
have 
\begin{equation}\hspace{3.5cm}\label{eq:P'}
P'(t' \geq t'_b) = P'_0 +t'_b\frac{\rho_0\sigma_0}{2} = -M_0(1+\sigma_0)u_1
+ M_0\Gamma_1\frac{\sigma_0}{2}\left[\sqrt{\frac{1+\sigma_0}{\sigma_0}}+\beta_1\right]\ .
\end{equation}
We are interested in the center of momentum frame in which by
definition $P' = 0$.  Note that here, in contrast to the previous
subsection, we evaluate the total momentum simultaneously in this frame,
rather than in the lab frame. According to Eq.~(\ref{eq:P'}), this
corresponds to
\begin{equation}\hspace{5.35cm}\label{eq:beta_CM'}
\beta_{\rm CM} = \frac{\sqrt{\sigma_0(1+\sigma_0)}}{2+\sigma_0}\ ,
\quad\quad
\Gamma_{\rm CM} = \frac{2+\sigma_0}{\sqrt{4+3\sigma_0}}\ .
\end{equation}
At the CM frame (which moves at $\beta_1 = \beta_{\rm CM}$ relative to
the lab frame), $P'(t'\geq t'_b) = 0$, so that the total momentum
remains zero from the time when the shell separates from the wall.  In
this sense, the shell as a whole simply does not accelerate in the CM
frame at $t'\geq t'_b$. Moreover, the energies of the front part and
the back part are comparable in this frame. The total energy at $t'=0$
is
\begin{equation}\hspace{3.42cm}
T^{\prime,00} = \rho_0\left[\Gamma_1^2(1+\sigma_0)-\frac{\sigma_0}{2}\right]\ ,
\quad\quad
E'_0 = l'_0T^{\prime,00} = M_0\left[\Gamma_1(1+\sigma_0)-\frac{\sigma_0}{2\Gamma_1}\right]\ ,
\end{equation}
and between $t'=0$ and $t' = t'_b$ it decreases due to the negative
work performed on it by the receding wall at its back, $dE'/dt' = \vec{F}\cdot\vec{v} = -\beta_1\rho_0\sigma_0/2$, so that
\begin{equation}
E'(t' \geq t'_b) = E'_0 -t'_b\beta_1\frac{\rho_0\sigma_0}{2} = 
M_0\left[\Gamma_1(1+\sigma_0)-\frac{\sigma_0}{2\Gamma_1}\right]
-M_0\frac{\sigma_0}{2}u_1\left[\sqrt{\frac{1+\sigma_0}{\sigma_0}}+\beta_1\right]
= M_0\Gamma_1\left[1+\frac{\sigma_0-\beta_1\sqrt{\sigma_0(1+\sigma_0)}}{2}\right]\ .
\end{equation}
For $\beta_1 = \beta_{\rm 1,CM}$ given by Eq.~(\ref{eq:beta_CM'}) this reduces to
\begin{equation}\hspace{5.6cm}
E'_{\rm CM}(t' \geq t'_b) = \frac{\sqrt{4+3\sigma_0}}{2}M_0 
= \frac{E_0}{\Gamma_{\rm 1,CM}}\ .
\end{equation} 

The self-similar solution describing the original rarefaction wave can
be expressed in a rest frame $S'$ moving at a velocity $\beta_w$ in
the negative $x$-direction relative to the lab frame (where the wall
is at rest), so that in this frame the wall is moving at a speed
$\beta_w$ in the positive $x'$-direction. This implies
\begin{equation}\hspace{2.0cm}
\xi' = \frac{x'}{t'} = \frac{\xi+\beta_w}{1+\beta_w\xi}\ ,\quad\quad
\xi = \frac{x}{t} = \frac{\xi'-\beta_w}{1-\beta_w\xi'}
\quad\quad\Longrightarrow\quad\quad
\delta_\xi = \frac{\delta_{\xi'}}{\delta_w}\ ,\quad\quad
\delta_v = \frac{\delta_{v'}}{\delta_w}\ ,
\end{equation}
\begin{equation}\hspace{0.25cm}
\label{deltav_w}
\delta_{v'} = \delta_{\rm c_{ms,0}}^{2/3}\delta_w^{1/3}\delta_{\xi'}^{2/3}\ ,\quad\quad
v' = \frac{\delta_{\rm c_{ms,0}}^{4/3}\delta_w^{2/3}\delta_{\xi'}^{4/3}-1}
{\delta_{\rm c_{ms,0}}^{4/3}\delta_w^{2/3}\delta_{\xi'}^{4/3}+1}\ ,\quad\quad
\Gamma' = \frac{\delta_{\rm c_{ms,0}}^{4/3}\delta_w^{2/3}\delta_{\xi'}^{4/3}+1}
{2\delta_{\rm c_{ms,0}}^{2/3}\delta_w^{1/3}\delta_{\xi'}^{2/3}}\ ,\quad\quad
u' = \Gamma' v' = \frac{\delta_{\rm c_{ms,0}}^{4/3}\delta_w^{2/3}\delta_{\xi'}^{4/3}-1}
{2\delta_{\rm c_{ms,0}}^{2/3}\delta_w^{1/3}\delta_{\xi'}^{2/3}}\ ,
\end{equation}
\begin{equation}\hspace{5.25cm}
\label{deltac_w}
\delta_{\rm c_{ms}} = \frac{\delta_{\rm c_{ms,0}}^{2/3}\delta_w^{1/3}}{\delta_{\xi'}^{1/3}}\ ,
\quad\quad
c_{\rm ms}^2 = \frac{\sigma}{1+\sigma} = 
\left[\frac{\delta_{\rm c_{ms,0}}^{4/3}\delta_w^{2/3}\delta_{\xi'}^{-2/3}-1}
{\delta_{\rm c_{ms,0}}^{4/3}\delta_w^{2/3}\delta_{\xi'}^{-2/3}+1}\right]^2\ ,
\end{equation}
\begin{equation}\hspace{5.5cm}
\label{B_w}
\frac{\sigma}{\sigma_0} = \frac{\rho}{\rho_0} =
\frac{1}{4\sigma_0}\left(\frac{\delta_{\rm c_{ms,0}}^{2/3}\delta_w^{1/3}}{\delta_{\xi'}^{1/3}}
-\frac{\delta_{\xi'}^{1/3}}{\delta_{\rm c_{ms,0}}^{2/3}\delta_w^{1/3}}\right)^2\ ,
\end{equation}
Using this result one can rewrite the integrals in \S~\ref{app:an-int}
in terms of quantities in frame $S'$. In particular, they retain the
same form up to the following simple substitutions:
\begin{equation}\hspace{2.0cm}
t\to t'\ ,\quad\quad
y\to y' = \delta_{\xi'}^{2/3}\ ,\quad\quad
a\to a' = \delta_{\rm c_{ms,0}}^{2/3}\delta_w^{1/3}\quad\quad
y_{\rm min}(t)\to y'_{\rm min}(t') = \delta_{\xi'_*(t')}^{2/3}\ .
\end{equation}
Finally, $\xi'_*(t')$ can either be computed either using the solution
for $\xi_*(t)$ and the relation $\xi'_*(t') =
[\xi_*(t)+\beta_w]/[1+\beta_w\xi_*(t)]$, or by directly generalizing
the derivation from \S~\ref{app:xi*}, as follows:
\begin{equation}\hspace{1.1cm}
\beta'_* \equiv \frac{dx'_*}{dt'} = 
\frac{v'(\xi'_*)+c_{\rm ms}(\xi'_*)}{1+v'(\xi'_*)c_{\rm ms}(\xi'_*)} = 
\frac{\delta_{\rm c_{ms,0}}^{8/3}\delta_w^{4/3}\delta_{\xi'_*}^{2/3}-1}
{\delta_{\rm c_{ms,0}}^{8/3}\delta_w^{4/3}\delta_{\xi'_*}^{2/3}+1}\ ,
\quad\quad\Longrightarrow\quad\quad
\frac{d\delta_{\xi'_*}^2}{d\ln t'} =
\delta_{\xi'_*}^2+1 - \frac{(\delta_{\xi'_*}^2+1)^2}
{\delta_{\rm c_{ms,0}}^{8/3}\delta_w^{4/3}(\delta_{\xi'_*}^2)^{1/3}+1}\ , 
\label{xs*2w}
\end{equation}   
which has the solution
\begin{equation}\label{t'(xi'_*)}\hspace{5.7cm}
\frac{t'}{t'_{0,*}} = (\delta_{\xi'_*}^2+1)
\left[1-\fracb{\delta_{\xi'_*}^2}{\delta_{\rm c_{ms,0}}^4\delta_w^2}^{2/3}\right]^{-3/2}\ ,
\end{equation}   
where $\delta_w\delta_{\rm c_{ms,0}}^{-1}\leq
\delta_{\xi'_*}<\delta_w\delta_{\rm c_{ms,0}}^2$ and
\begin{equation}\hspace{1.7cm}
\frac{t'_{0,*}}{t'_0} = \frac{(\delta_{\rm c_{ms,0}}^4-1)^{3/2}}
{\delta_{\rm c_{ms,0}}^4(\delta_{\rm c_{ms,0}}^2+\delta_w^2)}\ ,\quad\quad\quad\quad
\frac{t'_{0,*}}{t_0} = \frac{t_{0,*}}{\delta_wt_0} = \frac{(\delta_{\rm c_{ms,0}}^4-1)^{3/2}}
{\delta_w\delta_{\rm c_{ms,0}}^4(\delta_{\rm c_{ms,0}}^2+1)} =
\frac{4\sigma_0^{3/4}(1+\sigma_0)^{1/4}}
{\delta_w\left(\sqrt{1+\sigma_0}+\sqrt{\sigma_0}\right)^2}\ ,
\label{eq:t_0*w}
\end{equation}
where we use the notation $t'_0 = t'_b$ and the relation
\begin{equation}\hspace{5.8cm}
\frac{t'_0}{t_0} = \Gamma_w(1-\beta_w c_{{\rm ms},0}) = 
\frac{(\delta_{\rm c_{ms,0}}^2+\delta_w^2)}{\delta_w(\delta_{\rm c_{ms,0}}^2+1)}\ .
\end{equation}
For $\sigma_0 \gg 1$ (and $\delta_w \ll (\delta_{\rm c_{ms,0}}$) we
find that $t'_{0,*}\approx t'_0$ so that at $t' \gg t'_0$ we have
$\delta_{\xi'_*}^2 \approx 2/(1-\xi'_*) \gg 1$ and
\begin{equation}\hspace{0.3cm}
\frac{t'}{t'_0} \approx 
\frac{2}{1-\xi'_*}\left[1-\frac{1}{4\sigma_0^{4/3}\delta_w^{4/3}(1-\xi'_*)^{2/3}}\right]^{-3/2}\ ,
\quad\quad\quad
1-\xi'_* \approx \left\{\matrix{2t'_0/t'
\quad & t' \ll 16\sigma_0^2\delta_w^2t'_0 \ ,\cr\cr 
\frac{1}{8\sigma_0^2\delta_w^2}\left[1+\frac{3}{2}\fracb{t'}{16\sigma_0^2\delta_w^2t'_0}^{-2/3}\right] 
\quad & t' \gg 16\sigma_0^2\delta_w^2t'_0\ ,}\right. 
\label{xs*sol_2}
\end{equation} 
so that $t'_c/t'_0 \sim \sigma_0^2\delta_w^2$ or $t'_c/t'_0 \sim
\sigma_0$ for $\delta_w = \delta_{w,{\rm CM}} \sim
\sigma_0^{-1/2}$. Since $t'_0 \approx t_0/\delta_w$ this implies $t'_c
\approx \sigma_0^2\delta_w t_c$.

It is useful to calculate the ratio of magnetic to kinetic energies,
\begin{equation}\hspace{1.0cm}
\frac{E'_{\rm EM}[>\xi'_*(t')]}{E'_{\rm kin}[>\xi'_*(t')]} = 
\frac{3}{4}\left(\frac{g[(a')^2]-g[ y'_{\rm min}(t')]}
{f[(a')^2]-f[ y'_{\rm min}(t')]}\right)\ ,
\quad\quad\Longrightarrow\quad\quad
\frac{E'_{\rm EM}(t'_0)}{E'_{\rm kin}(t'_0)} = 
\frac{3}{4}
\left[\frac{g(\delta_w^{2/3}\delta_{\rm c_{ms,0}}^{4/3})
-g(\delta_w^{2/3}\delta_{\rm c_{ms,0}}^{-2/3})}
{f(\delta_w^{2/3}\delta_{\rm c_{ms,0}}^{4/3})
-f(\delta_w^{2/3}\delta_{\rm c_{ms,0}}^{-2/3})}\right]\ .
\end{equation}
For the CM frame we have 
\begin{equation}\hspace{1.0cm}\label{eq:delta_wCM} 
\beta_{w,{\rm CM}} = -\beta_{\rm CM} = -
\frac{\sqrt{\sigma_0(1+\sigma_0)}}{2+\sigma_0}\ , \quad\quad\quad\quad
\delta^2_{w,{\rm CM}} =
\frac{2+\sigma_0-\sqrt{\sigma_0(1+\sigma_0)}}{2+\sigma_0+\sqrt{\sigma_0(1+\sigma_0)}}
\approx \frac{3}{4\sigma_0} \approx (2\Gamma_{\rm CM})^{-2} 
\approx 3\delta_{\rm c_{ms,0}}^{-2}\ . \end{equation} 

We also generalize the result for the total energy in the flow,
\begin{equation}\hspace{0.7cm}
\frac{E'[>\xi'_*(t')]}{E'(t'_0)} = 
\frac{\sqrt{1+\sigma_0}}{16(2+\sigma_0)\sigma_0^{3/2}}\fracb{t'}{t'_0}
\frac{1-\beta_w\sqrt{\frac{\sigma_0}{1+\sigma_0}}}
{1+\beta_w\frac{\sqrt{\sigma_0(1+\sigma_0)}}{2+\sigma_0}}
\frac{(a^{\prime\,2}-y'_{\rm min})^3\left[y'_{\rm min}(1+a^{\prime\,6})
+3a^{\prime\,2}(1+a^{\prime\,2}y_{\rm min}^{\prime\,2})\right]}
{a^{\prime\,6}y'_{\rm min}(y_{\rm min}^{\prime\,3}+1)}\ ,
\end{equation}
and the same holds for the energy above some $\xi'>\xi'_*(t')$ where
$y'_{\rm min}(t') = \delta_{\xi'_*(t')}^{2/3}$ is replaced by $y'(t')
= \delta_{\xi'}^{2/3}$. One can verify that this ratio is indeed 1 at
$t' = t'_0$ when $y'_{\rm min} = (\delta_w/\delta_{\rm
c_{ms,0}})^{2/3} = \delta_w/a'$.

\begin{figure}
\includegraphics[width=55mm]{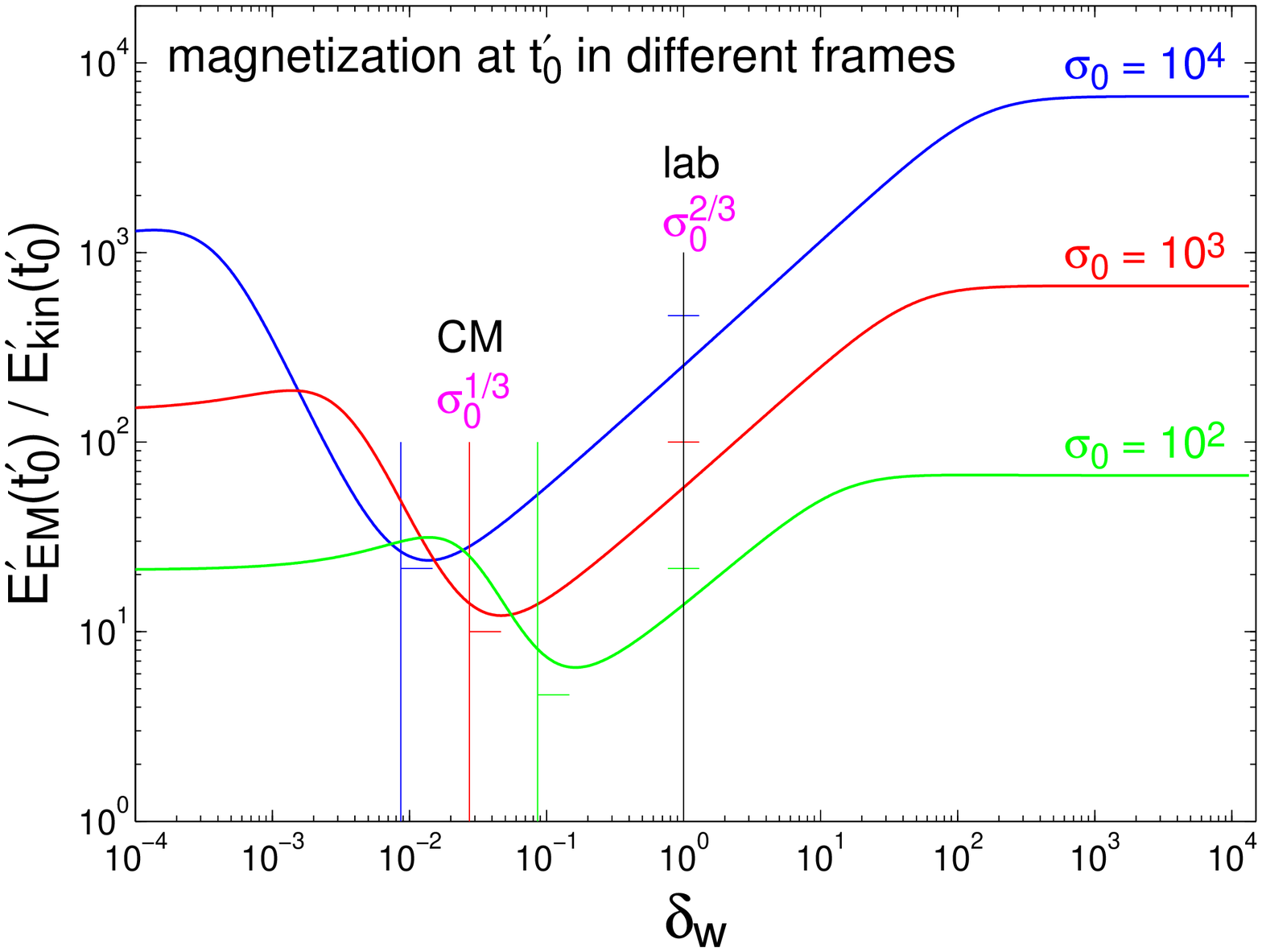}
\hspace{0.4cm}
\includegraphics[width=55mm]{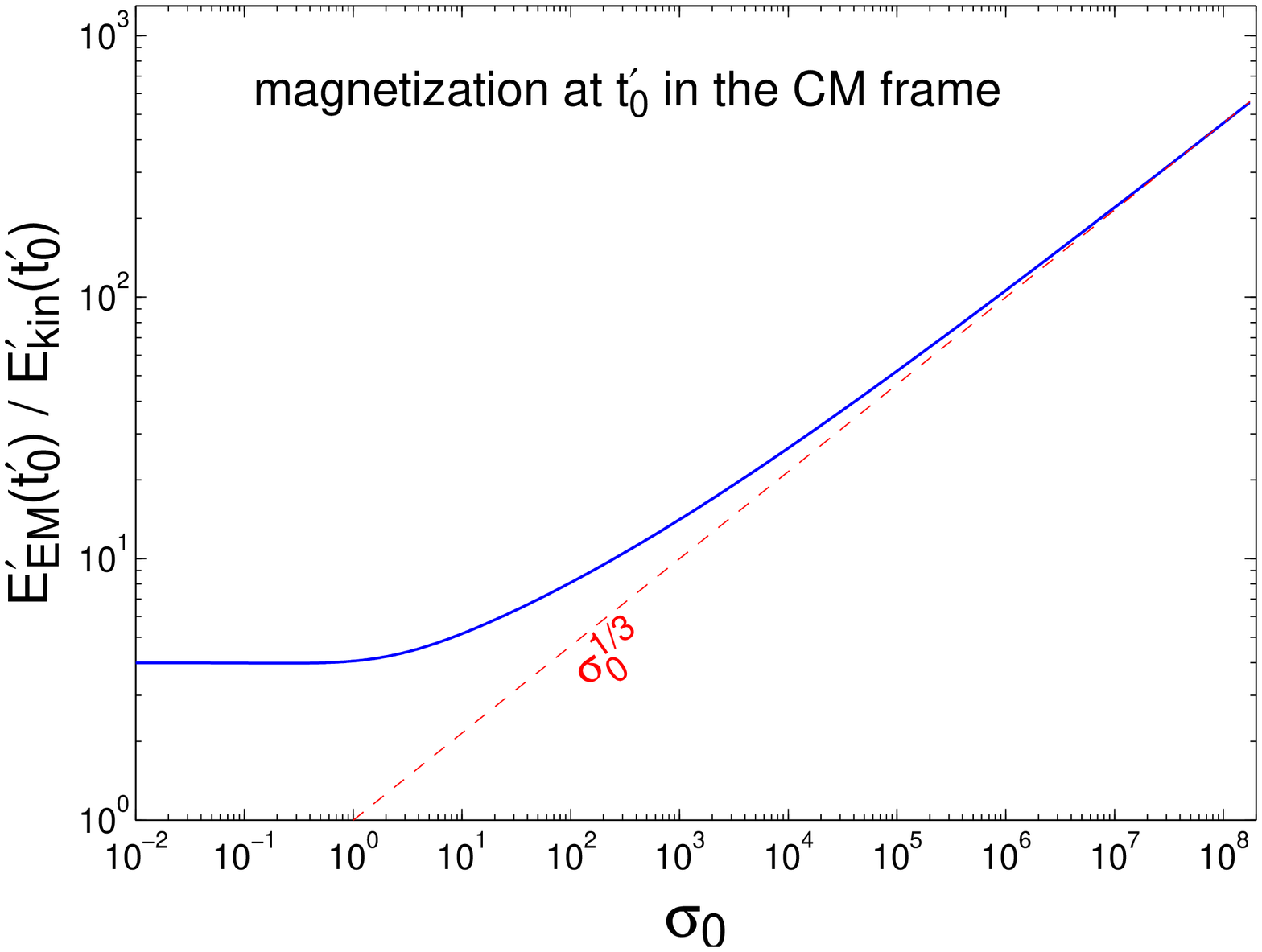}
\hspace{0.4cm}
\includegraphics[width=55mm]{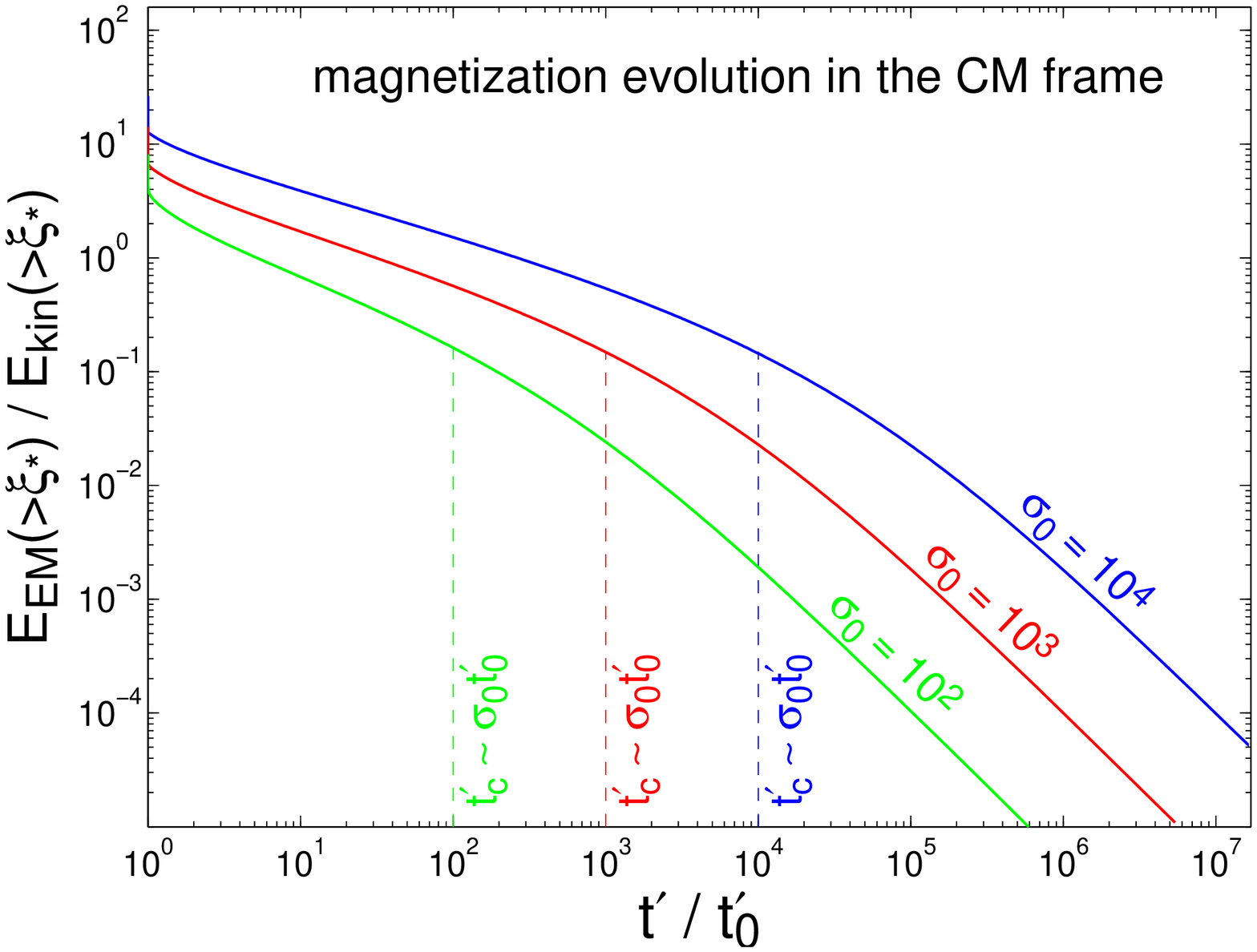}
\caption{{\bf Left panel:} The ratio of electromagnetic to kinetic
energy in a frame moving at $\beta_w$ in the negative $x$-direction
relative to the lab frame (i.e. the rest frame of the initial shell
and the wall) at the time $t'_0$ when the rarefaction wave reaches the
wall, as a function of $\delta_w$, for $\sigma_0 =
10^2,\,10^3,\,10^4$. Indicated by vertical lines are the lab frame and
the CM frame, where this ratio is $\sim\sigma_0^{2/3}$ and
$\sim\sigma_0^{1/3}$, respectively (indicated by short horizontal
lines). {\bf Middle panel:} the same for the CM frame as a function of
$\sigma_0$, using Eq.~(\ref{eq:delta_wCM}); this ratio approaches
$\sigma_0^{1/3}$ ({\it dashed red line}) for $\sigma_0\gg 1$. {\bf
Right panel:} the evolution of the same ratio as a
function of $t'/t'_0$ in the CM frame.}
\label{fig:CM-frame} \end{figure}

Figure~\ref{fig:CM-frame} shows $E'_{\rm EM}/E'_{\rm kin}$ first at
$t'_0$ as a function of the velocity of the primed frame of reference
({\it left panel}), then at $t'_0$ for the CM frame as a function of
$\sigma_0$ ({\it middle panel}), and finally for the CM frame as a
function of time $t'/t'_0$ ({\it right panel}).

\begin{figure}
\includegraphics[width=55mm]{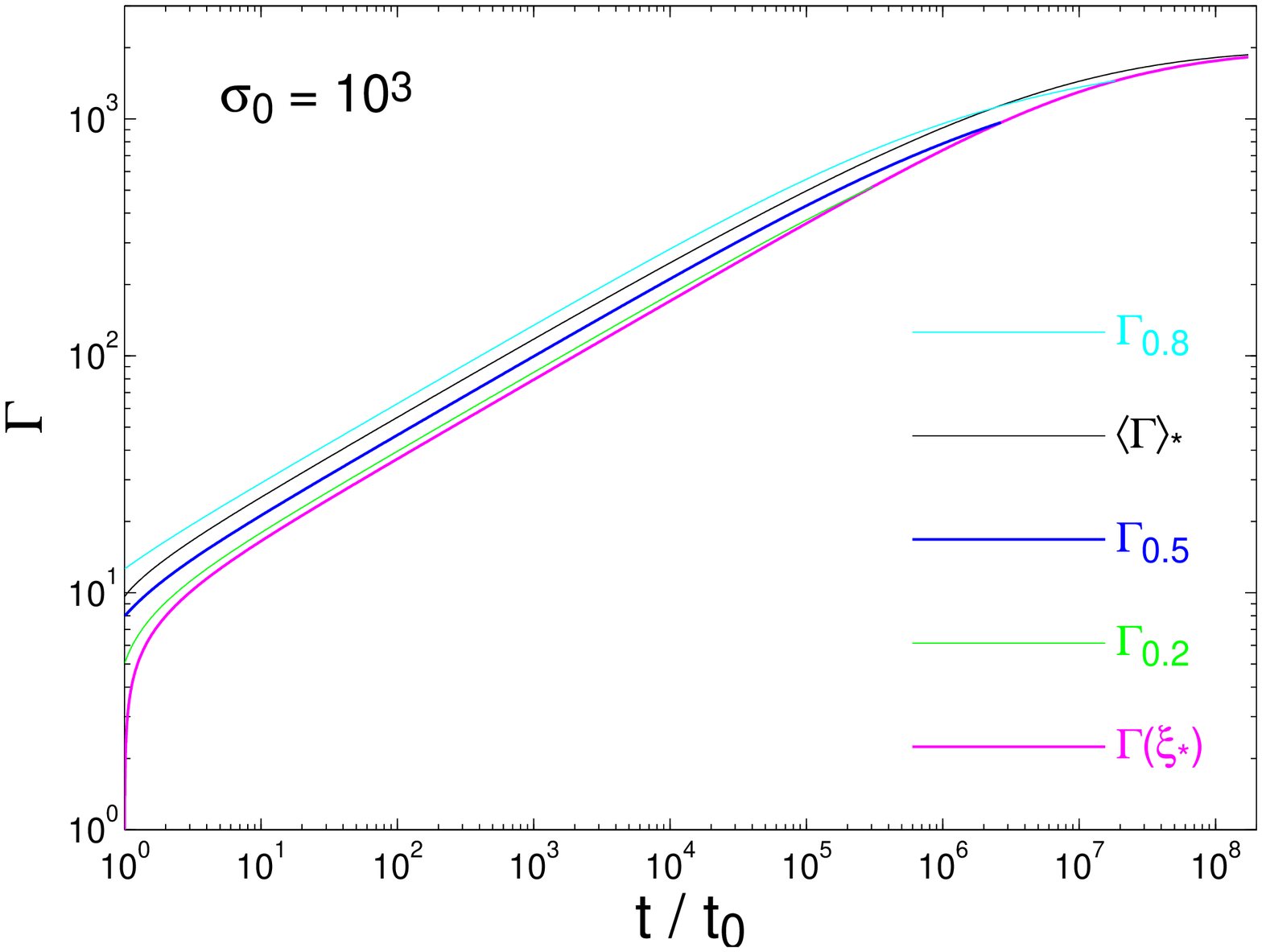}
\hspace{0.4cm}
\includegraphics[width=55mm]{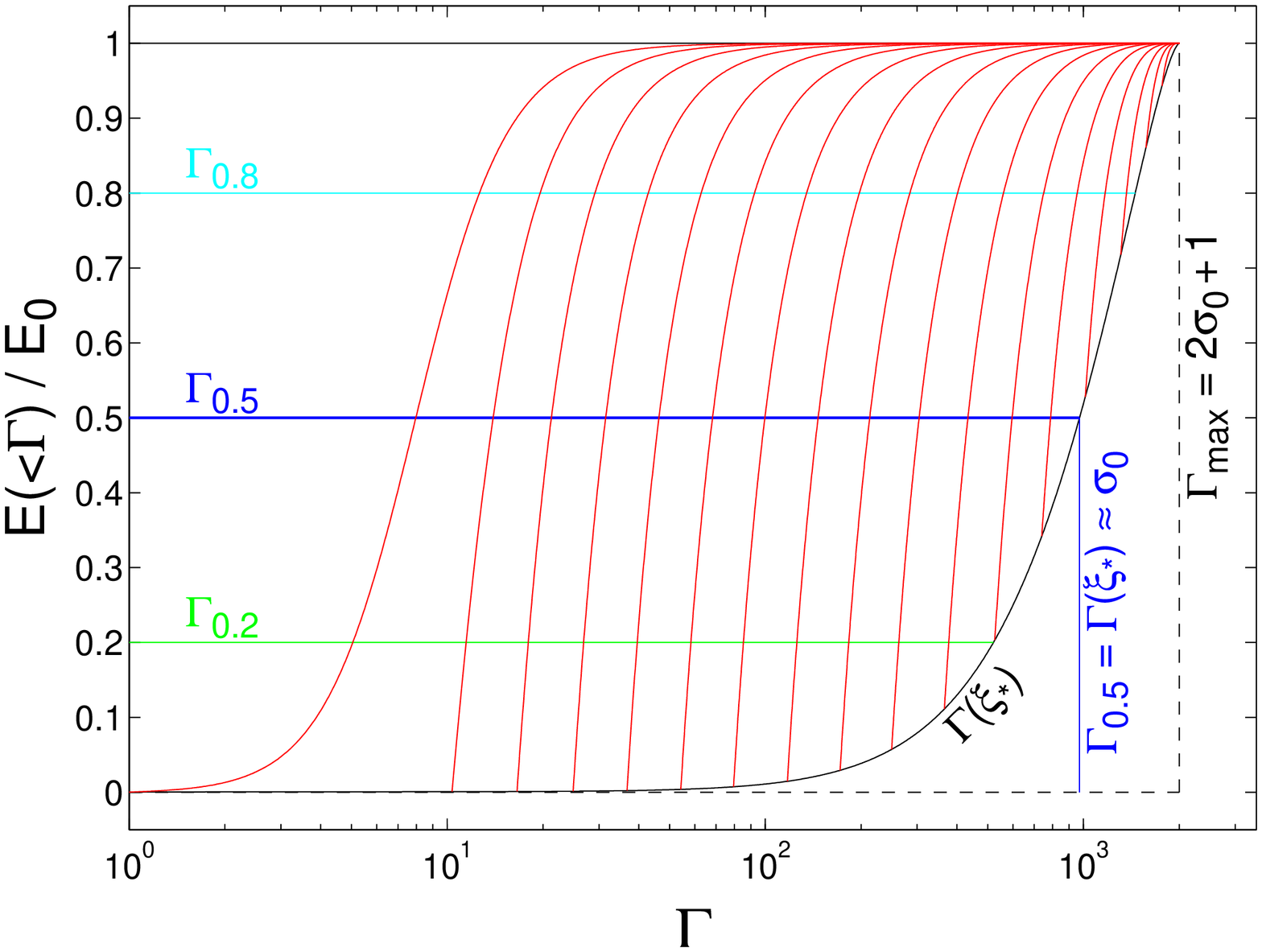}
\hspace{0.4cm}
\includegraphics[width=55mm]{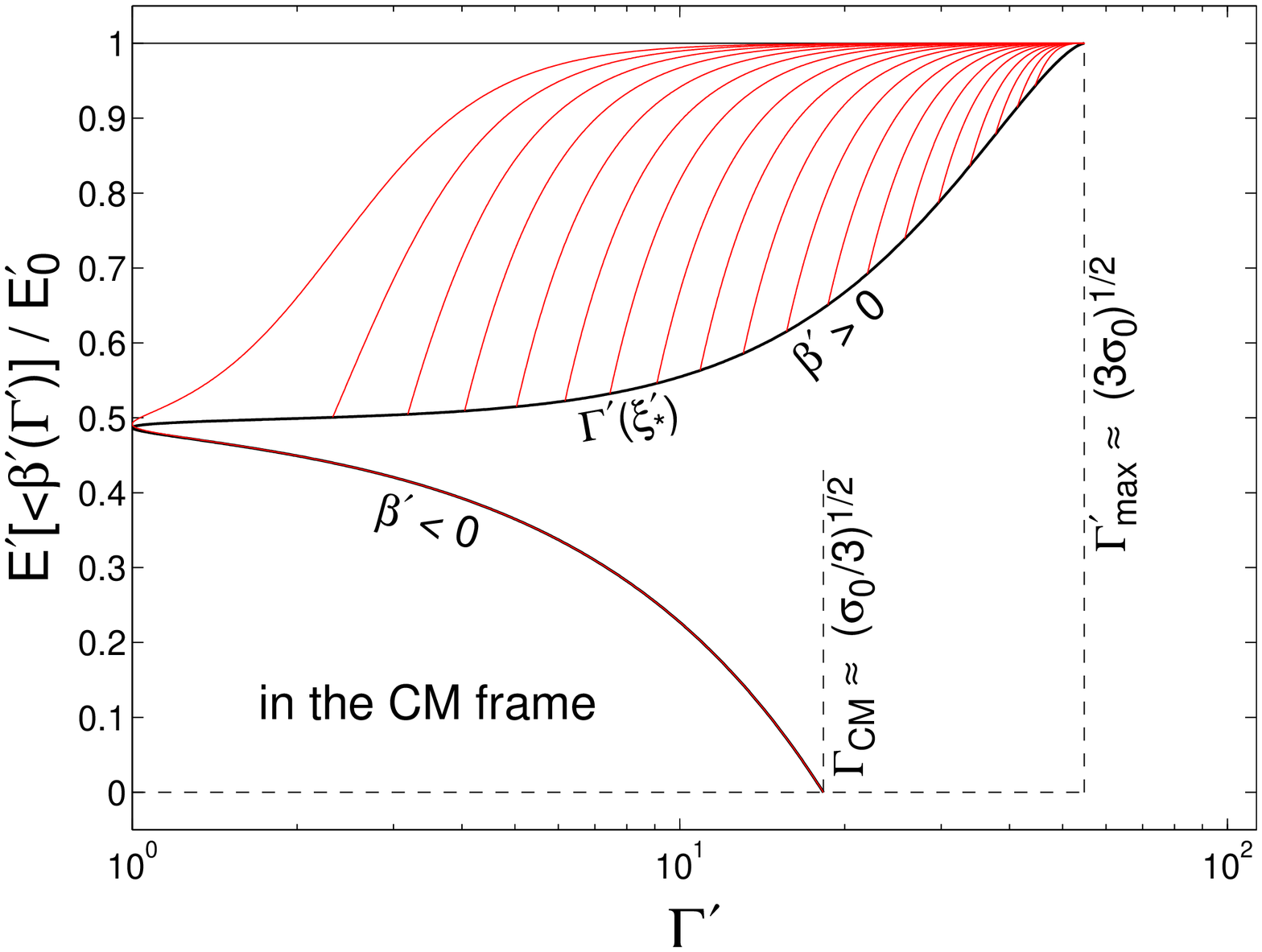}
\caption{
{\bf Left panel:} various estimates for the Lorentz factor as a
function of time for $\sigma_0 = 10^3$; $\Gamma_f$ denotes the Lorentz
factor below which there is a fraction $f$ of the total energy. For
example, $\Gamma_{0.5} = \langle\Gamma\rangle_{E,{\rm med}}$ is the
value for which there is equal energy in faster and slower
material. Also shown for reference are $\Gamma(\xi_*)$ and
$\langle\Gamma\rangle_*$ (which is $\langle\Gamma\rangle_E$ calculated
over the region $\xi>\xi_*(t)$). Note that $\Gamma_f$ can be
calculated analytically only as long as $\Gamma_f\geq\Gamma(\xi_*)$,
since otherwise we must include in the calculation the region $\xi
<\xi_*(t)$, which is not described by the self-similar solution.
{\bf Middle panel:} the corresponding cumulative distribution of the
fraction of energy in the flow as a function of $\Gamma$ for
$\log_{10}(t/t_0) = 0,\,0.5,\,1,...,\,8$ ({\it red lines}). The
lower-right boundary ({\it thin solid black curve}) corresponds to
$\Gamma(\xi_*)$. The horizontal blue line corresponds to $\Gamma_{0.5}
=
\langle\Gamma\rangle_{E,{\rm med}}$, and the vertical blue line shows
the value of $\Gamma$ where it meets with the curve for
$\Gamma(\xi_*)$: $\Gamma_{0.5} = \Gamma(\xi_*) \approx \sigma_0$.
{\bf Right panel:} the corresponding cumulative distribution
calculated in the center of momentum (CM) frame at $\log_{10}(t'/t'_0)
= 0,\,0.25,\,0.5,...,\,4.5$ ({\it red lines}). The bottom branch for
$\Gamma'(\xi'_*)$ ({\it thick solid black line}) and $t' = t'_0$
corresponds to material with a negative velocity in the CM frame,
which initially carries almost half of the total energy in this
frame.}  \label{fig:gamma_med} \end{figure}

Figure~\ref{fig:gamma_med} shows various estimates for the typical
Lorentz factor in the lab frame as a function of time ({\it left
panel}), along with the cumulative distribution of energy as a
function of the flow Lorentz factor (or velocity) at different times,
both in the lab frame ({\it middle panel}) and in the CM frame ({\it
right panel}). The first two panels help to quantitatively address an
important point that has been raised in \S~\ref{sec:BM-effect} in the
discussion around Eqs.~(\ref{eq:beta_E}) and (\ref{eq:vel_av}).
Taking the energy weighted average over $\Gamma$,
$\langle\Gamma\rangle_E$, is not a unique choice, and averaging over
the 4-velocity $u$, $\langle u\rangle_E$, would give a similar result.
However, as shown in \S~\ref{sec:BM-effect} using
$\langle\beta\rangle_E$ would give a very different result, where at
late times $\langle\beta\rangle_E \to \beta_{\rm CM}$ corresponding to
$\Gamma_{\rm CM} = (1-\beta_{\rm CM}^2)^{-1/2} \sim \sigma_0^{1/2}$
while $\langle\Gamma\rangle_E \sim \langle u\rangle_E \sim\sigma_0$.
Fortunately, we can also estimate the typical value of $\Gamma$ of the
material that carries most of the energy in the lab frame without
having to perform any averaging, thus avoiding the need to choose a
specific function of the flow velocity to average over. The {\it left
panel} of Fig.~\ref{fig:gamma_med} shows the median value of $\Gamma$,
$\langle\Gamma\rangle_{E,{\rm med}} =\Gamma_{0.5}$ ({\it thick solid
blue line}) according to Eq.~(\ref{Gamma_av4}), as well as the values
of $\Gamma$ below which there is a fraction $0.2$ ($\Gamma_{0.2}$;
{\it green line}) or $0.8$ ($\Gamma_{0.8}$; {\it cyan line}) of the
total energy. The {\it middle panel} shows the corresponding
cumulative distribution of the fraction of the energy in the flow as a
function of $\Gamma$ at different times.
Most of the energy in the flow is within a narrow range in $\Gamma$,
of less than a factor of 2, around $\langle\Gamma\rangle_{E,{\rm
med}}$. Note that $\langle\Gamma\rangle_{E,{\rm med}}$ is also very
close to $\langle\Gamma\rangle_*$, which is $\langle\Gamma\rangle_E$
calculated over the region $\xi>\xi_*$, and is close to
$\langle\Gamma\rangle_E$ calculated over the whole flow at $t < t_c =
t_0\sigma_0^2$. It can be seen that at the time when
$\langle\Gamma\rangle_{E,{\rm med}} = \Gamma(\xi_*)$ (after which we
can no longer calculate $\langle\Gamma\rangle_{E,{\rm med}}$
semi-analytically), we have $\langle\Gamma\rangle_{E,{\rm med}} =
\Gamma(\xi_*)\approx \sigma_0$, and the Lorentz factor of the plasma 
in the region $\xi>\xi_*$ is between $\approx\sigma_0$ and $\approx
2\sigma_0$. Since at that stage most of the magnetic energy is already
converted into kinetic energy, as $\sigma(\xi_*)\sim 0.1$ (see
Fig.~\ref{fig:shell-int-an}), then this should be close to the
asymptotic value of $\langle\Gamma\rangle_{E,{\rm med}}$ at late
times. Therefore, we can see that all along, from early to late times,
$\langle\Gamma\rangle_{E,{\rm med}}$ is very close to
$\langle\Gamma\rangle_{E}$. This supports the choice of
$\langle\Gamma\rangle_{E}$ as being representative of the typical
Lorentz factor of the material carrying most of the energy in the lab
frame.

The {\it right panel} of Fig.~\ref{fig:gamma_med} shows a calculation
in the CM frame of the cumulative energy ($E'$) in the flow as a
function of its velocity ($\beta'$) or Lorentz factor ($\Gamma'$), and
supports the picture described in \S~\ref{sec:BM-effect} below
Eq.~(\ref{eq:Gamma'_M}).  The evolution of the part of the flow ahead
of the secondary rarefaction wave ($\xi'>\xi'_*$), which is described
by an analytic self-similar solution, is followed from the time $t'_0$
when the original rarefaction wave reaches the wall, and the wall is
replaced by vacuum. At $t'=t'_0$ this region covers the whole flow,
and it is easy to see that in the CM frame almost half of the total
energy ($E'_0$) is carried by material with a negative velocity
($\beta' < 0$). This is expected, since by definition the total
momentum vanishes in the CM frame ($P'=0$), and remains so at later
times as well, when a good part of the total energy is at
$\xi'<\xi'_*$. In the CM frame we have $\langle\sigma'(t'=t'_0)\rangle
\sim \sigma_0^{1/3}$, since in the bulk of the flow $t'_0$ corresponds to 
$t \sim t_0\sigma_0$ in the lab frame and
$\langle\sigma(t=t_0\sigma_0)\rangle \sim \sigma_0^{1/3}$. This means
that the flow is still highly magnetized at $t'_0$, and subsequently
its front part accelerates while its back side decelerates, as the
magnetic energy is transformed into kinetic energy. At
$t'_0<t'<t'_c\sim t'_0\sigma_0$ we have
$\langle\Gamma'\rangle\langle\sigma'\rangle \sim E'_0/M_0 =
E_0/\Gamma_{\rm CM}M_0\sim\sigma_0^{1/2}$ so that the magnetization
drops as $\langle\sigma'\rangle \sim \sigma_0^{1/3}(t'/t'_0)^{-1/3}$
while the typical Lorentz factor increases as $\langle\Gamma'\rangle
\sim\sigma_0^{1/6}(t'/t_0)^{1/3}$. At $t'=t'_c$ the combined acceleration 
at the front and deceleration at the back saturates as the magnetic
and kinetic energies become comparable, $\langle\sigma'\rangle\sim 1$,
and the typical Lorentz factor approaches its asymptotic value,
$\langle\Gamma'\rangle\sim \sigma_0^{1/2}$. Similar to the lab frame,
at $t'>t'_c$ there is also a coasting phase in the CM frame, where
$\langle\Gamma'\rangle\sim\sigma_0^{1/2}$ while the magnetization
continues to drop as $\langle\sigma'\rangle\sim t'_0/t'$. However, in
the CM frame there is comparable mass and energy in material with
$\Gamma'\sim\sigma_0^{1/2}$ moving in the positive and negative
$x'$-directions, so that the total momentum adds up to zero.

\end{document}